\documentclass[AMA,Times1COL]{template} 

\articletype{Research Article}%

\begin{document}

\title{Causal machine learning methods and use of cross-fitting in settings with high-dimensional confounding}

\author[1,2]{Susan Ellul}


\author[3]{Stijn Vansteelandt}

\author[1,2]{John B. Carlin}

\author[1,2]{Margarita Moreno-Betancur}

\authormark{ELLUL \textsc{et al.}}
\titlemark{Causal machine learning methods and use of cross-fitting in settings with high-dimensional confounding}

\address[1]{\orgname{Murdoch Children’s Research Institute}, \orgaddress{\state{Victoria}, \country{Australia}}}

\address[2]{\orgdiv{Department of Paediatrics}, \orgname{University of Melbourne}, \orgaddress{\state{Victoria}, \country{Australia}}}

\address[3]{\orgdiv{Department of Mathematics, Computer Science and Statistics}, \orgname{Ghent University}, \orgaddress{\state{Ghent}, \country{Belgium}}}

\corres{Susan Ellul \email{s.ellul@student.unimelb.edu.au}}



\abstract[Abstract]{Observational epidemiological studies commonly seek to estimate the causal effect of an exposure on an outcome. Adjustment for potential confounding bias in modern studies is challenging due to the presence of high-dimensional confounding, which occurs when there are many confounders relative to sample size or complex relationships between continuous confounders and exposure and outcome. Doubly robust methods such as Augmented Inverse Probability Weighting (AIPW) and Targeted Maximum Likelihood Estimation (TMLE) have the potential to address these challenges, using data-adaptive approaches and cross-fitting, but despite recent advances limited evaluation and guidance are available on their implementation in realistic settings where high-dimensional confounding is present. Motivated by an early-life cohort study, we conducted an extensive simulation study to compare the relative performance of AIPW and TMLE using data-adaptive approaches for estimating the average causal effect (ACE). We evaluated the benefits of using cross-fitting with a varying number of folds, as well as the impact of using a reduced versus full (larger, more diverse) library in the Super Learner ensemble learning approach used for implementation. We found that AIPW and TMLE performed similarly in most cases for estimating the ACE, but TMLE was more stable. Cross-fitting improved the performance of both methods, but was more important for variance estimation and coverage than for point estimates, with the number of folds a less important consideration. Using a full Super Learner library was important to reduce bias and variance in complex scenarios typical of modern health research studies.}


\jnlcitation{\cname{%
\author{Taylor M.},
\author{Lauritzen P},
\author{Erath C}, and
\author{Mittal R}}.
\ctitle{On simplifying ‘incremental remap’-based transport schemes.} \cjournal{\it J Comput Phys.} \cvol{2021;00(00):1--18}.}

\maketitle

\renewcommand\thefootnote{}
\footnotetext{\textbf{Abbreviations:} ACE, average causal effect; AIPW, Augmented inverse probability weighting; CML, causal machine learning; TMLE, Targeted maximum likelihood estimation.}

\renewcommand\thefootnote{\fnsymbol{footnote}}
\setcounter{footnote}{1}

\section{Introduction}\label{sec1}

Estimating the causal effect of an intervention or exposure on an outcome in an observational study requires accounting for multiple sources of confounding. Modern day studies are data-intensive, often collecting a large amount of data, including background, demographic, and biological factors, which in principle should allow for stronger inferences by enabling more extensive control of confounding. However, exploiting these opportunities requires addressing what we call the problem of high-dimensional confounding, which arises when there is a large number of confounders relative to sample size, or a few continuous confounders that may have a complex relationship with the exposure and/or outcome of interest. Analytical adjustment for confounding in these settings can be challenging.

Commonly used methods for the estimation of causal effects, such as g-computation or inverse probability weighting (IPW) are singly robust (SR), meaning they rely upon a single model being correctly specified. \cite{Naimi2023ChallengesAlgorithms, Schuler2017TargetedStudies} G-computation involves the fitting of an outcome model (conditional expectation of the outcome given the exposure and confounders), and IPW an exposure model (probability of exposure given confounders). In the context of high-dimensional confounding, such methods are at risk of misspecification bias because the single model is commonly based on simplistic parametric functions. \cite{Naimi2023ChallengesAlgorithms, Zivich2021MachineEstimators} Specifically, in settings with a large number of confounders (or higher-order terms) and limited sample size, it can be challenging to apply the methods without some form of variable selection or model simplification, and attempting to do this naively can impact the validity of inferences. \cite{Dukes2020HowSelection, Leeb2006CanEstimators} Indeed, many studies reduce to a restricted, manageable adjustment set that may not be sufficient to control confounding, resulting in bias. Additionally, the failure to capture complex nonlinear relationships between variables can potentially lead to further bias.\cite{Schuler2017TargetedStudies, Diaz2019, VanderWeele2019, Groenwold2009QuantitativeStudies} 

To overcome these concerns, it is tempting to consider incorporating data-adaptive approaches to flexibly fit the single model required for SR methods. We define data-adaptive approaches as those where the modelling approach or algorithm used to fit the model is capable of learning from the data, in the sense that the full functional form is not fixed but can adapt. Data-adaptive approaches include machine learning (ML) approaches, and the terms are frequently used interchangeably \cite[(p~44)]{Rose2011}, although data-adaptive can also refer to parametric variable selection approaches like stepwise regression, an approach that we do not consider here. Data-adaptive approaches could potentially provide protection against misspecification bias as well as a way of avoiding inappropriate a priori variable selection in high-dimensional confounding settings. However, data-adaptive estimators typically have non-standard asymptotic behaviour, meaning that they are often non-normal, and their standard errors do not converge to zero at a rate of $1/\sqrt{n}$. As a result, the use of data-adaptive approaches with SR methods is problematic, with this behaviour leading to bias in point estimates and with no valid approach to obtain standard errors. \cite{VanderLaan2006TargetedLearning, Chernozhukov2017MACHINEEffects, Naimi2023ChallengesAlgorithms, Balzer2023InvitedResearch} For example, for g-computation with data-adaptive approaches, the non-parametric bootstrap has been shown to be invalid for variance estimation. \cite{Bickel1997ResamplingRemedies} 
In general, challenges in the construction of CIs \cite{vanderVaart2014HigherFunctions}, 
and underestimated variance and under-coverage have been reported, signalling that misleading inference is a concern. \cite{Naimi2023ChallengesAlgorithms, Balzer2023InvitedResearch}

As promising alternatives to SR methods, Augmented Inverse Probability Weighting (AIPW) \cite{Bang2005DoublyModels, Robins1994EstimationObservedb, Glynn2009} and Targeted Maximum Likelihood Estimation (TMLE) \cite{Rose2011, Lendle2013TargetedAnalysis, Rose2019MachineBiostatistics} involve the fitting of both an outcome and an exposure model. AIPW and TMLE provide consistent estimation if at least one of the two models is consistently estimated, which is why they are referred to as doubly robust (DR). These methods also achieve optimal semi-parametric efficiency if both models are consistently estimated. \cite{Luque-Fernandez2018, Robins2007Comment:Variable} Moreover, in contrast to SR methods, DR methods can validly incorporate the use of data-adaptive approaches to fit both models under some conditions. In particular, they are partially insulated against the slow convergence rates that affect data-adaptive estimators of the exposure and outcome model, provided that both are consistent. Consistency of both estimators is needed to ensure that AIPW and TMLE are root-n consistent when using data-adaptive approaches, meaning the double robustness property holds for consistency, but not for the stronger requirement of root-n-consistency (and this in particular for the validity of standard errors). Cross-fitting (CF), whereby the outcome and exposure models are fit in subsets (folds) of the data and the causal effect estimate is obtained from the remaining data, has been proposed to improve the estimation of standard errors and for valid inference in the context of DR methods.\cite{Chernozhukov2018, Zheng2011}

However, there is limited evaluation, comparison, and guidance on the implementation of what we will henceforth refer to as ``causal machine learning'' (CML) (AIPW and TMLE with data-adaptive approaches) with CF in realistic settings encountered in modern observational studies with high-dimensional confounding. \cite{Meng2022REFINE2:Studies}
Most methodological studies evaluate and compare the methods with and without CF in less realistic settings, with very few or binary confounders only. \cite{Naimi2023ChallengesAlgorithms, Zivich2021MachineEstimators} Empirical studies applying these methods often have large sample size, but many real-world studies have sample size limitations. In general, empirical studies applying these methods have been limited to settings where high-dimensional confounding has not been considered, when perhaps it should have been. \cite{Lendle2013TargetedAnalysis,Kreif2019MachineInference,Decruyenaere2020, Herrera2018} To our knowledge no studies have comprehensively evaluated and compared AIPW and TMLE, with and without CF, whilst considering varying sample sizes in the high-dimensional setting. In addition, there are limited studies that have evaluated how the number of folds used in CF may affect the performance of AIPW and/or TMLE.{\cite{Zivich2024CommentaryLearning}

Here, motivated by a real-world case study, we conduct an extensive simulation study to address these gaps in the context of estimating the average causal effect (ACE), to provide practical guidance on the application of CML methods in realistic settings. The manuscript is organised as follows. Section 2 introduces the motivating case study. Section 3 outlines relevant notation and assumptions and Section 4 provides details on AIPW and TMLE for the estimation of causal effects. Section 5 outlines key considerations for the implementation and evaluation of DR methods.
Section 6 and 7 describe the simulation study design and results. In Section 8, to illustrate, we apply the methods to the case study, following which in Section 9 we discuss our findings and their implications.

\section{Overview of the motivating case study}\label{sec2}

The motivating example draws on data from the Barwon Infant Study (BIS), a birth cohort study with antenatal recruitment, conducted in the south-east of Australia. \cite{Vuillermin2015CohortStudy} Infants included in the study were reviewed at multiple time points in early life (including at birth, 12 months, and 4 years of age). Further details regarding eligibility, recruitment criteria and measures obtained for BIS are provided elsewhere. \cite{Vuillermin2015CohortStudy}

One focus area for research in BIS is the developmental origins of cardiovascular disease (CVD). There is growing evidence for a role of inflammation in CVD. \cite{Sorriento2019InflammationFindings,DeWeerdt2021InflammationEnough} Chiesa et al \cite{Chiesa2022GlycoproteinYoung} found that inflammation (as measured by a biomarker called Glycoprotein acetyls (GlycA) obtained directly from blood samples) was associated with adverse cardiovascular profiles in adolescence, and also predicted future risk. In adults, pulse wave velocity (PWV), a measure of arterial stiffness (higher PWV can indicate greater arterial stiffness), has been shown to predict CVD events. \cite{Sutton-Tyrrell2005ElevatedAdults, Sequi-Dominguez2020AccuracyMeta-Analysis} It is therefore of interest to examine the  effect of early life inflammation on PWV at a later age in childhood (post infancy). The motivating example focusses specifically on the following research question: \textit{What is the effect of early life inflammation, as measured by GlycA in 1-year-old infants, on PWV at 4 years of age?} We standardise the outcome (PWV) in the BIS data, and consider GlycA as a dichotomous exposure, indicating either high or low inflammation. We dichotomise the exposure to make results relevant for the majority of applications that currently focus on binary exposures. The literature has not established a definitive cutoff for GlycA that would be indicative of high inflammation in infants so we examined the impact of having inflammation in the top quartile of the distribution, dichotomising GlycA at the 75th percentile.

To address the proposed causal question, one must consider the potential confounding role of background factors (e.g., demographic, environmental, familial, and perinatal) and other metabolomic factors, as depicted in the directed acyclic graph (DAG) of Figure~\ref{fig:BIS_DAG} developed using subject-matter expertise. In BIS, metabolomic measures at 1 year of age and PWV at 4 years of age were only obtained for a subset of the participants by design. Details on missing data are provided in Table S1 of Section 1 of the Supporting Information. Only participants with complete data (no missing data on any of the relevant variables, n=252) were included in the case study, as evaluation of challenges associated with the presence of missing data were not considered here (see Section 9). Characteristics of the infants included are outlined in Table~\ref{table:BIS_descriptive}. Here, we have clearly identified a problem of high-dimensional confounding because we have a sample size of 252 and up to 87 potential confounders, with 78 of these being continuous confounders. 


\begin{figure}[h]
\centering
\includegraphics[width=18cm]{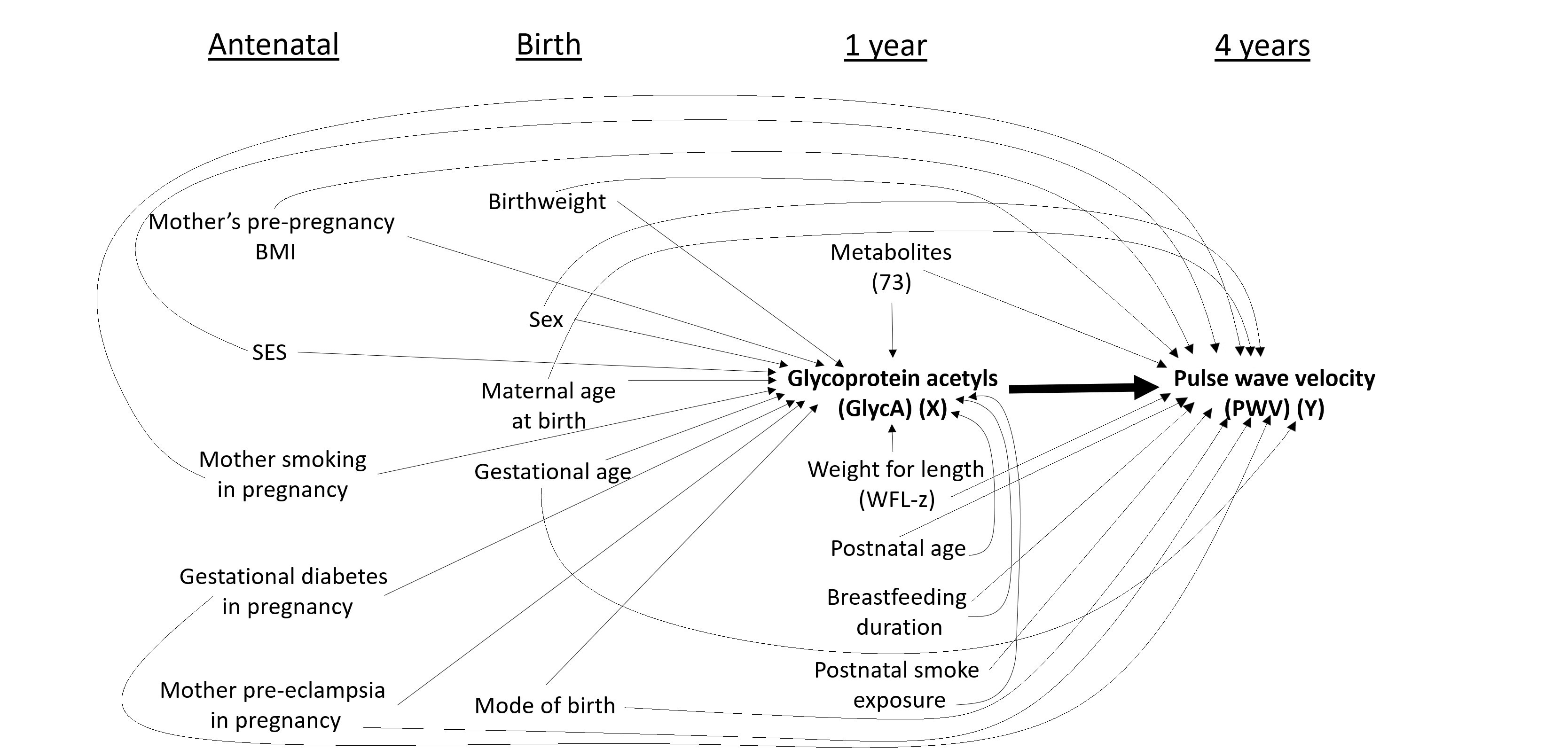}
\caption{Directed Acyclic Graph (BIS DAG) for the Barwon Infant Study motivating example, and as used for data-generation in the simulation study.}
\label{fig:BIS_DAG}
\end{figure}

\newpage


\scriptsize
\begin{longtable}{lccc}
\caption{Characteristics of participants in the Barwon Infant Study (BIS) analytic sample used in the case study.\label{table:BIS_descriptive}} \\
\toprule
& \multicolumn{2}{c}{\textbf{Inflammation\^{}}} & \\ 
 \cmidrule(lr){2-3}
\textbf{Characteristic} & \textbf{Low}, N = 189\textsuperscript{1} & \textbf{High}, N = 63\textsuperscript{1} & \textbf{Overall}, N = 252\textsuperscript{1} \\ 
\midrule
Pre-pregnancy BMI (kg/m\textsuperscript{2}) & 24.5 [21.7, 27.8] & 23.5 [21.3, 28.4] & 24.3 [21.7, 27.9] \\ 
Socio-Economic Indexes for Areas (SEIFA) &  &  &  \\ 
\hspace{0.5cm} Low & 65 (34\%) & 19 (30\%) & 84 (33\%) \\ 
\hspace{0.5cm} Med & 63 (33\%) & 21 (33\%) & 84 (33\%) \\ 
\hspace{0.5cm} High & 61 (32\%) & 23 (37\%) & 84 (33\%) \\ 
Mother smoking in pregnancy & 22 (12\%) & 10 (16\%) & 32 (13\%) \\ 
Gestational diabetes in pregnancy & 8 (4.2\%) & 3 (4.8\%) & 11 (4.4\%) \\ 
Pre-eclampsia in pregnancy & 7 (3.7\%) & 2 (3.2\%) & 9 (3.6\%) \\ 
Birthweight (grams) & 3,531 (521) & 3,442 (581) & 3,508 (537) \\ 
Infant sex &  &  &  \\ 
\hspace{0.5cm} Female & 82 (43\%) & 37 (59\%) & 119 (47\%) \\ 
\hspace{0.5cm} Male & 107 (57\%) & 26 (41\%) & 133 (53\%) \\ 
Maternal age at birth (years) & 32.1 (4.3) & 32.8 (3.9) & 32.3 (4.2) \\ 
Gestational age at birth &  &  &  \\ 
\hspace{0.5cm} 32-36 completed weeks & 9 (4.8\%) & 4 (6.3\%) & 13 (5.2\%) \\ 
\hspace{0.5cm} 37-42 completed weeks & 180 (95\%) & 59 (94\%) & 239 (95\%) \\ 
Mode of birth &  &  &  \\ 
\hspace{0.5cm} Caesarean & 75 (40\%) & 20 (32\%) & 95 (38\%) \\ 
\hspace{0.5cm} Vaginal & 114 (60\%) & 43 (68\%) & 157 (62\%) \\ 
Weight-for-length z-score at 12 months & 0.72 (1.09) & 0.75 (1.04) & 0.72 (1.08) \\ 
Age at 12-month measures (months) & 12.93 (0.80) & 12.97 (0.75) & 12.94 (0.79) \\ 
Breastfeeding duration (exclusive weeks*) & 8 [1, 22] & 4 [0, 24] & 7 [0, 22] \\ 
Postnatal smoke exposure & 29 (15\%) & 10 (16\%) & 39 (15\%) \\ 
\bottomrule
\end{longtable}
\vspace{-4.5mm}
\begin{minipage}{\linewidth}
\hspace{16mm}\textsuperscript{1}Mean (SD), Median {[}IQR{]} or Frequency (\%) as appropriate.
\end{minipage}
\begin{minipage}{\linewidth}
\vspace{2mm}\hspace{20mm}*Number of weeks that infant was exclusively breastfed (i.e. no supplementary feeding).\\
\end{minipage}
\begin{minipage}{\linewidth}
\vspace{-2mm}\hspace{20mm}\^{}GlycA at 1-year of age, dichotomised using the 75th percentile of the observed distribution as the threshold.\\
\end{minipage}

\section{Notation and assumptions}\label{sec3}

\normalsize

We consider a binary exposure indicator $X$, coded 1 for exposed (in the motivating example, 1=high GlycA), and 0 for the unexposed (0=low GlycA), continuous outcome $Y$ (in the example, $Y$ is the standardised PWV at 4 years of age) and a vector of confounders, $W$ (based on subject-matter expertise). Let $Y^{X=x}$ be the potential outcome under exposure $x$.

\subsection{Causal estimand} \label{estimand}

We focus on the average causal effect (ACE), defined as $E[Y^{X=1}] - E[Y^{X=0}]$, interpreted as the difference in average potential outcome in the population of interest when (a) everyone is exposed versus when (b) everyone is unexposed. \cite{Hernan2004AResearch, Rubin2005CausalDecisions} In the motivating example, this is the difference in average PWV that would be seen at 4 years of age if everyone in the target population were set to have high versus low GlycA at 1 year of age.

\subsection{Identification} \label{identification}

In the absence of missing data, under the assumptions of exchangeability ($Y^{X=x}$ $\indep$ $X|W$ for $x=0,1$), consistency ($Y^{X=x}=Y$ when $X=x$ and $x=0,1$), and positivity ($P[P(X=x|W) > 0]=1$ for $x=0,1$), we can identify the ACE from observable data as $E[E(Y|X=1,W)-E(Y|X=0,W)]$ which is the g-formula. \cite{Hernan2023, Schuler2017TargetedStudies} It is worth noting that some of these assumptions are debatable in the motivating example, particularly the consistency assumption given the lack of a well-defined intervention. However, we assume they hold for the remainder of the manuscript given our focus is on examining the performance of estimators under those conditions (also see Section 9).

\section{Doubly robust methods}
\label{sec4}

Under the assumptions outlined in Section \ref{sec3}, the ACE can be estimated using DR methods.  
We focus on two DR methods, AIPW and TMLE. 
For both of these methods, models are fitted for the conditional expectation of the outcome $Y$ given the exposure $X$ and confounders $W$, $E[Y|X, W]$ (the outcome model) and for the propensity score, ${P}(X=1|W)$ (the exposure model). The outcome and exposure models are often referred to as \textit{nuisance} models because they are not of intrinsic interest but instead are used within AIPW and TMLE to estimate the target parameter. \cite{Daniel2018DoubleRobustness}

DR estimators for the ACE are obtained by determining and then utilising an efficient influence function (EIF), which has a unique form that is determined by the target parameter of interest.\citep{Kennedy2016} DR estimators are constructed in a manner that allows them to attempt to correct for the bias (termed \emph{plug-in bias}) that is induced when using g-computation with the data-adaptive approaches. Further details regarding derivation and interpretation of the EIF, as well as plug-in bias can be found elsewhere. \cite{Hines2022DemystifyingFunctions}

\subsection{Augmented Inverse Probability Weighting (AIPW)}

AIPW is often referred to as a one-step correction or one-step
estimation approach \cite{Benkeser2017} and is based on directly subtracting an estimate of the bias term from the g-computation estimator. Based on the fitted outcome model, predicted values, $\hat{E}_1(w) = \hat{E}[Y|X=1,W=w]$ and $\hat{E}_0(w) = \hat{E}[Y|X=0,W=w]$ are obtained for each record, for $X=1$ and $X=0$ respectively. The estimate of the ACE is then calculated as

\normalsize
\begin{align} 
    \begin{split}
        \hat{\psi}_{\mathrm{AIPW}} &= \frac{1}{n}\sum_{i=1}^{n} 
        \Biggl[
        \hat{E}_1(w_i) - \hat{E}_0(w_i) +
        X_i\frac{(Y_i-\hat{E}_1(w_i))}{\hat{P}(w_i)} - (1-X_i)\frac{(Y_i-\hat{E}_0(w_i))}{1-\hat{P}(w_i)}
        \Biggr]. \label{AIPW_ACE_eqn}
    \end{split}
\end{align}
\normalsize

where $\hat{P}(w) = \hat{P}(X=1|W=w)$ is an estimate of the propensity score based on the fitted exposure model. The estimated variance of the AIPW estimator is obtained as follows: \cite{Wager2022STATSInference, Chernozhukov2018}

\normalsize
\begin{align} 
    \begin{split} 
  \widehat{var}(\hat{\psi}_{AIPW})
  = \frac{1}{n-1}\sum_{i=1}^{n}\Biggl[
   \hat{E}_1(w_i) - \hat{E}_0(w_i) +
        X_i\frac{(Y_i-\hat{E}_1(w_i))}{\hat{P}(w_i)} - (1-X_i)\frac{(Y_i-\hat{E}_0(w_i))}{1-\hat{P}(w_i)} - \hat{\psi}_{AIPW} 
  \Biggr]^2. \label{eqn:AIPW_var}
  \end{split} 
\end{align} 
\normalsize

\subsection{Targeted Maximum Likelihood Estimation}

In contrast to AIPW, after initial estimation of the outcome and exposure models, TMLE has a targeting or updating step, again with the purpose of correcting the bias of the g-computation estimator. Unlike AIPW, TMLE respects the bounds of the estimand’s parameter space. \cite{Zheng2011} TMLE can be implemented using a procedure already well described in the literature (e.g., Schuler and Rose\cite{Schuler2017TargetedStudies}, Luque-Fernandez et al \cite{Luque-Fernandez2018}), whereby the fitted exposure score model is used to generate a ``clever'' covariate. Using this, \emph{targeted} or updated predicted outcomes  $\hat{E}^*_1(w)$ and $\hat{E}^*_0(w)$ are constructed for $X=1$ and $X=0$, respectively, for each record, and these are used to estimate the ACE, with the estimator defined as

\normalsize
\begin{equation}\label{estimate_causal}
\hat{\psi}_{\mathrm{TMLE}} = \frac{1}{n}\left[\sum_{i=1}^{n} \hat{E}^*_1(w_i) - \sum_{i=1}^{n} \hat{E}^*_0(w_i)\right].
\end{equation}

The variance for TMLE is obtained similarly to AIPW by using (\ref{eqn:AIPW_var}), with $\hat{\psi}_{TMLE}$ in place of $\hat{\psi}_{AIPW}$, and with $\hat{E}^*_1(W)$ and $\hat{E}^*_0(W)$ in place of $\hat{E}_0(W)$ and $\hat{E}_0(W)$, respectively.

\section{Data-adaptive estimation of nuisance functions} \label{sec5}

When applying and evaluating the DR methods there are key implementation considerations. In this section, we briefly outline key considerations that are explored in this paper.

\subsection{Data-adaptive approaches}

Data-adaptive approaches can be used within the DR methods to estimate each of the nuisance functions, with the predictive performance of the approaches being the key criterion of interest. Therefore, nuisance function estimation can be viewed as a prediction modelling problem. 
One can consider parametric data-adaptive approaches, 
such as penalised regression methods (e.g., least absolute shrinkage and selection operator (LASSO) \cite{Tibshirani1996RegressionLasso} \citealp[(p~64-65)]{Hastie2001ThePrediction}) wherein coefficients can be shrunk towards zero (or in some cases, set to zero), and are controlled or influenced by decision rules of the method. 
One can also consider non-parametric data-adaptive approaches, which include tree-based (e.g., random forest (RF) \cite{Breiman2001RandomForests}) and non-tree-based (e.g. neural networks \cite[(p~141-145)]{Kuhn2013AppliedModeling}, support vector machines \cite[(p~337-389)]{Hastie2001ThePrediction}) methods, and often allow greater flexibility than the parametric approaches. In this work we consider both parametric and non-parametric data-adaptive approaches.

\subsubsection{Ensemble Learning} 

In seeking to improve predictive performance, it is common to use an ensemble learning approach \cite[(p~15)]{Zhou2012EnsembleAlgorithms}, wherein multiple data-adaptive approaches, termed learners, are considered simultaneously. The group of learners considered for the prediction problem is called a candidate library, and the learners within are used to solve the same prediction problem, but predictions from the learners are combined to arrive at a solution with strengthened predictive ability. 

In this work, we use the Super Learner (SL), a widely used ensembling approach in the context of DR estimation of causal effects. \citealp{Zivich2021MachineEstimators, Schuler2017TargetedStudies, Lendle2013TargetedAnalysis, Kreif2019MachineInference, Decruyenaere2020, Herrera2018, Meng2022REFINE2:Studies}

We consider two libraries of differing diversity containing non-adaptive parametric approaches (e.g. GLMs) as well as data-adaptive parametric and non-parametric approaches. Further details are provided in Section 6.3.

\subsection{Cross-fitting}

Cross-fitting (CF) has been proposed as an approach to help overcome issues that may arise from overfitting when using CML approaches. \cite{Kennedy2016, Zivich2021MachineEstimators} In this section, we outline the form of CF \cite{Chernozhukov2018, Kreif2019MachineInference, Jacob2020Cross-FittingEffects, Chernozhukov2020LocallyEstimation, Chernozhukov2017MACHINEEffects} that we consider in this work.

CF (or K-fold CF) is where the sample of size $n$ is split randomly into $K \geq 2$ parts (folds) of roughly equal size ($n/K$). The nuisance models are fitted on all but one fold (the complement), and the remaining fold is used to get predictions from these models. The process is repeated $K$ times, so that each fold $k=1,2,. . . ,K$ is used to obtain predictions once (by rotation). The predictions are then used within the given DR method to obtain a final estimate of the causal effect. In this study, when CF was applied, standard errors were calculated from the EIF (refer to Section 2 of the Supporting Information for implementation details). \cite{Chernozhukov2017MACHINEEffects} For TMLE with CF, targeting was performed in the remaining data that was not used to fit the nuisance models, an approach consistent with van der Laan and Rose \cite{Rose2011}, Motoya et al\cite{Montoya2023TheJustice} and Balzer and Westling \cite{Balzer2023InvitedResearch}, although we note that there are alternative implementations that retain the same theoretical properties. \cite{Levy2018AnCV-TMLE}

\section{Simulation study: design and methods}
\label{sec6}

\subsection{Aim}

We aimed to compare the performance of AIPW and TMLE using SL with different choices of library (reduced vs full) and different implementations of CF (no CF, and with 2, 5 and 10 folds) in realistic scenarios, largely informed by the BIS case study. 
We considered a range of scenarios, varying data generating mechanism complexity, confounder set size (p=14 or 87), and sample size (n=200, 500, 1000, 2000). The \textit{small} confounder set was modelled on demographic and background confounders in the BIS study, and the \textit{large} set additionally considered metabolites (see Section 3 of the Supporting Information (Tables S2-S4) for more details).

\subsection{Data-generating mechanisms}

The variables (confounders, exposure, and outcome) were generated in the order specified in Section 4 of the Supporting Information (Tables S5-S7).  Parametric regression models were used to simulate variables for all five data-generating mechanisms, three of which involved the small confounder set (\textit{simple-1}, \textit{complex-1a} and \textit{complex-1b}), and two of which involved the large confounder set (\textit{simple-2}, \textit{complex-2}), detailed below. 
Parameters for the data-generation models were, unless otherwise stated, obtained by fitting analogous models to the BIS data. 

Under the simple data-generating mechanisms (\textit{simple-1} and \textit{simple-2}), we considered exposure and outcome regression models with main effects only and a linear association for the continuous covariates. Complex data-generating mechanisms considered exposure and outcome models with main effects and two-way interactions (confounder-confounder and exposure-confounder as appropriate), with linear and non-linear associations for the continuous covariates. For the small confounder set, we used two complex scenarios, with coefficient values for interaction terms two times larger than observed in the BIS data (\textit{complex-1a}) and another with coefficients for the interaction terms four times larger (\textit{complex-1b}) than observed in the BIS data. For the large confounder set, we considered one complex scenario with coefficient values as observed in the BIS data (\textit{complex-2}). For full details regarding the data-generating mechanisms refer to Section 4 of the Supporting Information (Tables S6-S11). 

We simulated 2000 datasets for each data generation mechanism and sample size, with this number of simulation replications determined based on a requirement that the Monte Carlo SE for a coverage of 95\% be no greater than 0.5\%. Overall, 20 scenarios were considered (four sample sizes for five data-generating mechanisms). 
When generating the outcome in each scenario, we set the main effect to be of a size determined such that the null hypothesis of zero average causal effect was formally rejected (using the threshold of $p < 0.05$) in approximately 80\% of the simulated datasets for the given sample size. We chose to keep the power constant to ensure that coverage probabilities, our main metric of interest, were comparable across sample size scenarios. The intercepts in all outcome models were modified so that the mean of $Y$ remained at 0. Further details on the parameters used in the generation of confounders, exposure, and outcome are provided in Section 4 of the Supporting Information (Tables S12-S23).

\subsection{Estimand and methods compared}

We estimated the average causal effect (ACE) of $X$ on $Y$ via AIPW and TMLE with no CF and with CF, using varying folds ($2$, $5$ and $10$). SL was used with 10-fold cross-validation for outcome and exposure prediction modelling. Reduced (less flexible) and full (more flexible) candidate libraries for SL were considered using default hyperparameters for each learner. The reduced library was chosen to include less flexible approaches. Specifically, it included the following non-adaptive and data-adaptive parametric approaches: GLM, GLM with pairwise interactions, Bayesian GLM, generalised additive modelling and GLM with Lasso/Elastic net regularisation. The full library was comprised of the approaches in the reduced library and an additional set of more flexible, non-parametric data-adaptive approaches, including learners more likely to result in biased standard errors without the use of CF. Further details on the SL candidate libraries are provided in Section 5 of the Supporting Information (Table S24). To deal with extreme inverse probability weights, truncation of propensity score predictions at the 5th and 95th percentiles was undertaken.

\subsection{Performance measures}

For each scenario, to evaluate performance, we calculated the bias (the difference between the average of the ACE estimates across the 2000 simulations and the true value of the ACE), the relative bias (bias divided by the true ACE, as percent) as well as the empirical standard error, average model-based standard error (model SE) and relative error of the model-based SE compared to the empirical SE (\%). We also estimated the coverage probability of the 95\% Wald CI. To calculate performance measures, the true values of the ACE used in each scenario were obtained empirically in a single very large simulated dataset (full details and true values used are provided in Section 6 of the Supporting Information (Table S25)). We also estimated the Monte Carlo Standard Errors (MCSE's) for each performance measure. Morris et al \cite{Morris2019} provide more detail on the performance measures and how they are obtained. 

\section{Simulation study: results}
\label{sec7}

Our presentation and assessment of performance focus on the reporting of bias in point estimates (relative bias), the empirical SE, relative error in the model-based SE, and coverage probability. In general, patterns for the \textit{complex-1b} scenario were similar to complex-1a. Hence, Figures \ref{fig:Relbias_red}-\ref{fig:CP_red} show performance measures for AIPW and TMLE when the reduced and full libraries were applied in SL across sample sizes, for the small confounder set (\textit{simple-1} and \textit{complex-1a}) and the large confounder set (\textit{simple-2} and \textit{complex-2}). Tables S26-S34 in Section 7 of the Supporting Information contain results for both the reduced and full libraries across all scenarios, for the comprehensive panel of performance measures, including associated Monte Carlo standard errors. 
This section mainly describes the results, and a fuller discussion of the findings is provided in Section 9.


\begin{figure}[h!]
\centering
\includegraphics[width=15cm, height=15cm]{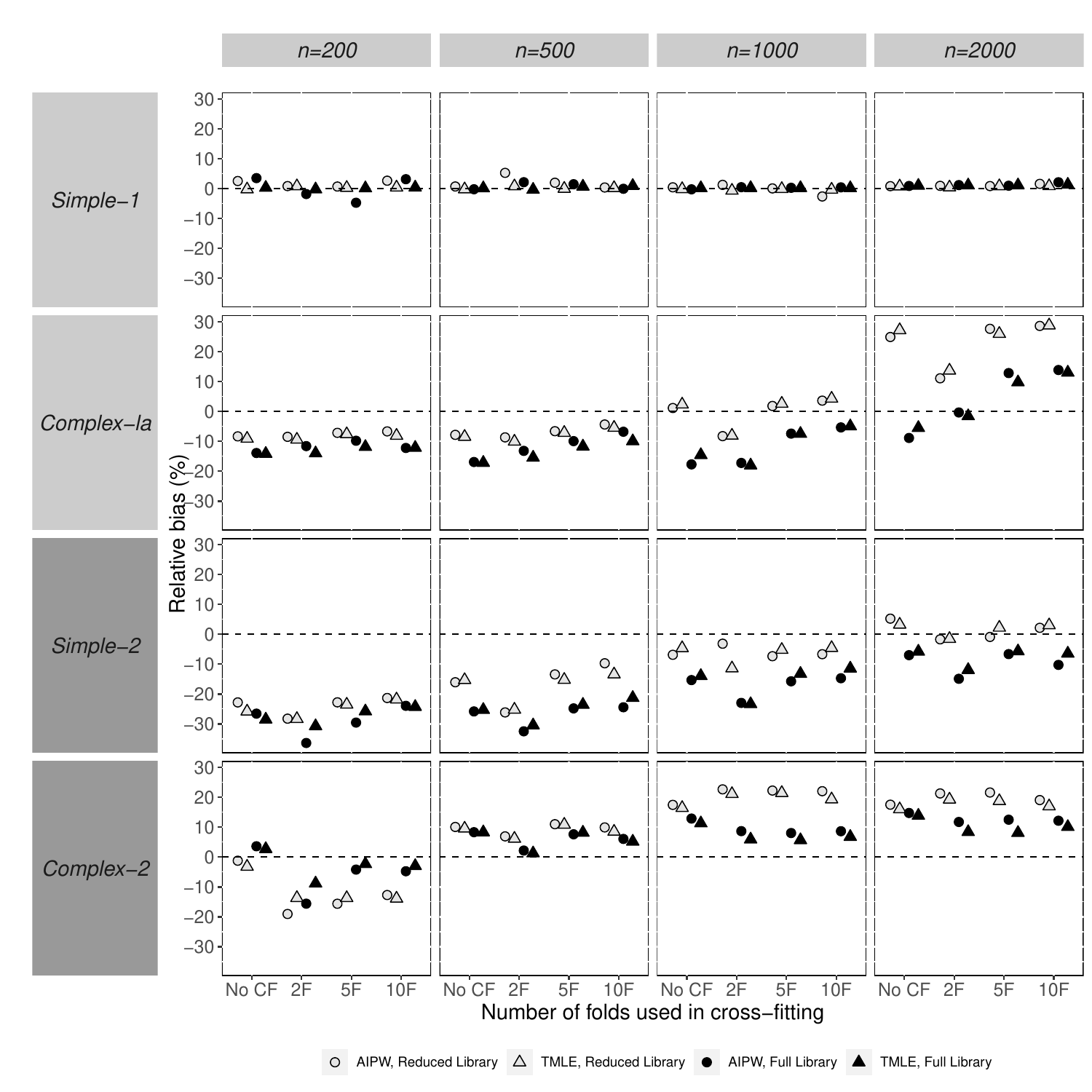}
\caption{Simulation study results for the relative bias of point estimates (\%) for AIPW and TMLE with varying use of cross-fitting, by scenario (data generating mechanism and sample size). Section 5 of the Supporting Information (Table S25) provides the true value of the ACE used for each scenario. Note: results from a small number of datasets were excluded for AIPW (but not TMLE) (refer to Section \ref{method_compare}, and Section 5 of the Supporting Information for more details).
}
\label{fig:Relbias_red}
\end{figure}

\newpage


\begin{figure}[h!]
\centering
\includegraphics[width=15cm, height=15cm]{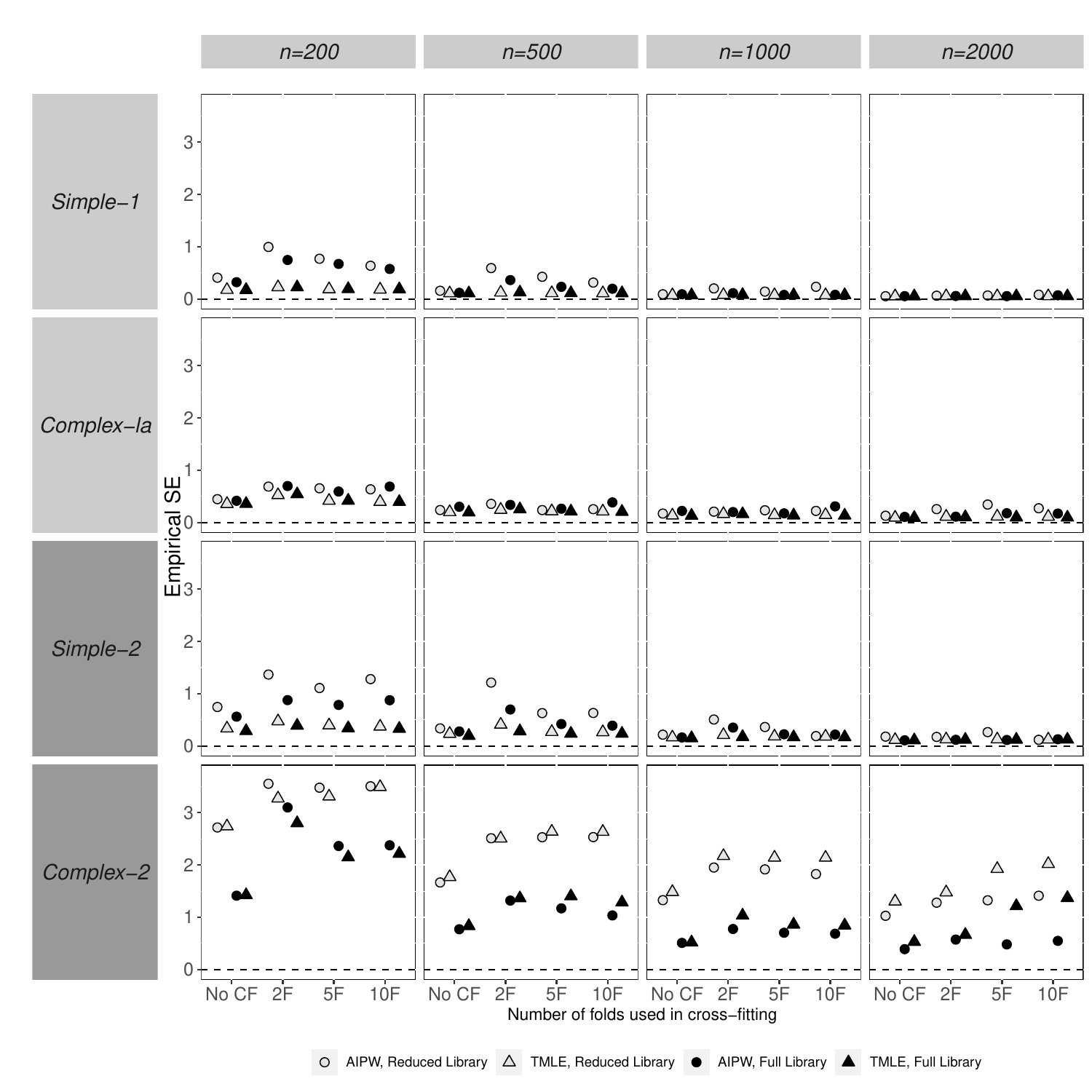}
\caption{Simulation study results for the Empirical SE for AIPW and TMLE with varying use of cross-fitting, by scenario (data generating mechanism and sample size). Note: results from a small number of datasets were excluded for AIPW (but not TMLE) due to the large effect that they had on some calculated performance measures (refer to Section \ref{method_compare}, and Section 5 of the Supporting Information for more details).
}
\label{fig:EmpSE_red}
\end{figure}

\newpage


\begin{figure}[h!]
\centering
\includegraphics[width=15cm, height=15cm]{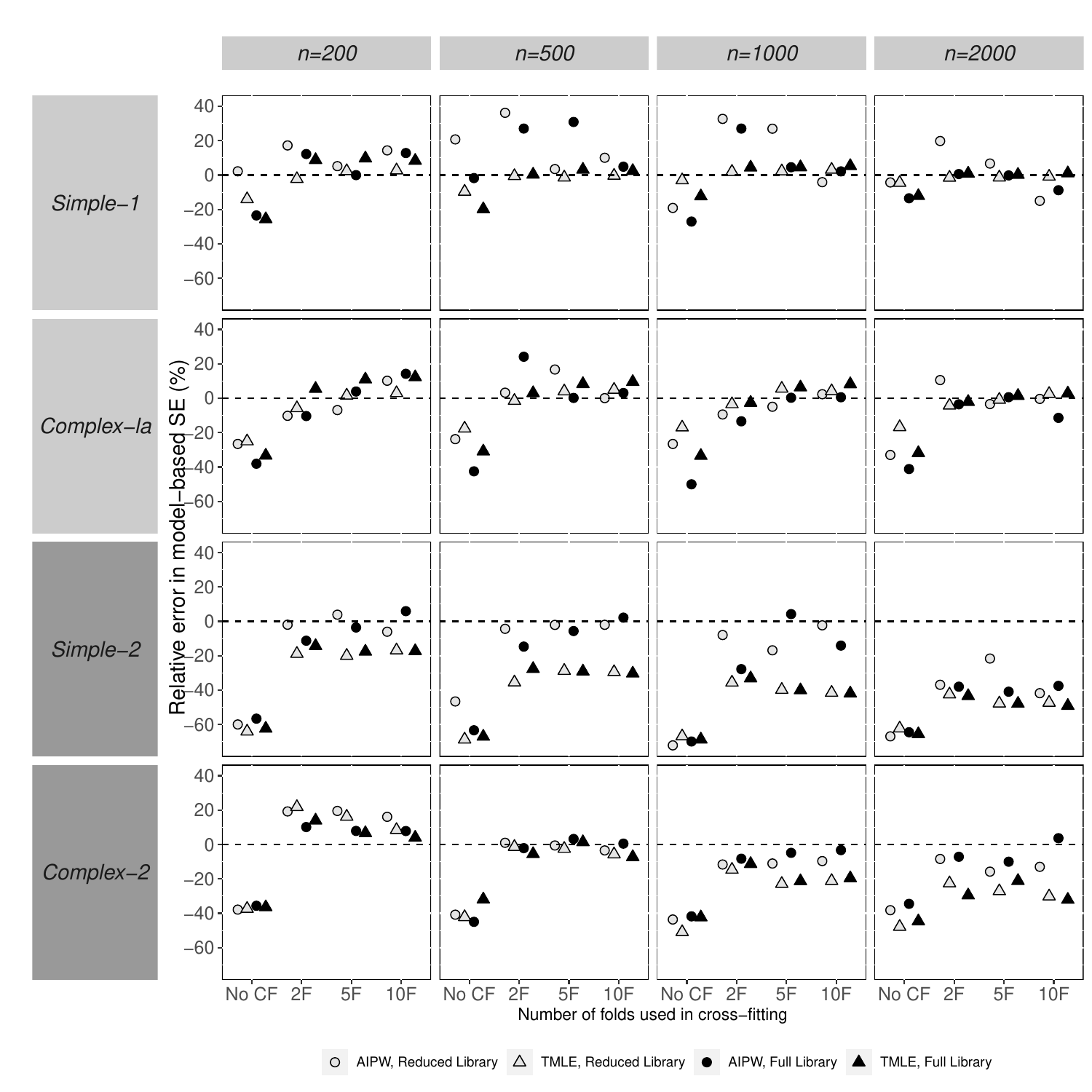}
\caption{Simulation study results for the bias in model-based SE (\%) for AIPW and TMLE with varying use of cross-fitting, by scenario (data generating mechanism and sample size). Note: results from a small number of datasets were excluded for AIPW (but not TMLE) due to the large effect that they had on some calculated performance measures (refer to Section \ref{method_compare}, and Section 5 of the Supporting Information for more details).
}
\label{fig:RelmodSE_red}
\end{figure}

\newpage


\begin{figure}[h!]
\centering
\includegraphics[width=15cm, height=15cm]{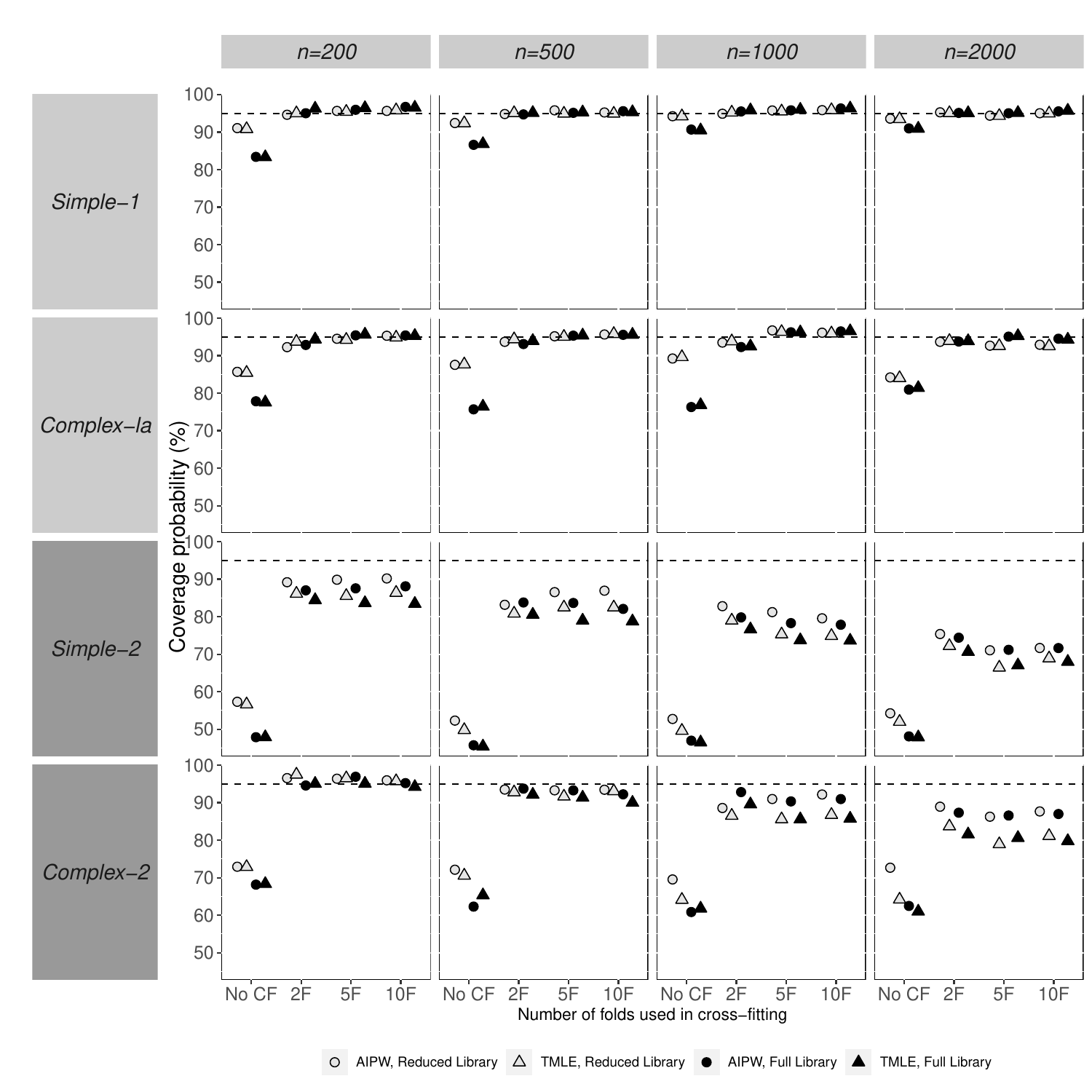}
\caption{Simulation study results for the coverage probability of the 95\% CI for AIPW and TMLE with varying use of cross-fitting, by scenario (data generating mechanism and sample size). Note: results from a small number of datasets were excluded for AIPW (but not TMLE) due to the large effect that they had on some calculated performance measures (refer to Section \ref{method_compare}, and Section 5 of the Supporting Information for more details).
}
\label{fig:CP_red}
\end{figure}

\subsection{Method comparison (AIPW vs. TMLE) } \label{method_compare}

In general, both DR methods performed similarly across the key performance measures, except that AIPW failed to produce sensible results for a few datasets when CF was applied (2-12 datasets across scenarios). In these settings, the standard error was $> 10$ times the median standard error or the absolute value of the point estimate was $> 5$ times the absolute value of the median point estimate. These results were omitted from the figures, with details provided in Figures \ref{fig:Relbias_red}-\ref{fig:CP_red} and in the Supporting Information (Section 7, Table S26).

Relative biases in point estimates were very similar between methods (Figure 2). A small number of exceptions were noted for the variance (empirical SE), and coverage probability (Figures 3,5). For example, at the smallest sample size of 200 with \textit{simple-1} and \textit{simple-2}, the empirical SE appeared larger for AIPW than TMLE, a difference that appeared to dissipate when larger sample sizes were considered. 

The relative error in model-SE was the indicator for which the most differences between methods were observed.
As mentioned above, AIPW results obtained on a small number of datasets were not sensible when CF was applied, and the relative error in model-SE was large with inclusion of these results. For example, for \textit{simple-1} with a sample size of 1000, full library in SL and 5-fold CF, the relative error in model-based SE was 87.5\% and 4.5\% prior to and after exclusion of 2 datasets respectively. The pattern of method differences for the relative error in model-SE was not clear and those observed may potentially be due to the instability of AIPW.

\subsection{Impact of using cross-fitting} \label{Fold_impact}

In general, relative bias appeared to be of similar magnitude whether CF was applied or not (Figure \ref{fig:Relbias_red}). Empirical SE was observed to be similar with or without CF, except for \textit{complex-2} at smaller sample sizes (n=200, 500), where it was larger with CF than without (Figure \ref{fig:EmpSE_red}). In general, the use of CF resulted in lower relative error in model-SE than without CF (Figure \ref{fig:RelmodSE_red}). The use of CF appeared to improve coverage compared to not using CF, resulting in coverage probabilities that were closer to nominal (Figure \ref{fig:CP_red}).

Lastly, when CF was applied in scenarios with a large confounder set (\textit{simple-2} and \textit{complex-2}), the relative error in model-SE appeared to grow with increasing sample size considered, which resulted in lower coverage (further from nominal) (Figure 4 and 5). The same pattern was not observed for scenarios that considered the small confounder set (\textit{simple-1} and \textit{complex-1a}), or when CF was not applied.
See Section 9 for discussion of these findings.

\subsection{Impact of varying the number of folds within cross-fitting} 

In general, method performance was similar regardless of the number of folds used within CF (Figures \ref{fig:Relbias_red}-\ref{fig:CP_red}). An exception was noted at larger sample sizes (e.g., n=2000) with \textit{complex-1a}, where using 2 folds appeared to result in lower bias compared to the use of $\geq$ 5 folds with CF (Figure 2). Across scenarios at smaller sample sizes, empirical SE was slightly smaller when using $\geq$ 5 folds than with 2 folds, but at larger sample sizes (e.g., n=2000), empirical SE was similar regardless of the number of folds used (Figure 3). Patterns were harder to discern for the relative error in model-SE (Figure 4), with other factors appearing more important than the number of folds (e.g., the method). For coverage, with the larger confounder set (\textit{simple-2} and \textit{complex-2}), fewer folds appeared slightly better as sample size increased (Figure \ref{fig:CP_red}).

\subsection{Library impact} \label{lib_general}

Here we outline general observations regarding method performance according to the library used in SL.
The full library produced substantially less bias in point estimates for \textit{complex-1a} at the largest sample size (n=2000), and for \textit{complex-2} across all sample sizes, compared to the reduced library. Otherwise, the reduced library performed as well as or better than the full library in terms of bias across all the sample sizes considered.
Empirical SE did not appear to differ by the library used in SL except for \textit{complex-2} where empirical SE appeared smaller with the full compared to the reduced library at all sample sizes. 

For relative error in model-SE and coverage, the impact of library on method performance appeared to differ according to whether CF was applied or not (Figure 4 and 5). In general, but with some exceptions, smaller library differences were observed for relative error in model-SE and coverage when CF was applied. When CF was not applied, larger library differences were observed for \textit{simple-1} and \textit{complex-1a}, where higher relative error in model-SE and lower coverage (further from nominal) was observed when the full library was used compared to the reduced library.

\section{Application to the BIS case study}
\label{sec8}

For the BIS case study example, we applied both AIPW and TMLE without CF and with CF, using 2, 5 and 10 folds. We used reduced and full libraries in SL (composition detailed in Section 5 of the Supporting Information (Table S24)\textcolor{red}{)}, and we considered the two confounder sets analogous with those that motivated the simulation study design (detailed in Section 3 of the Supporting Information (Tables S2-S4)). 
 Henceforth, we use the following terminology: partially adjusted refers to estimates obtained using a confounder set that contains demographic and background confounders only (p=14) and fully adjusted refers to an estimate adjusted for a set that contains demographic, background, and metabolomic confounders (p=87).  

\subsection{Results}

The results 
are presented in Figure \ref{Fig_res}. 
Full results are available in Section 8 of the Supporting Information (Table S35, and the distribution of estimated propensity scores by exposure group are provided in Figures S1 and S2).


\begin{figure}[h!]
\centering\includegraphics[scale=0.6]{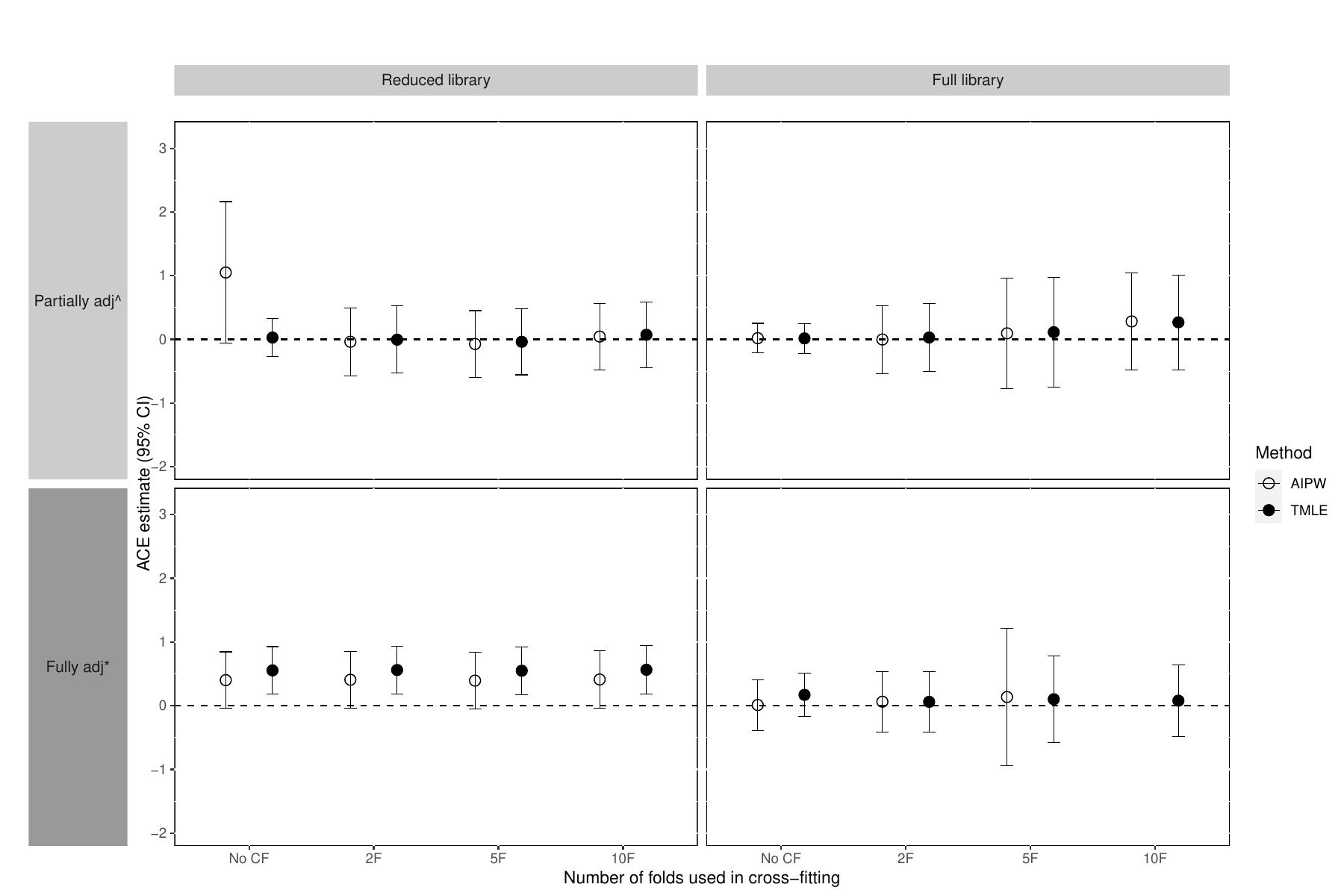}
\caption{Estimated average causal effect (ACE) with accompanying $95\%$ CI, of inflammation (GlycA) in 1-year old infants on Pulse Wave velocity (PWV) at 4 years of age (standardised), obtained by applying the methods to the BIS motivating example. $\hat{}$ demographic and background confounders only; $^*$demographic, background, and metabolomic confounders. \\ Note: Estimate for AIPW using fully adjusted confounder set and full library in SL with 10 fold cross-fitting not shown in figure as too large to be sensible (refer to Table 35 in Section 8 of the Supporting Information).}
\label{Fig_res}
\end{figure}

In general, estimated effect sizes were small for both methods, with and without CF and regardless of library used. A sensible fully adjusted estimate for the ACE was unable to be obtained for AIPW, with a full library and 10-fold CF, with the large and implausible estimate (-3.7; 95\% CI [-17.2, 9.8]) reminiscent of unstable results reported in the simulation study for AIPW in some datasets. In addition, fully adjusted point estimates when using the reduced library appeared larger than those obtained when using the full library, potentially explained by the larger bias in point estimates seen in the simulation study with the reduced library in the most complex scenario.

Otherwise, point estimates and the width of CIs were similar for both methods. One exception was for AIPW with partial adjustment when using the reduced library in SL without CF, where the estimate was larger and less precise than for TMLE (possibly due to the instability of AIPW mentioned above). Another exception was observed with full adjustment, where the CIs accompanying estimates (where available) were slightly wider for AIPW compared to TMLE, with this observation being more marked when the full library was used in SL.

In general, estimates were of similar magnitude with and without CF, but CIs were wider when implemented with CF compared to when no CF was applied.
When CF was used and sensible estimates were able to be obtained, point estimates were similar regardless of the number of folds used. The width of CIs were similar when 5 or 10 folds were used with CF. With the full library, but not the reduced library, CI widths were larger with 5 or 10 folds than when 2 folds were used with CF.

\subsection{Insights from the simulation study}

Next, we consider which of the several approaches used in the example would be considered most reliable in light of the simulation study results. Specifically, we consider simulation results for the complex scenarios, because in this example we think it is highly plausible that there are complex relationships amongst variables. With this in mind, simulation results suggest that for a larger
confounder set (full adjustment) and the small sample size in this motivating example, the use of the full library and CF may produce the most trustworthy and credible estimates. The findings of the simulation study suggest a slight preference for TMLE over AIPW due to stability. In contrast, for the small confounder set (partial adjustment), we found in the simulation study that method performance 
was best when using the reduced library rather than the full library, and when CF was applied. 

\section{Discussion}
\label{sec9}

We conducted a simulation study that was motivated by a realistic study to answer key questions regarding the relative performance of AIPW and TMLE, the use of CF, and the number of folds used within CF,  when estimating the ACE in the presence of high-dimensional confounding.
We found that in general in this setting, AIPW and TMLE performed similarly but AIPW exhibited some stability issues. CF did not greatly affect the bias of point estimates but 
was important to reduce the bias in the model-based SE, and for improving the coverage of the methods. In general, there was no substantial difference between using 2, 5 or 10 folds when using CF with the CML methods. In complex scenarios, the full, more flexible library was important for bias in point estimates, especially at larger sample sizes, and for empirical SE.

In the high-dimensional setting our findings suggest that, although either AIPW or TMLE may be sensible choices of method to estimate the ACE, some caution is warranted with AIPW. Occasional instability of AIPW is likely explained by limitations of the method, where estimates produced may not always be within the natural bounds of the parameter space, in contrast to TMLE where this is not an issue because it is a substitution estimator.\citep{Diaz2019} With a complex scenario (regardless of the number of confounders), the full library appeared important, and this is as we would expect, with the more flexible algorithms needed to capture the complexity.
Regardless of sample size or the number of confounders used, CF should be used with CML methods, and this is even more important if a full library is used. These results are in line with expectations that CF helps reduce the bias in variance estimation arising from the use of flexible ML algorithms. Interestingly, the number of folds used in CF did not have much impact on method performance, with other factors appearing more important (e.g. method and library choice). 

Our findings align with previous methodological studies evaluating and comparing these CML methods with and without CF, although, unlike our study, most have considered low-dimensional confounding only (with very few or binary confounders only), with little focus on strong high-dimensional confounding \citep{Naimi2023ChallengesAlgorithms, Zivich2021MachineEstimators} or motivation from realistic case studies. One exception is a recent study by Meng and Huang\cite{Meng2022REFINE2:Studies} which considered both a realistic setting and high-dimensional confounding ($\approx$1000 observations, 331 covariates), and like us, observed that AIPW and TMLE with CF performed similarly in general. Li et al\cite{Li2022EvaluatingTrials} evaluated CML methods, finding that both AIPW and TMLE with CF are favourable for robust inferences, as we found, but their study was in the context of a randomised experiment, with sample sizes in general much larger than we considered. 
In the literature it has been suggested that the number of folds considered with CF should be higher for smaller sample sizes than for larger sample sizes. \cite{Balzer2023InvitedResearch} In simple settings without high-dimensional confounding, Chernozhukov et al \cite{Chernozhukov2018} report that 4 or 5 may work better than 2 folds. However, until our study, there has been no existing guidance on the number of folds to use in CF in realistic settings that feature high-dimensional confounding. We found that in general, CML method performance was similar regardless of the number of folds. 
 In line with previous recommendations, other factors (library, method) appeared more important than number of folds used across the full range of performance indicators.

Meng and Huang \cite{Meng2022REFINE2:Studies} found that parametric learners generally performed similarly to or outperformed non-parametric learners in a realistic setting and had lower computational times. 
Our findings in simple scenarios are consistent with that study, but in complex scenarios (e.g., \textit{complex-2}), we found that the use of the full, more flexible library was important to reduce bias in point estimates and variance, 
particularly for larger sample sizes, with variation depending on number of confounders used. An explanation for this could be 
that 
our full library considered learners that extended beyond tree-based methods, with the importance of doing so highlighted in a recent commentary. \citep{Balzer2023InvitedResearch} It could also be due to differences in the design of our simulation study such as our data generating mechanisms,  because conclusions in regard to the use of parametric versus non-parametric learners will likely be very specific to the chosen data-generating mechanisms. Another possible explanation is the variant of CF used (single CF) as opposed to other variants of CF available including what is referred to as double CF. \citep{Zivich2021MachineEstimators}
Phillips et al \cite{Phillips2023PracticalLearner} provide guidance regarding the choices that need to be made when using SL, including library construction. They state the importance of using a diverse library in SL though emphasise that the number of learners included might need to be limited in the high-dimensional confounding setting. Our study adds to this important literature, by re-emphasising the need to carefully consider the setting\textcolor{red}{-}specific constraints (e.g., sample size, number of confounders) when choosing a library.

Our research strongly suggests that caution is warranted when applying these methods, with careful consideration of SL library composition required. We found indications that performance does not necessarily improve with increasing sample sizes. For example, in scenarios where we considered the large confounder set and used CF, we found increasing bias in model-SE and decreasing coverage, alongside decreasing bias in point estimates, as larger sample sizes were considered. For these scenarios (simple-2, complex-2), we noted larger SL coefficients were assigned to the glmnet learner at smaller sample sizes than at larger sample sizes (Supplementary Section 9, Figures S3 and S4). A possible explanation for this observation could be that simpler algorithms like glmnet are prioritised at smaller sample sizes, which are in turn less prone to overfitting, or less likely to give extreme propensity score estimates. In keeping with this suggestion, we observed that the estimated propensity scores were more variable and had a much smaller minimum (closer to zero) for the larger sample sizes than for the smaller sample sizes (Supplementary Section 9, Figure S5 and S6). A possible explanation for this observation is that at the lower sample size, the screening within glmnet reduces the variable set, inducing bias in point estimates but dampening variability in the propensity scores. At larger sample sizes the retention of variables helps reduce bias in point estimates, but induces greater variability, with estimated propensity scores closer to zero resulting in biased model-SE's and poorer coverage. 
This phenomenon suggests that careful selection of the learners within SL is important in the sample sizes commonly encountered in modern observational studies. Future research should examine inclusion of ridge regression or high-dimensional regression \citep{Sur2019ARegression} as learners, which are expected to be better at handling a large number of variables with non-null effects. It is important to note that in practice estimated propensity scores close to zero or one may indicate a lack of overlap, which raises conceptual issues, and therefore it is important for researchers to always inspect estimated propensity scores.

Strengths of our simulation study include that it was informed by a realistic case study, where we attempted to simulate confounders with dependencies consistent with associations observed between them in our realistic dataset. We considered several scenarios, representing both simple and complex data generating mechanisms, and simultaneously evaluated performance of the methods across a range of modest sample sizes that are likely to be encountered in observational studies.  
However, although we considered a wide range of scenarios, further data generating mechanisms could be considered to control the heterogeneity of the ACE across scenarios. Indeed,
in the complex scenarios, for the outcome generation models, while the main effect coefficient value for the exposure was modified depending on sample size to control power (and thus,
ensure comparability of coverage probability estimates across scenarios), coefficients for the interaction terms
were not modified depending on sample size. This could have induced greater heterogeneity in the causal effect across strata in some scenarios, making it more challenging to compare bias of point and SE estimates across scenarios. 
Further aspects of SL implementation that could potentially affect CML method performance \citep{Bach2024HyperparameterStudy} were not fully explored. For example, we did not consider the tuning of hyperparameters for the learners and did not compare the use of screening vs no screening in SL implementations as we only considered libraries without separate explicit screening.

A further limitation of our study is that we did not examine the method for CF that aggregates over different random seeds to do the split, suggested by Chernozhukov et al \cite{Chernozhukov2018}, which could be more robust than the methods we assessed that relied on a single seed/split. Recent studies have highlighted challenges in the practical implementation of DR methods with CF due to dependence on particular splits \cite{Schader2024DontInference, Naimi2024Pseudo-randomLearning, Zivich2024CommentaryLearning}, and shown that increasing the number of folds in CF can potentially reduce the dependence on the seed.\cite{Zivich2024CommentaryLearning}
We examined this in additional simulations, focusing on the most complex scenario (complex-2), at both the smallest (n=200) and largest sample sizes (n=2000). Specifically, we ran simulations under these settings to obtain  estimates aggregated over multiple splits and examined the sensitivity of our simulation conclusions to this factor. Results are provided in Supplementary Material, Section 10, Figures S7-S8 and although there was some variation in results for these scenarios (e.g. for both sample sizes when the reduced library was used, the aggregate method had greater bias in model-SE and lower coverage than that observed in the original simulation study), our main conclusions regarding the impact of increasing number of folds  held.
We also compared the results for the case-study application with results obtained by aggregating estimates over multiple splits (e.g. applying Chernozhukov et al \cite{Chernozhukov2018} suggested approach) (Supplementary Material, Section 10, Figure S9). Conclusions were similar in this case regardless of approach, but confidence intervals for the aggregated estimates were wider. We note that recent literature has suggested that an alternative way of reducing seed dependence is to increase folds in CF to a sensible number dependent on the dimensionality of the data. \cite{Williams2025Re:Inference} This is less computationally demanding than aggregating. Careful consideration to the dependence on splits should be given in the practical application of these methods, with further research in this area warranted.

Finally, the simulation was based on one realistic motivating example, which may limit the guidance that is able to be provided from this study. Further, this case study had its own limitations.  The exposure was dichotomised at an arbitrary cut-off. An alternative approach to tackle such settings is to consider the exposure as continuous, which would require defining the estimand in a different way (see below). In addition, we acknowledge that there are many interventions that may bring
GlycA concentration below the 75th percentile and each of these could have different effects on cardiovascular health, that is, the intervention remains ill-defined, which would challenge the causal consistency assumption. While these complexities arise often in discovery-phase studies of this kind, it does not justify ignoring confounding bias.

Future research directions could explore the CML methods in applications with a rare treatment or a binary, possibly rare outcome, or a large number of categorical confounders, all of which could bring about challenges with the application of CF. Exploring the impact of CF variant (e.g., double CF compared to single CF (the variant of CF that we used)) on performance of the methods would also provide insight.
Double CF may possibly result in weaker rate of convergence requirements for the nuisance parameters compared to single CF. \cite{Meng2022REFINE2:Studies} However, Zivich and Breskin \cite{Zivich2021MachineEstimators} discuss the potential challenge of using double CF with small sample sizes and suggest that single CF may overcome some of the difficulty. Limited sample sizes are common in realistic settings, and it remains for future research to determine whether double CF may be preferable to single CF in these situations. 
We have shown that further research is required into the composition of SL libraries, particularly as there is a need to balance the requirement of sufficient diversity of learners with performance across an extensive range of data-generating mechanisms, with or without sparsity. In our simulations we observed that SE estimation was more sensitive to bias when using a large confounder set, reflecting a scenario that may be too complex. Therefore, it would also be helpful to explore any potential benefits of applying explicit screening within SL itself in a high-dimensional confounding setting, which potentially could assist with trimming down the size of the confounder set. 

Further possibilities include extending this work beyond dichotomous exposures (e.g. continuous exposures) as they are often of interest. Possibilities for continuous exposures include exploring a shifting exposure distribution \cite{Munoz2012PopulationInterventions, Diaz2013TargetedCurve, Hejazi2020txshift:R, Hejazi2021EfficientTrials} or estimation of standard regression parameters via an influence curve approach \cite{Vansteelandt2022Assumption-leanParameters}. The latter approach focusses on a summary of the exposure effect with reduced modelling assumptions, rather than ascertaining how the outcome changes with step increases or changes in exposure. It will also be important to explore challenges posed by multivariable missing data, which require consideration of the missingness mechanism as well as the identifiability or “recoverability” of the estimand. \cite{Lee2023AssumptionsClassification, Mohan2021GraphicalData, Moreno-Betancur2018CanonicalStudies, MohanGraphicalData} In the high-dimensional confounding setting, evaluation, and development of the CML methods in the presence of missing data is important. This is a growing area of research. \cite{Dashti2024HandlingEstimation}

To conclude, in a high-dimensional confounding setting, we found that on most occasions AIPW and TMLE performed similarly for estimating the ACE, with a slight preference toward the use of TMLE, given its greater stability. In cases where complex confounding mechanisms are suspected, a more diverse and flexible SL library may be beneficial to reduce bias in point estimates and variance. CF is important for the estimation of SE and should be used, as without it we observed underestimation of model-SE and undercoverage, particularly with a more diverse library. Our study suggests that in the high-dimensional confounding setting, the number of folds used with CF is less important than the use of CF itself.

\bmsection*{Author contributions}

All authors contributed to the planning and design of the simulation study. SE conducted the simulation study and performed the analyses, and prepared the first draft of the manuscript. JC, SV and MM-B reviewed the manuscript, and SE revised the manuscript accordingly. All authors reviewed, sighted and approved the final version of the manuscript.

\bmsection*{Acknowledgments}
We thank the BIS investigator group for providing access to the case-study data for illustrative purposes in this work.
In addition, the authors thank David Burgner and Toby Mansell for sharing their expertise in the growing cardiovascular research area. 

\bmsection*{Financial disclosure}
This work was supported by the Australian National Health and Medical Research Council (NHMRC) Investigator Grant Emerging Leadership Level 2 (grant 2009572 awarded to MM-B). SE is funded by an Australian Government Research Training Program Scholarship. Research at the Murdoch Children’s Research Institute is supported by the Victorian Government’s Operational Infrastructure Support Program.

\bmsection*{Conflict of interest}

The authors declare no potential conflict of interests.

\bmsection*{Supporting information}

Additional supporting information may be found in the
online version of the article at the publisher’s website.

\bibliography{Manuscript}

\newpage

\vspace{10cm}\hspace{-4mm}\LARGE{SUPPLEMENTARY MATERIAL: Causal machine learning methods and use of cross-fitting in settings with high-dimensional confounding}

\normalsize
\title{Causal machine learning methods and use of sample splitting in settings with high-dimensional confounding}

\authormark{ELLUL \textsc{et al.}}
\titlemark{Supporting Information}

\beginsupplement

\section{Further details and summary of motivating example}  \label{sec:si1}

\begin{longtable}{lcc}
\caption{Characteristics of participants in the BIS inception birth cohort (n=1,074).} \refstepcounter{SItab}\label{Miss_data_summary} \\
\toprule
\textbf{Characteristic} & \textbf{Inception birth cohort\textsuperscript{1,\#}} & \textbf{Missing data (\%)\textsuperscript{$\wedge$}} \\ 
\midrule
Pre-pregnancy BMI (kg/m\textsuperscript{2}) & 24.0 [21.5, 27.9] & 149 (14\%) \\ 
Socio-Economic Indexes for Areas (SEIFA) &  & 3 (0.3\%) \\ \hspace{0.5cm} Low & 357 (33\%) &  \\ 
\hspace{0.5cm} Med & 357 (33\%) &  \\ 
\hspace{0.5cm} High & 357 (33\%) &  \\ 
Mother smoking in pregnancy & 169 (16\%) & 13 (1.2\%) \\ 
Gestational diabetes in pregnancy & 44 (4.8\%) & 166 (15\%) \\ 
Pre-eclampsia in pregnancy & 35 (3.3\%) & 4 (0.4\%) \\ 
Birthweight (grams) & 3,527 (519) & 0 (0\%) \\ 
Infant sex &  & 0 (0\%) \\ 
\hspace{0.5cm} Female & 519 (48\%) &  \\ 
\hspace{0.5cm} Male & 555 (52\%) &  \\ 
Maternal age at birth (years) & 32.1 (4.8) & 0 (0\%) \\ 
Gestational age at birth &  & 0 (0\%) \\ 
\hspace{0.5cm} 32-36 completed weeks & 47 (4.4\%) &  \\ 
\hspace{0.5cm} 37-42 completed weeks & 1,027 (96\%) &  \\ 
Mode of birth &  & 2 (0.2\%) \\ 
\hspace{0.5cm} Caesarean & 333 (31\%) &  \\ 
\hspace{0.5cm} Vaginal & 739 (69\%) &  \\ 
Weight-for-length z-score at 12 months & 0.72 (1.04) & 214 (20\%) \\ 
Age at 12-month measures (months) & 13.03 (0.82) & 194 (18\%) \\ 
Breastfeeding duration (exclusive weeks*) & 4 [1, 22] & 295 (27\%) \\ 
Postnatal smoke exposure & 130 (15\%) & 197 (18\%) \\ 
Inflammation (GlycA) at 1-year of age (mmol/l) & 1.32 (0.22) & 332 (31\%) \\ 
Pulse wave velocity at 4-years of age (m/s) & 3.97 (0.45) & 486 (45\%) \\ 
\bottomrule
\end{longtable}
\vspace{-8mm}
\footnotesize
\begin{minipage}{\linewidth}
\vspace{5mm}\hspace{13mm}\textsuperscript{1}Mean (SD), Median {[}IQR{]} or Frequency (\%) as appropriate, \textsuperscript{\#}{n=1074}, \textsuperscript{$\wedge$}{Indicates \% of n=1074},\\
\end{minipage}
\begin{minipage}{\linewidth}
\vspace{-2mm}\hspace{17mm}*Number of weeks that infant was exclusively breastfed (i.e. no supplementary feeding)\\
\end{minipage}
\begin{minipage}{\linewidth}
\vspace{-2mm}\hspace{17mm} GlycA: Glycoprotein Acetyls \\
\end{minipage}

\normalsize

\newpage

\section{Doubly Robust methods, implementation with Cross-fitting \label{sec:si2}}

In this study, when CF was applied, point and variance estimates were obtained in the following manner for each DR method using different libraries in SL to fit the models. \\

Assume a dataset of dimension $[n$ x $p]$, where $n$ refers to the number of observations (or rows) and $p$ refers to the number of analysis variables. 
Let $I_k$ denote the observations (a vector of indices) in fold $k$ ($k=1,....,K)$ , and $I_k^C$ denote the indices corresponding to observations not in $I_k$. For fold $k$, outcome and exposure models are fit on $I_k^C$, and using these, predicted outcome and exposure values are obtained for those observations in $I_k$. We denote predicted values for outcome for fold $k$ under $X=x$ as $\hat{E}_x^{(-k(i))}(W_i)$, and for exposure as $\hat{P}^{(-k(i))}(W_i)$, to indicate that predicted values for an individual $i$ in fold $k$ are obtained by using nuisance models that were fit on the complement of fold $k$ ($I_k^C$). For AIPW with CF, the estimate of the ACE is calculated using,

\footnotesize
\begin{align} 
    \begin{split} 
        \hat{\psi}_{\mathrm{AIPW_{CF}}} &=
        \frac{1}{n}\sum_{i=1}^n \left[\hat{E}_1^{(-k(i))}(W_i) - \hat{E}_0^{(-k(i))}(W_i) + X_i \frac{Y_i-\hat{E}_1^{(-k(i))}(W_i)}{\hat{P}^{(-k(i))}(W_i)} - (1-X_i)\frac{Y_i-\hat{E}_0^{(-k(i))}(W_i)}{1-\hat{P}^{(-k(i))}(W_i)}\right] \\
        \label{AIPW_ACE_eqn_CF}
    \end{split}
\end{align}
\normalsize

and the variance is calculated for the ACE using,

\footnotesize
\begin{align} 
    \begin{split} 
  \widehat{var}(\hat{\psi}_{AIPW_{CF}})
  &= \frac{1}{n-1}\sum_{i=1}^{n}\Biggl[\hat{E}_1^{(-k(i))}(W_i) - \hat{E}_0^{(-k(i))}(W_i) + X_i \frac{Y_i-\hat{E}_1^{(-k(i))}(W_i)}{\hat{P}^{(-k(i))}(W_i)} - (1-X_i)\frac{Y_i-\hat{E}_0^{(-k(i))}(W_i)}{1-\hat{P}^{(-k(i))}(W_i)} - \hat{\psi}_{AIPW_{CF}} 
  \Biggr]^2 \label{eqn:AIPW_var_CF}
  \end{split} 
\end{align} 
\normalsize

For TMLE with CF the estimate of the ACE is calculated using,
\begin{equation}\label{estimate_causal_CF}
\hat{\psi}_{\mathrm{TMLE}_{CF}} = \frac{1}{n}\left[\sum_{i=1}^{n} \hat{E}_1^{*(-k(i))}(W_i) - \sum_{i=1}^{n}\hat{E}_0^{*(-k(i))}(W_i)\right]
\end{equation}

Similarly to AIPW, the variance for TMLE is as before (\ref{eqn:AIPW_var_CF}) but with $\hat{E}_1^{*(-k(i))}(W)$ and $\hat{E}_0^{*(-k(i))}(W)$ in place of $\hat{E}_0^{(-k(i))}(W)$ and $\hat{E}_0^{(-k(i))}(W)$ respectively.

\newpage

\section{Further details regarding confounders \label{sec:si3}}

\begin{longtable}{clc}
\caption{Description and summary of background and demographic confounders in the \\ motivating case study sample (n=252)} \refstepcounter{SItab}\label{conf_data_summary} \\
\toprule
\textbf{Background/demographic} & \textbf{Confounder} & \textbf{Summary statistics}\textsuperscript{1} \\ 
\midrule
{} & Pre pregnancy BMI (kg/m$^2$) & 24.3 [21.7, 27.9] \\ 
{} & Socio-Economic Indexes for Areas (SEIFA) \\ 
{} & \hspace{1cm} Low & 73 (29\%) \\ 
\textbf{Antenatal} & \hspace{1cm} Med & 93 (37\%) \\ 
{} & \hspace{1cm} High & 86 (34\%) \\ 
{} & Mother smoking in pregnancy & 32 (13\%) \\ 
{} & Gestational diabetes in pregnancy & 11 (4.4\%) \\ 
{} & Pre-eclampsia in pregnancy & 9 (3.6\%)  \\ 
\midrule
{} & Birthweight (grams) & 3,508 (537) \\ 
{} & Infant sex &  {}  \\ 
{} & \hspace{1cm} Female & 119 (47\%) \\ 
{} & \hspace{1cm} Male & 133 (53\%) \\ 
\textbf{Birth} & Maternal age at birth (years) & 32.3 (4.2) \\ 
{} & Gestational age at birth \\ 
{} & \hspace{1cm} 32-36 completed weeks & 13 (5.2\%) \\ 
{} & \hspace{1cm} 37-42 completed weeks & 239 (95\%) \\ 
{} & Mode of birth  \\ 
{} & Caesarean & 95 (38\%)  \\ 
{} & Vaginal & 157 (62\%) \\ 
\midrule
{} & Weight-for-length z-score at 12 months & 0.72 (1.08) \\ 
\textbf{1 year} & Age at 12-month measures (months) & 12.94 (0.79) \\ 
{} & Breastfeeding duration (exclusive weeks) & 7 [0, 22] \\ 
{} & Postnatal smoke exposure & 39 (15\%)  \\ 
 \bottomrule
\end{longtable}
\footnotesize
\vspace{-3mm}
\begin{minipage}{\linewidth}
\hspace{13mm}\textsuperscript{1}Mean (SD), Median {[}IQR{]} or Frequency (\%) as appropriate \\ 
\end{minipage}

\normalsize


\clearpage\begin{longtable}{clc}
\caption{Description and summary of metabolite confounders} \refstepcounter{SItab}\label{metab_conf_data_summary} \\\toprule
\textbf{Metabolite*} & \textbf{Confounder} & \textbf{Summary statistics}\textsuperscript{1} \\ 
\midrule
\endfirsthead
\caption*{Table S\ref{metab_conf_data_summary}: Description and summary of metabolite confounders (continued)}\\\toprule
\textbf{Metabolite*} & \textbf{Confounder} & \textbf{Summary statistics}\textsuperscript{1} \\ 
\midrule
\endhead
\midrule
\endfoot
\bottomrule
\endlastfoot
{} & Total lipids in chylomicrons and extremely large VLDL  (mmol/l) & -4.44 (-8.11, -3.76) \\ 
{} & Total lipids in very large VLDL  (mmol/l) & -3.31 (-5.74, -2.77) \\ 
{} & Total lipids in large VLDL  (mmol/l) & -1.66 (-2.19, -1.29) \\ 
{} & Total lipids in medium VLDL  (mmol/l) & -0.80 (-1.10, -0.55) \\ 
{} & Total lipids in small VLDL  (mmol/l) & -0.83 (-1.00, -0.60) \\ 
{} & Total lipids in very small VLDL  (mmol/l) & -1.02 (-1.15, -0.86) \\ 
{} & Total lipids in IDL  (mmol/l) & -0.20 (-0.38, -0.07) \\ 
{} & Total lipids in large LDL  (mmol/l) & -0.06 (-0.26, 0.07) \\ 
{} & Total lipids in medium LDL  (mmol/l) & -0.61 (-0.81, -0.46) \\ 
{} & Total lipids in small LDL  (mmol/l) & -1.03 (-1.19, -0.88) \\ 
{} & Total lipids in very large HDL  (mmol/l) & -0.96 (-1.24, -0.72) \\ 
{} & Total lipids in large HDL  (mmol/l) & -0.61 (-0.87, -0.36) \\ 
{} & Total lipids in medium HDL  (mmol/l) & -0.33 (-0.44, -0.23) \\ 
{} & Total lipids in small HDL  (mmol/l) & 0.01 (-0.05, 0.06) \\ 
{} & Mean diameter for VLDL particles (nm) & 3.61 (3.58, 3.63) \\ 
{} & Mean diameter for LDL particles (nm) & 3.158 (3.155, 3.161) \\ 
{} & Mean diameter for HDL particles (nm) & 2.289 (2.276, 2.303) \\ 
{} & Serum total cholesterol (mmol/l) & 1.19 (1.04, 1.31) \\ 
{} & Total cholesterol in VLDL (mmol/l) & -0.84 (-1.09, -0.61) \\ 
{} & Remnant cholesterol (non-HDL, non-LDL -cholesterol) (mmol/l) & -0.10 (-0.28, 0.08) \\ 
{} & Total cholesterol in LDL (mmol/l) & 0.14 (-0.10, 0.31) \\ 
{} & Total cholesterol in HDL (mmol/l) & 0.18 (0.06, 0.29) \\ 
{} & Total cholesterol in HDL2 (mmol/l) & -0.31 (-0.48, -0.15) \\ 
{} & Total cholesterol in HDL3 (mmol/l) & -0.78 (-0.81, -0.74) \\ 
\textbf{1 year} & Esterified cholesterol (mmol/l) & 0.83 (0.68, 0.96) \\ 
{} & Free cholesterol (mmol/l) & 0.00 (-0.13, 0.11) \\ 
{} & Serum total triglycerides (mmol/l) & 0.08 (-0.13, 0.28) \\ 
{} & Triglycerides in VLDL (mmol/l) & -0.33 (-0.63, -0.08) \\ 
{} & Triglycerides in LDL (mmol/l) & -1.86 (-2.04, -1.72) \\ 
{} & Triglycerides in HDL (mmol/l) & -2.17 (-2.35, -2.01) \\ 
{} & Total phosphoglycerides (mmol/l) & 0.41 (0.32, 0.49) \\ 
{} & Ratio of triglycerides to phosphoglycerides & -0.53 (-0.74, -0.32) \\ 
{} & Phosphatidylcholine and other cholines (mmol/l) & 0.45 (0.38, 0.54) \\ 
{} & Sphingomyelins (mmol/l) & -1.14 (-1.24, -1.02) \\ 
{} & Total cholines (mmol/l) & 0.63 (0.54, 0.71) \\ 
{} & Apolipoprotein A-I (g/l) & 0.25 (0.19, 0.31) \\ 
{} & Apolipoprotein B (g/l) & -0.39 (-0.52, -0.27) \\ 
{} & Ratio of apolipoprotein B to apolipoprotein A-I & -0.63 (-0.77, -0.52) \\ 
{} & Total fatty acids (mmol/l) & 2.17 (2.06, 2.29) \\ 
{} & Estimated degree of unsaturation & 0.16 (0.13, 0.19) \\ 
{} & 22:6, docosahexaenoic acid (mmol/l) & -2.66 (-2.96, -2.44) \\ 
{} & 18:2, linoleic acid (mmol/l) & 0.79 (0.65, 0.89) \\ 
{} & Omega-3 fatty acids (mmol/l) & -1.29 (-1.49, -1.10) \\ 
{} & Omega-6 fatty acids (mmol/l) & 0.99 (0.88, 1.09) \\ 
{} & Polyunsaturated fatty acids (mmol/l) & 1.09 (0.97, 1.20) \\ 
{} & Monounsaturated fatty acids; 16:1, 18:1 (mmol/l) & 0.95 (0.82, 1.09) \\ 
{} & Saturated fatty acids (mmol/l) & 1.16 (1.06, 1.26) \\ 
{} & Ratio of 22:6 docosahexaenoic acid to total fatty acids (\%) & -0.22 (-0.48, -0.05) \\ 
{} & Ratio of 18:2 linoleic acid to total fatty acids (\%) & 3.21 (3.14, 3.28) \\ 
{} & Ratio of omega-3 fatty acids to total fatty acids (\%) & 1.15 (1.00, 1.27) \\ 
{} & Ratio of omega-6 fatty acids to total fatty acids (\%) & 3.42 (3.37, 3.47) \\ 
{} & Ratio of polyunsaturated fatty acids to total fatty acids (\%) & 3.53 (3.47, 3.57) \\ 
{} & Ratio of monounsaturated fatty acids to total fatty acids (\%) & 3.38 (3.33, 3.44) \\ 
{} & Ratio of saturated fatty acids to total fatty acids (\%) & 3.60 (3.56, 3.64) \\ 
{} & Glucose (mmol/l) & 1.16 (0.80, 1.24) \\ 
{} & Lactate (mmol/l) & 0.71 (0.47, 1.34) \\ 
{} & Pyruvate (mmol/l) & -1.94 (-2.17, -1.60) \\ 
{} & Citrate (mmol/l) & -2.11 (-2.19, -2.02) \\ 
\textbf{1 year} & Glycerol (mmol/l) & -2.27 (-2.44, -2.06) \\ 
{} & Alanine (mmol/l) & -1.03 (-1.16, -0.88) \\ 
{} & Glutamine (mmol/l) & -0.84 (-0.90, -0.75) \\ 
{} & Glycine (mmol/l) & -1.49 (-1.60, -1.37) \\ 
{} & Histidine (mmol/l) & -2.75 (-2.89, -2.60) \\ 
{} & Isoleucine (mmol/l) & -2.61 (-2.84, -2.42)  \\ 
{} & Leucine (mmol/l) & -2.40 (-2.66, -2.20) \\ 
{} & Valine (mmol/l) & -1.57 (-1.81, -1.35) \\ 
{} & Phenylalanine (mmol/l) & -2.69 (-2.83, -2.60) \\ 
{} & Tyrosine (mmol/l) & -2.61 (-2.86, -2.42) \\ 
{} & Acetate (mmol/l) & -3.31 (-3.50, -3.03) \\ 
{} & Acetoacetate (mmol/l) & -2.77 (-3.22, -2.47) \\ 
{} & 3-hydroxybutyrate (mmol/l) & -1.90 (-2.10, -1.65) \\ 
{} & Albumin (signal area) & -2.46 (-2.51, -2.41)  \\ 
{} & Creatinine (mmol/l) & -3.95 (-4.08, -3.82) \\ 
\end{longtable}
\vspace{-4mm}
\footnotesize
\begin{minipage}{\linewidth}
\hspace{9mm}\textsuperscript{*}Log transformed (zero converted to half minimum (non-zero) detected prior to transformation) \\ 
\end{minipage}
\begin{minipage}{\linewidth}
\vspace{-3mm}\hspace{13mm}\textsuperscript{1}Mean (95\% CI) \\ 
\end{minipage} \\\\\\\\

\normalsize

\begin{longtable}{cc}
\caption{Confounders included in each confounder set considered} \refstepcounter{SItab}\label{confounder_set_summary} \\\toprule
\textbf{Confounder set} & \textbf{Confounders included} \\ 
\midrule
\endhead
\bottomrule
\endlastfoot
 Small (1) & \hspace{5mm}\makecell[l]{All confounders listed in Table S\ref{conf_data_summary}.} \\
 \midrule
 Large (2) & All confounders listed in Table S\ref{conf_data_summary} and Table S\ref{metab_conf_data_summary}. \\
\end{longtable}

\normalsize
\newpage

\section{Details regarding data-generation mechanisms} \label{sec:si4}

\begin{longtable}[l]{cll}
\caption{Details (and order) of data generation for the confounders} \refstepcounter{SItab}\label{conf_data_gen} \\\toprule
\textbf{Confounder} & \textbf{BIS variable} & \textbf{Generating distribution and details} \\ 
\midrule
\endfirsthead
\caption*{Table S\ref{conf_data_gen}: Details (and order) of data generation for the confounders (continued)}\\\toprule
\textbf{Confounder} & \textbf{BIS variable} & \textbf{Generating distribution and details} \\ 
\midrule
\endhead
\midrule
\endfoot
\bottomrule
\endlastfoot
{$C_1$} & \makecell[l]{Socio-Economic Indexes \\ for Areas (SEIFA)} & {$C_1\sim \text{Categorical}(\boldsymbol{\theta} = \theta_1, \theta_2, \theta_3) $} \\\\
\vspace{2mm}{$C_2$} & \makecell[l]{Pre-pregnancy \\ BMI (kg/m$^2$)} & 
{\makecell[l]{$C_2 \sim \text{Lognormal}\left(log\left(\frac{\tau^2}{\sqrt{\rho^2 + \tau^2}}\right), \sqrt{\log\left(\frac{\rho^2}{\tau^2} + 1 \right)} \right)$ \\
$log(C_2) \sim \text{Normal}(\tau, \rho)$ \\
$\tau: E(log(C_2)) = \beta_0 + \beta_1C_{1,low}+ \beta_2C_{1,med}$}} \\
\vspace{2mm}{$C_3$} & \makecell[l]{Mother smoking \\ in pregnancy} & {$C_3 \sim \text{Binomial}(1, expit(\beta_0 + \beta_1C_{1,low} + \beta_2C_{1,med}))$} \\
\vspace{2mm}{$C_4$} & \makecell[l]{Gestational diabetes \\ in pregnancy} & {$C_4 \sim \text{Binomial}(1, expit(\beta_0 + \beta_1C_2))$} \\
\vspace{2mm}{$C_5$} & \makecell[l]{Pre-eclampsia \\ in pregnancy} & {$C_5 \sim \text{Binomial}(1, expit(\beta_0 + \beta_1C_2 + \beta_2C_3))$} \\
\midrule
\vspace{2mm}{$C_6$} & {Infant sex} & {$C_6 \sim \text{Binomial}(1, p=0.53)$} \\\\
\vspace{2mm}{$C_7$} & \makecell[l]{Maternal age at \\ birth (years)} & {$C_7 = \text{Normal}(E(C_7)) = \beta_0 + \beta_1C_{1,low} + \beta_2C_{1,med}, \sigma = s_{C_7})$} \\
\vspace{2mm}{$C_8$} & \makecell[l]{Gestational age at \\ birth (weeks)} & {$C_8 \sim \text{Binomial}(1, expit(\beta_0 + \beta_1C_2 + \beta_2C_3 + \beta_3C_4 + \beta_4C_5 + \beta_5C_7))$} \\
\vspace{2mm}{$C_9$} & {Mode of birth} & {$C_9 \sim \text{Binomial}(1, expit(\beta_0 + \beta_1C_2 + \beta_sC_3 + \beta_3C_4 + \beta_4C_5 + \beta_5C_7 + \beta_6C_8))$} \\
\vspace{2mm}{$C_{10}$} & {Birthweight (grams)} & \makecell[l]{$C_{10} \sim \text{Normal}(E(C_{10})=\beta_0 + \beta_1C_2 + \beta_2C_3 + \beta_3C_4 + \beta_4C_5 + \beta_6C_7 + $ \\ \hspace{3.5cm} $  \beta_7C_8, \sigma=s_{C_10})$} \\
\vspace{2mm}{$C_{11}$} & \makecell[l]{Age at 12-month \\ measures (months)} &
{\makecell[l]{$C_{11} \sim \text{Lognormal}\left(log\left(\frac{\tau^2}{\sqrt{(\rho^2 + \tau^2)}}\right), \sqrt{log\left(\frac{\rho^2}{\tau^2} + 1\right)}\right)$ \\
$log(C_{11}) \sim \text{Normal}(\tau, \rho)$}} \\
\vspace{2mm}{$C_{12}$} & \makecell[l]{Breastfeeding \\ duration \\ (exclusive weeks)} &
{\makecell[l]{$C_{12} \sim \text{(zero-inflated) Poisson} (\mu, \pi)$ \\
$log(\mu) = \beta_0 + \beta_1C_{1,low} + \beta_2C_{1,med} + \beta_3C_7 + \beta_4C_8 \text{ [counts]}$ \\
$log(\pi) = \varphi_0 + \varphi_1C_{1,low} + \varphi_2C_{1,med} + \varphi_3C_7 + \varphi_4C_8 \text{ [zero-inflation]}$}} \\
\vspace{2mm}{$C_{13}$} & \makecell[l]{Postnatal smoke \\ exposure} & {$C_{13} \sim \text{Binomial}(1, expit(\beta_0 + \beta_1C_{1,low} + \beta_2C_{1, med} + \beta_3C_3)$} \\
\vspace{2mm}{$C_{14}$} & \makecell[l]{Weight-for-length \\ z-score at 12 months} & {$C_{14} \sim \text{Normal}(E(C_{14} = \beta_0 + \beta_1C_6 + \beta_2C_{10} + \beta_3C_{11} + \beta_4C_{12}, \sigma = s_{C_{14}})$} \\
{$\boldsymbol{M} = M_1,.,M_{73}$} & {Metabolites} &
{\makecell[l]{$\boldsymbol{M} \sim \text{MVN}[\boldsymbol{m}, \boldsymbol{\Sigma}]$ \\
$\boldsymbol{M} = (M_1, ...., M_{73})^T$ \\
$\boldsymbol{m} = (E(M_1), ...., E(M_{73}))^T$ \\
$\Sigma_{i,j} = Cov(M_i, M_j)$ \\
$ 1 \leq i,j \leq 73$}}
\end{longtable}


\clearpage\begin{longtable}[l]{cccl}
\caption{Details (and order) of data generation for the exposure} \refstepcounter{SItab}\label{exp_data_gen} \\\toprule
\textbf{Confounder set} & \textbf{Mechanism} & \textbf{Generating distribution and details} \\ 
\midrule
\endfirsthead
\caption*{Table S\ref{exp_data_gen}: Details (and order) of data generation for the exposure (continued)}\\\toprule
\textbf{Confounder set} & \textbf{Mechanism} & \textbf{Generating distribution and details} \\ 
\midrule
\endhead
\midrule
\endfoot
\bottomrule
\endlastfoot
{Small (1)} & {Simple-1} & 
\makecell[l]{$X \sim \text{N}\left(\gamma_0 + \sum_{j=1}^{15}(\gamma_jC_j), sd_x\right)$ \\
\vspace{2mm}with $X$ then dichotomised using the 75th percentile.} \\

\midrule

{Small (1)} & {Complex-1a} & 
\hspace{5mm}\makecell[l]{$X \sim \text{N}(\gamma_0 + \sum_{j=1}^{15}(\gamma_jC_j) + 
\sum_{1 \leq i < j \leq 14}\gamma_{(13+i+j)}C_iC_j, sd_x)$ \\
$\Rightarrow$ but set coefficient terms for the interaction terms \\
to \textbf{two} times that observed in the BIS data: \\
\vspace{2mm}$\gamma_{\left(13+i+j\right)} \to 2\gamma_{\left(13+i+j\right)}$ \\
with $X$ then dichotomised using the 75th percentile.} \\

\midrule

{Small (1)} & {Complex-1b} & 
\hspace{5mm}\makecell[l]{$X \sim \text{N}(\gamma_0 + \sum_{j=1}^{15}(\gamma_jC_j) + \sum_{1 \leq i < j \leq 14}\gamma_{(13+i+j)}C_iC_j, sd_x)$ \\
$\Rightarrow$ but set coefficient terms for the interaction terms \\
to \textbf{four} times that observed in the BIS data: \\
\vspace{2mm}$\gamma_{\left(13+i+j\right)} \to 4\gamma_{\left(13+i+j\right)}$ \\
with $X$ then dichotomised using the 75th percentile.} \\

\midrule

{Large (2)} & {Simple-2} & 
\makecell[l]{$X \sim \text{N}\left(\gamma_0 + \sum_{j=1}^{88}(\gamma_jC_j), sd_x\right)$ \\
\vspace{1mm}with $X$ then dichotomised using the 75th percentile.} \\
\midrule
{Large (2)} & {Complex-2} & 
\makecell[l]{$X \sim \text{N}\left(\gamma_0 + \sum_{j=1}^{88}(\gamma_jC_j) + (a), sd_x\right)$ \\
Where (a) is interaction and squared terms as detailed \\ in Tables \ref{exp_data_gen_complex_1} and \ref{exp_data_gen_complex_2}, with $X$ then dichotomised using \\
the 75th percentile.} \\
\end{longtable}
\vspace{-6mm}
\begin{minipage}{30cm}
\footnotesize
\hspace{-3.5mm}{\textbf{Note:} For Small confounder set (1), $C_j$ refers to background and demographic confounder $j, (j=1,…,14)$}
\end{minipage}
\begin{minipage}{30cm}
\footnotesize
\vspace{-4mm}\hspace{0mm}{and for Large confounder set (2), $C_j$ refers to background, demographic and metabolite confounders $j, (j=1,…,87)$}
\end{minipage}
\begin{minipage}{30cm}
\footnotesize
\vspace{-8mm}\hspace{0mm}{as outlined in Table S\ref{confounder_set_summary}. $\gamma_0, \gamma_1,....\gamma_{14},....,sd_x$ are parameter values, estimated using the Barwon Infant Study (BIS) data.}
\end{minipage}

\newpage

\clearpage
\begin{longtable}[l]{ccl}
\caption{Details (and order) of data generation for the outcome}
\label{outcome_data_gen} \\
\toprule
\textbf{Confounder set} & \textbf{Mechanism} & \textbf{Generating distribution and details} \\
\midrule
\endfirsthead
\midrule
\endfoot
\bottomrule
\endlastfoot

{Small (1)} & {Simple-1} & 
\makecell[l]{
$Y \sim \text{N}\left(\alpha_0 + \alpha_{1}X + \sum_{j=2}^{16} \alpha_j C_{j-1},\ \text{sd}_y = 1\right)$
} \\

\midrule

{Small (1)} & {Complex-1a} & 
\makecell[l]{
$Y \sim \text{N}\left(
\begin{array}{l}
\alpha_0 + \alpha_1 X + \sum_{j=2}^{16} \alpha_j C_{j-1} + \sum_{1 \leq i < j \leq 14} \alpha_{14+i+j} C_i C_j + \\
\alpha_{79} C_{10}^2 + \alpha_{80} C_{12}^2 + \alpha_{81} C_{14}^2 + \sum_{j=1}^{14} \alpha_{82+j} C_j X,\ \text{sd}_y = 1
\end{array}
\right)$ \\
$\Rightarrow$ but set coefficient terms for the interaction terms \\
to \textbf{two} times those observed in the BIS data: \\
$\alpha_{14+i+j} \to 2\alpha_{14+i+j}$ \quad and \quad $\alpha_{82+j} \to 2\alpha_{82+j}$
} \\

\midrule

{Small (1)} & {Complex-1b} & 
\makecell[l]{
$Y \sim \text{N}\left(
\begin{array}{l}
\alpha_0 + \alpha_1 X + \sum_{j=2}^{16} \alpha_j C_{j-1} + \sum_{1 \leq i < j \leq 14} \alpha_{14+i+j} C_i C_j + \\
\alpha_{79} C_{10}^2 + \alpha_{80} C_{12}^2 + \alpha_{81} C_{14}^2 + \sum_{j=1}^{14} \alpha_{82+j} C_j X,\ \text{sd}_y = 1
\end{array}
\right)$ \\
$\Rightarrow$ but set coefficient terms for the interaction terms \\
to \textbf{four} times those observed in the BIS data: \\
$\alpha_{14+i+j} \to 4\alpha_{14+i+j}$ \quad and \quad $\alpha_{82+j} \to 4\alpha_{82+j}$
} \\

\midrule

{Large (2)} & {Simple-2} & 
\makecell[l]{
$Y \sim \text{N}\left(\alpha_0 + \alpha_{1}X + \sum_{j=2}^{89} \alpha_j C_{j-1},\ \text{sd}_y = 1\right)$
} \\

\midrule

{Large (2)} & {Complex-2} & 
\makecell[l]{
$Y \sim \text{N}\left(\alpha_0 + \alpha_{1}X + \sum_{j=2}^{89} \alpha_j C_{j-1} + (b),\ \text{sd}_y = 1\right)$ \\
Where (b) is interaction and squared terms as detailed \\
in Tables~\ref{out_data_gen_complex_1} and~\ref{out_data_gen_complex_2}
} \\

\end{longtable}
\vspace{-6mm}
\begin{minipage}{30cm}
\footnotesize
\hspace{-3mm}{\textbf{Note:} For Small confounder set (1), $C_j$ refers to background and demographic confounder $j, (j=1,…,14)$}
\end{minipage}
\begin{minipage}{30cm}
\footnotesize
\vspace{-4mm}\hspace{0mm}{ and for Large confounder set (2), $C_j$ refers to background, demographic and metabolite confounders $j, (j=1,…,87)$ }
\end{minipage}
\begin{minipage}{30cm}
\footnotesize
\vspace{-8mm}\hspace{0.5mm}{as outlined in Table S\ref{confounder_set_summary}. $\alpha_0, \alpha_1,....\alpha_{15},....$ are parameter values, estimated using the Barwon Infant Study (BIS) data.}
\end{minipage}

\newpage

\begin{table}[h!]
\centering
\caption{Interaction and squared terms for the data-generation of exposure for Complex-2 (a)}
\label{exp_data_gen_complex_1}
\begin{tikzpicture}[element/.style={minimum width=1cm,minimum height=0.5cm}]
\matrix (m) [matrix of nodes,nodes={element},column sep=-\pgflinewidth, row sep=-\pgflinewidth,]{
         & $C_1$  & $C_2$ & $C_3$  & $C_4$ & $C_5$  & $C_6$ & $C_7$  & $C_8$ & $C_9$ & $C_{10}$ & $C_{11}$ & $C_{12}$ & $C_{13}$ & $C_{14}$ \\
$C_1$ & |[draw]|\phantom{$\times$} & |[draw]|$\times$ & |[draw]|$\times$ & |[draw]|\phantom{$\times$} & |[draw]|\phantom{$\times$} & |[draw]|$\times$ & |[draw]|$\times$ & |[draw]|\phantom{$\times$} & |[draw]|$\times$ & |[draw]|\phantom{$\times$} & |[draw]|$\times$ & |[draw]|$\times$ & |[draw]|$\times$ & |[draw]|$\times$ \\
$C_2$ & |[draw]|\phantom{$\times$} & |[draw]|\vspace{-1mm}\mytimes & |[draw]|\phantom{$\times$} & |[draw]|\phantom{$\times$} & |[draw]|\phantom{$\times$} & |[draw]|$\times$ & |[draw]|$\times$ & |[draw]|\phantom{$\times$} & |[draw]|$\times$ & |[draw]|$\times$ & |[draw]|$\times$ & |[draw]|$\times$ & |[draw]|$\times$ & |[draw]|$\times$ \\
$C_3$ & |[draw]|\phantom{$\times$} & |[draw]|\phantom{$\times$} & |[draw]|\phantom{$\times$} & |[draw]|\phantom{$\times$} & |[draw]|\phantom{$\times$} & |[draw]|$\times$ & |[draw]|$\times$ & |[draw]|\phantom{$\times$} & |[draw]|$\times$ & |[draw]|$\times$ & |[draw]|$\times$ & |[draw]|$\times$ & |[draw]|$\times$ & |[draw]|$\times$ \\
$C_4$ & |[draw]|\phantom{$\times$} & |[draw]|\phantom{$\times$} & |[draw]|\phantom{$\times$} & |[draw]|\phantom{$\times$} & |[draw]|\phantom{$\times$} & |[draw]|\phantom{$\times$} & |[draw]|\phantom{$\times$} & |[draw]|\phantom{$\times$} & |[draw]|\phantom{$\times$} & |[draw]|\phantom{$\times$} & |[draw]|\phantom{$\times$} & |[draw]|\phantom{$\times$} & |[draw]|\phantom{$\times$} & |[draw]|\phantom{$\times$} \\ 
$C_5$ & |[draw]|\phantom{$\times$} & |[draw]|\phantom{$\times$} & |[draw]|\phantom{$\times$} & |[draw]|\phantom{$\times$} & |[draw]|\phantom{$\times$} & |[draw]|\phantom{$\times$}& |[draw]|\phantom{$\times$} & |[draw]|\phantom{$\times$} & |[draw]|\phantom{$\times$} & |[draw]|\phantom{$\times$} & |[draw]|\phantom{$\times$} & |[draw]|\phantom{$\times$} & |[draw]|\phantom{$\times$} & |[draw]|\phantom{$\times$} \\  
$C_6$ & |[draw]|\phantom{$\times$} & |[draw]|\phantom{$\times$} & |[draw]|\phantom{$\times$} & |[draw]|\phantom{$\times$} & |[draw]|\phantom{$\times$} & |[draw]|\phantom{$\times$} & |[draw]|$\times$ & |[draw]|\phantom{$\times$} & |[draw]|$\times$ & |[draw]|\phantom{$\times$} & |[draw]|$\times$ & |[draw]|$\times$ & |[draw]|$\times$ & |[draw]|$\times$ \\  
$C_7$ & |[draw]|\phantom{$\times$} & |[draw]|\phantom{$\times$} & |[draw]|\phantom{$\times$} & |[draw]|\phantom{$\times$} & |[draw]|\phantom{$\times$} & |[draw]|\phantom{$\times$} & |[draw]|\phantom{$\times$} & |[draw]|\phantom{$\times$} & |[draw]|$\times$ & |[draw]|$\times$ & |[draw]|$\times$ & |[draw]|$\times$ & |[draw]|$\times$ & |[draw]|$\times$ \\  
$C_8$ & |[draw]|\phantom{$\times$} & |[draw]|\phantom{$\times$} & |[draw]|\phantom{$\times$} & |[draw]|\phantom{$\times$} & |[draw]|\phantom{$\times$} & |[draw]|\phantom{$\times$} & |[draw]|\phantom{$\times$} & |[draw]|\phantom{$\times$} & |[draw]|\phantom{$\times$} & |[draw]|\phantom{$\times$} & |[draw]|\phantom{$\times$} & |[draw]|\phantom{$\times$} & |[draw]|\phantom{$\times$} & |[draw]|\phantom{$\times$} \\  
$C_9$ & |[draw]|\phantom{$\times$} & |[draw]|\phantom{$\times$} & |[draw]|\phantom{$\times$} & |[draw]|\phantom{$\times$} & |[draw]|\phantom{$\times$} & |[draw]|\phantom{$\times$} & |[draw]|\phantom{$\times$} & |[draw]|\phantom{$\times$} & |[draw]|\phantom{$\times$} & |[draw]|$\times$ & |[draw]|$\times$ & |[draw]|$\times$ & |[draw]|$\times$ & |[draw]|$\times$ \\  
$C_{10}$ & |[draw]|\phantom{$\times$} & |[draw]|\phantom{$\times$} & |[draw]|\phantom{$\times$} & |[draw]|\phantom{$\times$} & |[draw]|\phantom{$\times$} & |[draw]|\phantom{$\times$} & |[draw]|\phantom{$\times$} & |[draw]|\phantom{$\times$} & |[draw]|\phantom{$\times$} & |[draw]|\vspace{-1mm}\mytimes & |[draw]|$\times$ & |[draw]|$\times$ & |[draw]|$\times$ & |[draw]|$\times$ \\ 
$C_{11}$ & |[draw]|\phantom{$\times$} & |[draw]|\phantom{$\times$} & |[draw]|\phantom{$\times$} & |[draw]|\phantom{$\times$} & |[draw]|\phantom{$\times$} & |[draw]|\phantom{$\times$} & |[draw]|\phantom{$\times$} & |[draw]|\phantom{$\times$} & |[draw]|\phantom{$\times$} & |[draw]|\phantom{$\times$} & |[draw]|\phantom{$\times$} & |[draw]|$\times$ & |[draw]|$\times$ & |[draw]|$\times$ \\  
$C_{12}$ & |[draw]|\phantom{$\times$} & |[draw]|\phantom{$\times$} & |[draw]|\phantom{$\times$} & |[draw]|\phantom{$\times$} & |[draw]|\phantom{$\times$} & |[draw]|\phantom{$\times$} & |[draw]|\phantom{$\times$} & |[draw]|\phantom{$\times$} & |[draw]|\phantom{$\times$} & |[draw]|\phantom{$\times$} & |[draw]|\phantom{$\times$} & |[draw]|\vspace{-1mm}\mytimes & |[draw]|$\times$ & |[draw]|$\times$ \\ 
$C_{13}$ & |[draw]|\phantom{$\times$} & |[draw]|\phantom{$\times$} & |[draw]|\phantom{$\times$} & |[draw]|\phantom{$\times$} & |[draw]|\phantom{$\times$} & |[draw]|\phantom{$\times$} & |[draw]|\phantom{$\times$} & |[draw]|\phantom{$\times$} & |[draw]|\phantom{$\times$} & |[draw]|\phantom{$\times$} & |[draw]|\phantom{$\times$} & |[draw]|\phantom{$\times$} & |[draw]|\phantom{$\times$} & |[draw]|$\times$ \\  
$C_{14}$ & |[draw]|\phantom{$\times$} & |[draw]|\phantom{$\times$} & |[draw]|\phantom{$\times$} & |[draw]|\phantom{$\times$} & |[draw]|\phantom{$\times$} & |[draw]|\phantom{$\times$} & |[draw]|\phantom{$\times$} & |[draw]|\phantom{$\times$} & |[draw]|\phantom{$\times$} & |[draw]|\phantom{$\times$} & |[draw]|\phantom{$\times$} & |[draw]|\phantom{$\times$} & |[draw]|\phantom{$\times$} & |[draw]|\vspace{-1mm}\mytimes \\  
{} & {} & {} & {}  &  {} & {} & {} & {}  &  {}  \\ 
$M_1$ & |[draw]|\phantom{$\times$} & |[draw]|\phantom{$\times$} & |[draw]|\phantom{$\times$} & |[draw]|\phantom{$\times$} & |[draw]|\phantom{$\times$} & |[draw]|\phantom{$\times$} & |[draw]|\phantom{$\times$} & |[draw]|\phantom{$\times$} & |[draw]|$\times$ & |[draw]|\phantom{$\times$} & |[draw]|\phantom{$\times$} & |[draw]|\phantom{$\times$} & |[draw]|\phantom{$\times$} & |[draw]|\phantom{$\times$} \\
$M_2$ & |[draw]|\phantom{$\times$} & |[draw]|\phantom{$\times$} & |[draw]|\phantom{$\times$} & |[draw]|\phantom{$\times$} & |[draw]|\phantom{$\times$} & |[draw]|\phantom{$\times$} & |[draw]|\phantom{$\times$} & |[draw]|\phantom{$\times$} & |[draw]|$\times$ & |[draw]|\phantom{$\times$} & |[draw]|\phantom{$\times$} & |[draw]|\phantom{$\times$} & |[draw]|\phantom{$\times$} & |[draw]|\phantom{$\times$} \\  
$M_4$ & |[draw]|\phantom{$\times$} & |[draw]|\phantom{$\times$} & |[draw]|\phantom{$\times$} & |[draw]|\phantom{$\times$} & |[draw]|\phantom{$\times$} & |[draw]|\phantom{$\times$} & |[draw]|\phantom{$\times$} & |[draw]|$\times$ & |[draw]|\phantom{$\times$} & |[draw]|\phantom{$\times$} & |[draw]|\phantom{$\times$} & |[draw]|\phantom{$\times$} & |[draw]|\phantom{$\times$} & |[draw]|\phantom{$\times$} \\  
$M_5$ & |[draw]|\phantom{$\times$} & |[draw]|\phantom{$\times$} & |[draw]|\phantom{$\times$} & |[draw]|\phantom{$\times$} & |[draw]|\phantom{$\times$} & |[draw]|\phantom{$\times$} & |[draw]|\phantom{$\times$} & |[draw]|\phantom{$\times$} & |[draw]|$\times$ & |[draw]|\phantom{$\times$} & |[draw]|\phantom{$\times$} & |[draw]|\phantom{$\times$} & |[draw]|\phantom{$\times$} & |[draw]|\phantom{$\times$} \\  
$M_6$ & |[draw]|\phantom{$\times$} & |[draw]|\phantom{$\times$} & |[draw]|\phantom{$\times$} & |[draw]|\phantom{$\times$} & |[draw]|\phantom{$\times$} & |[draw]|\phantom{$\times$} & |[draw]|\phantom{$\times$} & |[draw]|\phantom{$\times$} & |[draw]|$\times$ & |[draw]|\phantom{$\times$} & |[draw]|\phantom{$\times$} & |[draw]|\phantom{$\times$} & |[draw]|\phantom{$\times$} & |[draw]|\phantom{$\times$} \\ 
$M_{11}$ & |[draw]|\phantom{$\times$} & |[draw]|\phantom{$\times$} & |[draw]|\phantom{$\times$} & |[draw]|\phantom{$\times$} & |[draw]|\phantom{$\times$} & |[draw]|\phantom{$\times$} & |[draw]|\phantom{$\times$} & |[draw]|\phantom{$\times$} & |[draw]|$\times$ & |[draw]|\phantom{$\times$} & |[draw]|\phantom{$\times$} & |[draw]|\phantom{$\times$} & |[draw]|\phantom{$\times$} & |[draw]|\phantom{$\times$} \\
$M_{12}$ & |[draw]|\phantom{$\times$} & |[draw]|\phantom{$\times$} & |[draw]|\phantom{$\times$} & |[draw]|\phantom{$\times$} & |[draw]|\phantom{$\times$} & |[draw]|\phantom{$\times$} & |[draw]|\phantom{$\times$} & |[draw]|\phantom{$\times$} & |[draw]|\phantom{$\times$} & |[draw]|$\times$ & |[draw]|\phantom{$\times$} & |[draw]|\phantom{$\times$} & |[draw]|\phantom{$\times$} & |[draw]|\phantom{$\times$} \\  
$M_{15}$ & |[draw]|\phantom{$\times$} & |[draw]|\phantom{$\times$} & |[draw]|\phantom{$\times$} & |[draw]|\phantom{$\times$} & |[draw]|\phantom{$\times$} & |[draw]|\phantom{$\times$} & |[draw]|\phantom{$\times$} & |[draw]|\phantom{$\times$} & |[draw]|$\times$ & |[draw]|\phantom{$\times$} & |[draw]|\phantom{$\times$} & |[draw]|\phantom{$\times$} & |[draw]|\phantom{$\times$} & |[draw]|\phantom{$\times$} \\  
$M_{24}$ & |[draw]|\phantom{$\times$} & |[draw]|\phantom{$\times$} & |[draw]|\phantom{$\times$} & |[draw]|\phantom{$\times$} & |[draw]|\phantom{$\times$} & |[draw]|\phantom{$\times$} & |[draw]|\phantom{$\times$} & |[draw]|\phantom{$\times$} & |[draw]|\phantom{$\times$} & |[draw]|\phantom{$\times$} & |[draw]|$\times$ & |[draw]|\phantom{$\times$} & |[draw]|\phantom{$\times$} & |[draw]|\phantom{$\times$} \\  
$M_{29}$ & |[draw]|\phantom{$\times$} & |[draw]|\phantom{$\times$} & |[draw]|\phantom{$\times$} & |[draw]|\phantom{$\times$} & |[draw]|\phantom{$\times$} & |[draw]|\phantom{$\times$} & |[draw]|\phantom{$\times$} & |[draw]|\phantom{$\times$} & |[draw]|\phantom{$\times$} & |[draw]|\phantom{$\times$} & |[draw]|$\times$ & |[draw]|\phantom{$\times$} & |[draw]|\phantom{$\times$} & |[draw]|\phantom{$\times$} \\ 
$M_{30}$ & |[draw]|\phantom{$\times$} & |[draw]|$\times$ & |[draw]|\phantom{$\times$} & |[draw]|\phantom{$\times$} & |[draw]|\phantom{$\times$} & |[draw]|\phantom{$\times$} & |[draw]|\phantom{$\times$} & |[draw]|\phantom{$\times$} & |[draw]|$\times$ & |[draw]|\phantom{$\times$} & |[draw]|\phantom{$\times$} & |[draw]|\phantom{$\times$} & |[draw]|\phantom{$\times$} & |[draw]|\phantom{$\times$} \\ 
$M_{32}$ & |[draw]|\phantom{$\times$} & |[draw]|\phantom{$\times$} & |[draw]|\phantom{$\times$} & |[draw]|\phantom{$\times$} & |[draw]|\phantom{$\times$} & |[draw]|\phantom{$\times$} & |[draw]|\phantom{$\times$} & |[draw]|\phantom{$\times$} & |[draw]|\phantom{$\times$} & |[draw]|\phantom{$\times$} & |[draw]|$\times$ & |[draw]|\phantom{$\times$} & |[draw]|\phantom{$\times$} & |[draw]|\phantom{$\times$} \\ 
$M_{56}$ & |[draw]|\phantom{$\times$} & |[draw]|\phantom{$\times$} & |[draw]|\phantom{$\times$} & |[draw]|\phantom{$\times$} & |[draw]|\phantom{$\times$} & |[draw]|\phantom{$\times$} & |[draw]|\phantom{$\times$} & |[draw]|\phantom{$\times$} & |[draw]|\phantom{$\times$} & |[draw]|\phantom{$\times$} & |[draw]|$\times$ & |[draw]|\phantom{$\times$} & |[draw]|\phantom{$\times$} & |[draw]|\phantom{$\times$} \\ 
$M_{57}$ & |[draw]|\phantom{$\times$} & |[draw]|\phantom{$\times$} & |[draw]|\phantom{$\times$} & |[draw]|\phantom{$\times$} & |[draw]|\phantom{$\times$} & |[draw]|\phantom{$\times$} & |[draw]|$\times$ & |[draw]|\phantom{$\times$} & |[draw]|\phantom{$\times$} & |[draw]|\phantom{$\times$} & |[draw]|\phantom{$\times$} & |[draw]|\phantom{$\times$} & |[draw]|\phantom{$\times$} & |[draw]|\phantom{$\times$} \\ 
$M_{60}$ & |[draw]|\phantom{$\times$} & |[draw]|\phantom{$\times$} & |[draw]|\phantom{$\times$} & |[draw]|\phantom{$\times$} & |[draw]|\phantom{$\times$} & |[draw]|$\times$ & |[draw]|\phantom{$\times$} & |[draw]|\phantom{$\times$} & |[draw]|\phantom{$\times$} & |[draw]|\phantom{$\times$} & |[draw]|\phantom{$\times$} & |[draw]|\phantom{$\times$} & |[draw]|\phantom{$\times$} & |[draw]|\phantom{$\times$} \\ 
$M_{70}$ & |[draw]|\phantom{$\times$} & |[draw]|$\times$ & |[draw]|\phantom{$\times$} & |[draw]|\phantom{$\times$} & |[draw]|\phantom{$\times$} & |[draw]|\phantom{$\times$} & |[draw]|\phantom{$\times$} & |[draw]|\phantom{$\times$} & |[draw]|\phantom{$\times$} & |[draw]|\phantom{$\times$} & |[draw]|\phantom{$\times$} & |[draw]|\phantom{$\times$} & |[draw]|\phantom{$\times$} & |[draw]|\phantom{$\times$} \\ 
$M_{71}$ & |[draw]|\phantom{$\times$} & |[draw]|\phantom{$\times$} & |[draw]|\phantom{$\times$} & |[draw]|\phantom{$\times$} & |[draw]|\phantom{$\times$} & |[draw]|\phantom{$\times$} & |[draw]|\phantom{$\times$} & |[draw]|\phantom{$\times$} & |[draw]|\phantom{$\times$} & |[draw]|\phantom{$\times$} & |[draw]|\phantom{$\times$} & |[draw]|\phantom{$\times$} & |[draw]|\phantom{$\times$} & |[draw]|$\times$ \\  };

\node[above=0.25cm, xshift=6cm] at ($(m-1-2)!0.5!(m-1-3)$){\textbf{Background/demographic confounders}};
\node[rotate=90, xshift=-7.5cm] at ($(m-2-1)!0.5!(m-3-1)+(-1.25,0)$){\textbf{Background/demographic/metabolite confounders}};
\end{tikzpicture}
\end{table}

\vspace{-10mm}
\begin{minipage}{\linewidth}
\hspace{25mm}\footnotesize{\textbf{Note:} $\times$ indicates that interaction or squared term (bold) included.} \\ 
\end{minipage}


\begin{table}[h!]
\centering
\footnotesize
\caption{Interaction and squared terms for the data-generation of exposure for Complex-2 (a) (continued) \\ \footnotesize{\phantom{SpaceToCentreFootNote}[\textbf{Note:} $\times$ indicates that interaction or squared term (bold) included.]}}
\label{exp_data_gen_complex_2}

\end{table}

\newpage

\begin{table}[h!]
\centering
\caption{Interaction and squared terms for the data-generation of outcome for Complex-2 (b)}
\label{out_data_gen_complex_1}
%
\end{table}

\vspace{-8mm}
\begin{minipage}{\linewidth}
\hspace{32mm}\footnotesize{\textbf{Note:} $\times$ indicates that interaction or squared term (bold) included.} \\ 
\end{minipage}

\newpage


\begin{table}[h!]
\centering
\caption{Interaction and squared terms for the data-generation of outcome for Complex-2 (b) (continued)}
\label{out_data_gen_complex_2}

\end{table}

\vspace{-8mm}
\begin{minipage}{\linewidth}
\hspace{43mm}\footnotesize{\textbf{Note:} $\times$ indicates that interaction or squared term (bold) included.} \\ 
\end{minipage}

\newpage

\begin{table}[ht]
\centering
\caption{Parameter (coefficient) values used in the simulation study} 
\refstepcounter{SItab}\label{param_vals_1}
\begin{footnotesize}
\begin{tabular}{clr}
  \hline
Variable generated & Parameters & Values \\ 
  \hline
C1 & $\theta_1$ & 0.333 \\ 
&   $\theta_2$ & 0.333 \\ 
& 	 $\theta_3$ & 0.333 \\ 
\hline
C2 &  $\beta_0$ (Intercept) & 3.237 \\ 
 &  $\beta_1$ & -0.004 \\ 
 &   $\beta_2$ & -0.056 \\ 
 &  $\sigma_{Log bmi}$ & 0.198 \\
\hline
\end{tabular}
\end{footnotesize}
\end{table}

\newpage

\begin{table}[h!]
\centering
\caption{Parameter values (coefficients) used in the simulation study for the exposure (Simple-1 scenarios)}
\refstepcounter{SItab}\label{param_vals_2}
\begin{tabular}{cr}
  \hline
Parameters & Simple-1 \\ 
  \hline
$\gamma_{0}$ & 1.297 \\ 
  $\gamma_{1}$ & -0.019 \\ 
  $\gamma_{2}$ & -0.038 \\ 
  $\gamma_{3}$ & -0.002 \\ 
  $\gamma_{4}$ & 0.021 \\ 
  $\gamma_{5}$ & 0.032 \\ 
  $\gamma_{6}$ & -0.037 \\ 
  $\gamma_{7}$ & -0.051 \\ 
  $\gamma_{8}$ & 0.005 \\ 
  $\gamma_{9}$ & -0.043 \\ 
  $\gamma_{10}$ & 0.037 \\ 
  $\gamma_{11}$ & 0.0001 \\ 
  $\gamma_{12}$ & -0.002 \\ 
  $\gamma_{13}$ & -0.0001 \\ 
  $\gamma_{14}$ & -0.042 \\ 
  $\gamma_{15}$ & -0.003 \\ 
   \hline
\end{tabular}
\end{table}


\begin{table}[ht]
\centering
\caption{Parameter values (coefficients) used in the simulation study for the exposure (Complex-1a and Complex-1b scenarios)}
\refstepcounter{SItab}\label{param_vals_3}
\begin{tabular}[t]{crr}
  \hline
Parameters & Complex-1a & Complex-1b \\
  \hline
$\gamma_{0}$ & 1.750 & 1.750 \\ 
  $\gamma_{1}$ & -0.998 & -0.998 \\ 
  $\gamma_{2}$ & -1.098 & -1.098 \\ 
  $\gamma_{3}$ & 0.035 & 0.035 \\ 
  $\gamma_{4}$ & -1.983 & -1.983 \\ 
  $\gamma_{5}$ & -0.089 & -0.089 \\ 
  $\gamma_{6}$ & 0.014 & 0.014 \\ 
  $\gamma_{7}$ & -0.083 & -0.083 \\ 
  $\gamma_{8}$ & 0.057 & 0.057 \\ 
  $\gamma_{9}$ & -0.181 & -0.181 \\ 
  $\gamma_{10}$ & 1.024 & 1.024 \\ 
  $\gamma_{11}$ & 0.0001 & 0.0001 \\ 
  $\gamma_{12}$ & -0.091 & -0.091 \\ 
  $\gamma_{13}$ & -0.045 & -0.045 \\ 
  $\gamma_{14}$ & -0.609 & -0.609 \\ 
  $\gamma_{15}$ & -0.267 & -0.267 \\
  \addlinespace \addlinespace 
  $\gamma_{16}$ & -0.008 & -0.015 \\ 
  $\gamma_{17}$ & -0.011 & -0.023 \\ 
  $\gamma_{18}$ & -0.528 & -1.056 \\ 
  $\gamma_{19}$ & -0.874 & -1.748 \\ 
  $\gamma_{20}$ & 0.091 & 0.182 \\ 
  $\gamma_{21}$ & 0.047 & 0.095 \\ 
  $\gamma_{22}$ & 0.012 & 0.023 \\ 
  $\gamma_{23}$ & -0.040 & -0.080 \\ 
  $\gamma_{24}$ & 0.110 & 0.219 \\ 
  $\gamma_{25}$ & -0.047 & -0.093 \\ 
  $\gamma_{26}$ & 0.0001 & 0.0001 \\ 
  $\gamma_{27}$ & 0.0001 & 0.001 \\ 
  $\gamma_{28}$ & 0.105 & 0.210 \\ 
  $\gamma_{29}$ & 0.222 & 0.445 \\ 
  $\gamma_{30}$ & 0.001 & 0.002 \\ 
  $\gamma_{31}$ & -0.012 & -0.024 \\ 
  $\gamma_{32}$ & 0.629 & 1.259 \\ 
  $\gamma_{33}$ & 0.290 & 0.579 \\ 
  $\gamma_{34}$ & 0.165 & 0.329 \\ 
  $\gamma_{35}$ & -0.069 & -0.138 \\ 
  $\gamma_{36}$ & 0.025 & 0.050 \\ 
  $\gamma_{37}$ & -0.013 & -0.026 \\ 
  $\gamma_{38}$ & -0.006 & -0.013 \\ 
  $\gamma_{39}$ & -0.003 & -0.006 \\ 
  \hline
  \end{tabular}
  \begin{tabular}[t]{crr}
  \hline
Parameters & Complex-1a & Complex-1b \\
  \hline
  $\gamma_{40}$ & 0.0001 & 0.0001 \\ 
  $\gamma_{41}$ & 0.006 & 0.013 \\ 
  $\gamma_{42}$ & 0.001 & 0.002 \\ 
  $\gamma_{43}$ & -0.029 & -0.058 \\ 
  $\gamma_{44}$ & -0.010 & -0.020 \\ 
  $\gamma_{45}$ & -0.729 & -1.458 \\ 
  $\gamma_{46}$ & 0.067 & 0.135 \\ 
  $\gamma_{47}$ & -0.107 & -0.215 \\ 
  $\gamma_{48}$ & 0.0001 & 0.0001 \\ 
  $\gamma_{49}$ & 0.153 & 0.305 \\ 
  $\gamma_{50}$ & -0.029 & -0.059 \\ 
  $\gamma_{51}$ & -0.290 & -0.580 \\ 
  $\gamma_{52}$ & -0.007 & -0.015 \\ 
  $\gamma_{53}$ & 0.003 & 0.007 \\ 
  $\gamma_{54}$ & -0.072 & -0.145 \\ 
  $\gamma_{55}$ & 0.0001 & 0.0001 \\ 
  $\gamma_{56}$ & 0.042 & 0.085 \\ 
  $\gamma_{57}$ & -0.002 & -0.004 \\ 
  $\gamma_{58}$ & 0.411 & 0.823 \\ 
  $\gamma_{59}$ & -0.014 & -0.029 \\ 
  $\gamma_{60}$ & 0.004 & 0.009 \\ 
  $\gamma_{61}$ & 0.0001 & 0.0001 \\ 
  $\gamma_{62}$ & -0.002 & -0.005 \\ 
  $\gamma_{63}$ & 0.0001 & 0.0001 \\  
  $\gamma_{64}$ & -0.007 & -0.015 \\ 
  $\gamma_{65}$ & -0.005 & -0.010 \\ 
  $\gamma_{66}$ & 0.0001 & 0.0001 \\ 
  $\gamma_{67}$ & -0.102 & -0.203 \\ 
  $\gamma_{68}$ & -0.001 & -0.002 \\ 
  $\gamma_{69}$ & 0.101 & 0.202 \\ 
  $\gamma_{70}$ & 0.052 & 0.105 \\ 
  $\gamma_{71}$ & 0.0001 & 0.0001 \\ 
  $\gamma_{72}$ & 0.0001 & 0.0001 \\ 
  $\gamma_{73}$ & 0.0001 & 0.0001 \\ 
  $\gamma_{74}$ & 0.0001 & 0.0001 \\ 
  $\gamma_{75}$ & 0.006 & 0.012 \\ 
  $\gamma_{76}$ & 0.081 & 0.162 \\ 
  $\gamma_{77}$ & 0.088 & 0.176 \\ 
  $\gamma_{78}$ & 0.004 & 0.008 \\ 
  $\gamma_{79}$ & 0.003 & 0.007 \\ 
  $\gamma_{80}$ & -0.012 & -0.024 \\ 
   \hline
\end{tabular}
\end{table}

\newpage

\begin{table}[ht]
\centering
\caption{Parameter values (coefficients) used in the simulation study for the exposure (Simple-2 scenarios)}
\refstepcounter{SItab}\label{param_vals_4}
\begin{tabular}[t]{cr}
  \hline
Parameters & Simple-2 \\ 
  \hline
$\gamma_{0}$ & 1.315 \\ 
  $\gamma_{1}$ & 0.038 \\ 
  $\gamma_{2}$ & 0.035 \\ 
  $\gamma_{3}$ & 0.001 \\ 
  $\gamma_{4}$ & -0.001 \\ 
  $\gamma_{5}$ & -0.036 \\ 
  $\gamma_{6}$ & 0.040 \\ 
  $\gamma_{7}$ & -0.016 \\ 
  $\gamma_{8}$ & -0.003 \\ 
  $\gamma_{9}$ & 0.027 \\ 
  $\gamma_{10}$ & -0.0003  \\ 
  $\gamma_{11}$ & 0.0001 \\ 
  $\gamma_{12}$ & 0.002 \\ 
  $\gamma_{13}$ & 0.0005  \\ 
  $\gamma_{14}$ & 0.005 \\ 
  $\gamma_{15}$ & 0.012 \\ 
 \addlinespace \addlinespace 
  $\gamma_{16}$ & -0.023 \\ 
  $\gamma_{17}$ & 0.014 \\ 
  $\gamma_{18}$ & -0.021 \\ 
  $\gamma_{19}$ & -0.243 \\ 
  $\gamma_{20}$ & 0.020 \\ 
  $\gamma_{21}$ & 0.421 \\ 
  $\gamma_{22}$ & -0.425 \\ 
  $\gamma_{23}$ & 0.937 \\ 
  $\gamma_{24}$ & 0.155 \\ 
  $\gamma_{25}$ & -0.166 \\ 
  $\gamma_{26}$ & 0.004 \\ 
  $\gamma_{27}$ & 0.038 \\ 
  $\gamma_{28}$ & -0.151 \\ 
  $\gamma_{29}$ & -0.243 \\ 
  $\gamma_{30}$ & 0.551 \\ 
  $\gamma_{31}$ & -0.021 \\ 
  $\gamma_{32}$ & -0.404 \\ 
  $\gamma_{33}$ & 3.978 \\ 
  $\gamma_{34}$ & -0.193\\ 
  $\gamma_{35}$ & 0.218 \\ 
  $\gamma_{36}$ & -0.296 \\ 
  $\gamma_{37}$ & -1.029 \\ 
  $\gamma_{38}$ & 0.946 \\ 
  $\gamma_{39}$ & 0.147 \\ 
  $\gamma_{40}$ & -3.534 \\ 
  $\gamma_{41}$ & -1.129 \\
  $\gamma_{42}$ & 0.388 \\
  $\gamma_{43}$ & -0.451 \\ 
    \hline
  \end{tabular}
  \begin{tabular}[t]{cr}
   \hline
Parameters & Simple-2 \\ 
  \hline
  $\gamma_{44}$ & -0.103 \\ 
  $\gamma_{45}$ & 0.054 \\ 
  $\gamma_{46}$ & -0.342 \\ 
  $\gamma_{47}$ & -0.143 \\ 
  $\gamma_{48}$ & 0.423 \\ 
  $\gamma_{49}$ & 0.079 \\ 
  $\gamma_{50}$ & 0.117 \\ 
  $\gamma_{51}$ & 1.198 \\ 
  $\gamma_{52}$ & -1.898 \\ 
  $\gamma_{53}$ & 1.540 \\ 
  $\gamma_{54}$ & -8.103 \\ 
  $\gamma_{55}$ & -0.022 \\ 
  $\gamma_{56}$ & -0.780 \\ 
  $\gamma_{57}$ & 1.755 \\ 
  $\gamma_{58}$ & -4.55 \\ 
  $\gamma_{59}$ & 7.848 \\ 
  $\gamma_{60}$ & -7.035 \\ 
  $\gamma_{61}$ & -1.566 \\ 
  $\gamma_{62}$ & 9.572 \\ 
  $\gamma_{63}$ & 0.790 \\ 
  $\gamma_{64}$ & -1.338 \\ 
  $\gamma_{65}$ & 3.400 \\ 
  $\gamma_{66}$ & -3.107 \\ 
  $\gamma_{67}$ & 3.091 \\ 
  $\gamma_{68}$ & 0.726 \\ 
  $\gamma_{69}$ & -3.034 \\ 
  $\gamma_{70}$ & -0.026 \\ 
  $\gamma_{71}$ & 0.051 \\ 
  $\gamma_{72}$ & 0.058 \\ 
  $\gamma_{73}$ & -0.002 \\ 
  $\gamma_{74}$ & 0.032 \\ 
  $\gamma_{75}$ & -0.007 \\ 
  $\gamma_{76}$ & 0.013 \\ 
  $\gamma_{77}$ & 0.030 \\ 
  $\gamma_{78}$ & -0.034 \\ 
  $\gamma_{79}$ & 0.001 \\ 
  $\gamma_{80}$ & 0.040 \\ 
  $\gamma_{81}$ & -0.012 \\ 
  $\gamma_{82}$ & 0.035 \\ 
  $\gamma_{83}$ & 0.002 \\ 
  $\gamma_{84}$ & 0.024 \\ 
  $\gamma_{85}$ & -0.017 \\ 
  $\gamma_{86}$ & 0.015 \\ 
  $\gamma_{87}$ & 0.004 \\ 
  $\gamma_{88}$ & 0.008 \\ 
   \hline
\end{tabular}
\end{table}

\newpage

\begin{table}[ht]
\centering
\begin{footnotesize}
\caption{Parameter values (coefficients) used in the simulation study for the exposure (Complex-2 scenarios)}
\refstepcounter{SItab}\label{param_vals_5}
\begin{tabular}[t]{cr}
  \hline
Parameters & Complex-2 \\ 
  \hline
$\gamma_{0}$ & 5.885 \\ 
  $\gamma_{1}$ & 0.238 \\ 
  $\gamma_{2}$ & -0.595 \\ 
  $\gamma_{3}$ & 0.012 \\ 
  $\gamma_{4}$ & 0.912 \\ 
  $\gamma_{5}$ & 0.002 \\ 
  $\gamma_{6}$ & 0.013 \\ 
  $\gamma_{7}$ & 0.561 \\ 
  $\gamma_{8}$ & -0.061 \\ 
  $\gamma_{9}$ & -0.030 \\ 
  $\gamma_{10}$ & 0.427 \\ 
  $\gamma_{11}$ & -0.001 \\ 
  $\gamma_{12}$ & -0.323 \\ 
  $\gamma_{13}$ & 0.0001 \\ 
  $\gamma_{14}$ & -0.846 \\ 
  $\gamma_{15}$ & -0.775 \\ 
  $\gamma_{16}$ & -0.086 \\ 
  $\gamma_{17}$ & 0.135 \\ 
  $\gamma_{18}$ & 0.020 \\ 
  $\gamma_{19}$ & -0.441 \\ 
  $\gamma_{20}$ & 0.066 \\ 
  $\gamma_{21}$ & -0.789 \\ 
  $\gamma_{22}$ & -0.621 \\ 
  $\gamma_{23}$ & -0.864 \\ 
  $\gamma_{24}$ & -1.071 \\ 
  $\gamma_{25}$ & 1.167 \\ 
  $\gamma_{26}$ & 0.106 \\ 
  $\gamma_{27}$ & -0.296 \\ 
  $\gamma_{28}$ & 0.086 \\ 
  $\gamma_{29}$ & -0.145 \\ 
  $\gamma_{30}$ & -0.910 \\
  $\gamma_{31}$ & 0.382 \\ 
  $\gamma_{32}$ & -0.051 \\ 
  $\gamma_{33}$ & 15.641 \\ 
  $\gamma_{34}$ & 0.249 \\ 
  $\gamma_{35}$ & -1.150 \\ 
  $\gamma_{36}$ & 0.502 \\ 
  $\gamma_{37}$ & -2.311 \\ 
  $\gamma_{38}$ & 0.741\\ 
  $\gamma_{39}$ & -0.275 \\ 
  $\gamma_{40}$ & -10.073 \\ 
  $\gamma_{41}$ & -3.425 \\ 
  $\gamma_{42}$ & 0.372 \\ 
  $\gamma_{43}$ & 1.053\\ 
  $\gamma_{44}$ & 0.110 \\ 
  $\gamma_{45}$ & -0.115 \\ 
  $\gamma_{46}$ & -0.127 \\ 
  $\gamma_{47}$ & -0.334 \\ 
  $\gamma_{48}$ & 0.126 \\ 
  $\gamma_{49}$ & 0.177 \\ 
  $\gamma_{50}$ & 0.298 \\
  $\gamma_{51}$ & 2.783 \\ 
  $\gamma_{52}$ & -4.859 \\ 
  $\gamma_{53}$ & 4.431 \\ 
  $\gamma_{54}$ & -1.421 \\ 
  $\gamma_{55}$ & -0.032 \\ 
  $\gamma_{56}$ & 3.530\\ 
              \hline
  \end{tabular}
    \begin{tabular}[t]{cr}
   \hline
Parameters & Complex-2 \\ 
  \hline
  $\gamma_{57}$ & 14.641 \\ 
  $\gamma_{58}$ & -9.326 \\ 
  $\gamma_{59}$ & -14.181 \\ 
  $\gamma_{60}$ & 18.410 \\
  $\gamma_{61}$ & -2.712 \\ 
  $\gamma_{62}$ & -7.694 \\ 
  $\gamma_{63}$ & -2.915 \\ 
  $\gamma_{64}$ & -8.897 \\ 
  $\gamma_{65}$ & 7.075 \\ 
  $\gamma_{66}$ & 8.473 \\ 
  $\gamma_{67}$ & -9.826 \\ 
  $\gamma_{68}$ & 1.187 \\ 
  $\gamma_{69}$ & 2.511 \\ 
  $\gamma_{70}$ & 0.029 \\ 
  $\gamma_{71}$ & -0.332 \\ 
  $\gamma_{72}$ & 0.192 \\ 
  $\gamma_{73}$ & -0.029 \\ 
  $\gamma_{74}$ & 0.044 \\ 
  $\gamma_{75}$ & -0.065 \\ 
  $\gamma_{76}$ & 0.019 \\ 
  $\gamma_{77}$ & 0.006 \\ 
  $\gamma_{78}$ & 0.007 \\ 
  $\gamma_{79}$ & 0.037 \\ 
  $\gamma_{80}$ & -0.011 \\ 
  $\gamma_{81}$ & 0.001 \\ 
  $\gamma_{82}$ & 0.004 \\ 
  $\gamma_{83}$ & -0.001 \\ 
  $\gamma_{84}$ & 0.009 \\ 
  $\gamma_{85}$ & -0.050 \\ 
  $\gamma_{86}$ & 0.028 \\ 
  $\gamma_{87}$ & -0.022 \\ 
  $\gamma_{88}$ & 0.020 \\
 $\gamma_{89}$ & 0.012 \\ 
  $\gamma_{90}$ & -0.028 \\
  $\gamma_{91}$ & 0.006 \\ 
  $\gamma_{92}$ & -0.022\\ 
  $\gamma_{93}$ & -0.017 \\ 
  $\gamma_{94}$ & 0.022 \\ 
  $\gamma_{95}$ & 0.072 \\ 
  $\gamma_{96}$ & 0.013 \\ 
  $\gamma_{97}$ & 0.039 \\ 
  $\gamma_{98}$ & 0.004 \\ 
  $\gamma_{99}$ & -0.060\\ 
  $\gamma_{100}$ & -0.037 \\
  $\gamma_{101}$ & 0.001 \\ 
  $\gamma_{102}$ & 0.122 \\ 
  $\gamma_{103}$ & 0.047 \\ 
  $\gamma_{104}$ & -0.017 \\
  $\gamma_{105}$ & -0.020 \\ 
  $\gamma_{106}$ & -0.057 \\ 
  $\gamma_{107}$ & -0.189 \\ 
  $\gamma_{108}$ & -0.031 \\ 
  $\gamma_{109}$ & 0.087\\ 
  $\gamma_{110}$ & -0.005 \\ 
  $\gamma_{111}$ & -0.030 \\ 
  $\gamma_{112}$ & 0.080 \\
   $\gamma_{113}$ & 0.009 \\ 
        \hline
  \end{tabular}
 
\end{footnotesize}
\end{table}

\newpage

\begin{table}[ht]
\centering
\begin{footnotesize}
\caption{Parameter values (coefficients) used in the simulation study for the exposure (Complex-2 scenarios) (continued)}
\refstepcounter{SItab}\label{param_vals_6}
\begin{tabular}[t]{cr}
      \hline
Parameters & Complex-2\\
   \hline 
  $\gamma_{114}$ & 0.253 \\ 
  $\gamma_{115}$ & -0.312 \\ 
  $\gamma_{116}$ & 109.536 \\ 
  $\gamma_{117}$ & -79.449 \\ 
  $\gamma_{118}$ & -30.781 \\ 
  $\gamma_{119}$ & -107.067 \\ 
  $\gamma_{120}$ & 77.438\\ 
  $\gamma_{121}$ & 30.413 \\ 
  $\gamma_{122}$ & 0.103 \\ 
  $\gamma_{123}$ & -0.125 \\ 
  $\gamma_{124}$ & -0.049 \\ 
  $\gamma_{125}$ & 0.069 \\ 
  $\gamma_{126}$ & -0.043 \\ 
  $\gamma_{127}$ & -0.004 \\ 
  $\gamma_{128}$ & 0.064 \\ 
  $\gamma_{129}$ & 0.047 \\ 
  $\gamma_{130}$ & -0.019 \\ 
  $\gamma_{131}$ & -0.091 \\ 
  $\gamma_{132}$ & 0.056 \\ 
  $\gamma_{133}$ & -0.099 \\ 
  $\gamma_{134}$ & 0.067 \\ 
  $\gamma_{135}$ & -0.016 \\ 
  $\gamma_{136}$ & 0.012 \\ 
  $\gamma_{137}$ & -0.277 \\ 
  $\gamma_{138}$ & -0.052 \\ 
  $\gamma_{139}$ & 0.007 \\ 
  $\gamma_{140}$ & -0.014 \\ 
  $\gamma_{141}$ & 0.192 \\ 
  $\gamma_{142}$ & -0.014 \\ 
  $\gamma_{143}$ & -0.062 \\ 
  $\gamma_{144}$ & 0.064 \\ 
  $\gamma_{145}$ & -0.027 \\ 
  $\gamma_{146}$ & 0.016 \\ 
   $\gamma_{147}$ & 0.028 \\ 
  $\gamma_{148}$ & -0.035 \\ 
  $\gamma_{149}$ & -0.006 \\ 
  $\gamma_{150}$ & 0.016 \\ 
  $\gamma_{151}$ & -0.053 \\ 
  $\gamma_{152}$ & 0.042 \\ 
  $\gamma_{153}$ & 0.034 \\ 
  $\gamma_{154}$ & -0.123 \\ 
  $\gamma_{155}$ & 0.063 \\ 
  $\gamma_{156}$ & -0.018 \\ 
  $\gamma_{157}$ & 0.0001 \\ 
  $\gamma_{158}$ & 0.038 \\ 
  $\gamma_{159}$ & 0.030 \\ 
  $\gamma_{160}$ & -0.015 \\ 
  $\gamma_{161}$ & 0.002 \\ 
  $\gamma_{162}$ & 0.042 \\ 
  $\gamma_{163}$ & 0.035 \\ 
  $\gamma_{164}$ & 0.030 \\ 
  $\gamma_{165}$ & -0.002 \\ 
  $\gamma_{166}$ & 0.046 \\ 
  $\gamma_{167}$ & 0.003 \\ 
  $\gamma_{168}$ & -0.003 \\ 
  $\gamma_{169}$ & 0.002 \\ 
  $\gamma_{170}$ & 0.012 \\
    $\gamma_{171}$ & -0.004 \\ 
  $\gamma_{172}$ & -0.062 \\ 
  $\gamma_{173}$ & -0.068 \\ 
          \hline
  \end{tabular}  
\begin{tabular}[t]{cr}
   \hline
Parameters & Complex-2 \\ 
  \hline
  $\gamma_{174}$ & -0.061 \\ 
  $\gamma_{175}$ & -0.001 \\ 
  $\gamma_{176}$ & -0.010 \\ 
  $\gamma_{177}$ & -0.154 \\ 
  $\gamma_{178}$ & -0.144 \\ 
  $\gamma_{179}$ &  0.0001 \\ 
  $\gamma_{180}$ &  0.0001 \\ 
  $\gamma_{181}$ & -0.018 \\ 
  $\gamma_{182}$ & 0.068 \\ 
  $\gamma_{183}$ & 0.001 \\ 
  $\gamma_{184}$ & -0.0004 \\ 
  $\gamma_{185}$ & -0.212 \\ 
  $\gamma_{186}$ & -0.206 \\ 
  $\gamma_{187}$ & 0.041 \\ 
  $\gamma_{188}$ & 0.002 \\ 
  $\gamma_{189}$ & -0.004 \\ 
  $\gamma_{190}$ & -0.013 \\ 
  $\gamma_{191}$ & -0.0003 \\ 
  $\gamma_{192}$ & -0.002 \\ 
  $\gamma_{193}$ &  0.0001 \\ 
  $\gamma_{194}$ &  0.0001 \\ 
  $\gamma_{195}$ &  0.0001 \\ 
  $\gamma_{196}$ & -0.004 \\ 
  $\gamma_{197}$ & 0.003 \\ 
  $\gamma_{198}$ & -0.054 \\ 
  $\gamma_{199}$ & -0.012 \\ 
  $\gamma_{200}$ & 0.388 \\ 
  $\gamma_{201}$ & -0.0001 \\ 
  $\gamma_{202}$ & -0.022 \\ 
  $\gamma_{203}$ & 0.008 \\ 
  $\gamma_{204}$ & -0.098 \\ 
  $\gamma_{205}$ & 0.082 \\ 
  $\gamma_{206}$ & -0.005 \\ 
  $\gamma_{207}$ & -0.013 \\ 
  $\gamma_{208}$ & 0.0001 \\ 
  $\gamma_{209}$ & -0.003 \\ 
  $\gamma_{210}$ & 0.002 \\ 
  $\gamma_{211}$ & -0.078 \\ 
  $\gamma_{212}$ & 0.017 \\ 
  $\gamma_{213}$ & -0.004 \\ 
  $\gamma_{214}$ & 0.0001 \\ 
  $\gamma_{215}$ & 0.004 \\ 
  $\gamma_{216}$ & -0.0001 \\ 
  $\gamma_{217}$ & -0.006 \\ 
  $\gamma_{218}$ & 0.008 \\ 
  $\gamma_{219}$ & 0.0001 \\ 
  $\gamma_{220}$ & -0.025 \\ 
  $\gamma_{221}$ & -0.002 \\ 
  $\gamma_{222}$ & -0.073 \\ 
  $\gamma_{223}$ & 0.030 \\ 
  $\gamma_{224}$ & 0.0001 \\ 
  $\gamma_{225}$ & 0.0001 \\ 
  $\gamma_{226}$ & 0.0001 \\ 
  $\gamma_{227}$ & 0.0001 \\ 
  $\gamma_{228}$ & -0.001 \\ 
  $\gamma_{229}$ & 0.109 \\ 
  $\gamma_{230}$ & 0.026 \\ 
  $\gamma_{231}$ & -0.006 \\ 
  $\gamma_{232}$ & 0.0004 \\ 
  $\gamma_{233}$ & -0.117 \\ 
 \hline
\end{tabular}
\end{footnotesize}
\end{table}

\newpage


\begin{table}[ht]
\centering
\caption{Parameter values (coefficients) used in the simulation study for the outcome (Simple-1 scenarios)}
\refstepcounter{SItab}\label{param_vals_7}
\begin{tabular}{crrrr}
  \hline
\textbf{Parameters} & \multicolumn{4}{c}{\textbf{Values}} \\
\cmidrule{2-5} & n=200 & n=500 & n=1000 & n=2000 \\
\hline
$\alpha_{0}$ & -0.670* & -0.670* & -0.670* & -0.670* \\ 
  $\alpha_{1}$ & 0.450 & 0.300 & 0.200 & 0.150 \\ 
  $\alpha_{2}$ & -0.159 & -0.159 & -0.159 & -0.159 \\ 
  $\alpha_{3}$ & 0.021 & 0.021 & 0.021 & 0.021 \\ 
  $\alpha_{4}$ & 0.011 & 0.011 & 0.011 & 0.011 \\ 
  $\alpha_{5}$ & 0.276 & 0.276  & 0.276 & 0.276 \\ 
  $\alpha_{6}$ & 0.039 & 0.039 & 0.039 & 0.039 \\ 
  $\alpha_{7}$ & -0.071 & -0.071 & -0.071 & -0.071 \\ 
  $\alpha_{8}$ & -0.109 & -0.109  & -0.109  & -0.109  \\ 
  $\alpha_{9}$ & -0.019 & -0.019 & -0.019 & -0.019 \\ 
  $\alpha_{10}$ & -0.056 & -0.056 & -0.056 & -0.056 \\ 
  $\alpha_{11}$ & 0.049 & 0.049 & 0.049 & 0.049 \\ 
  $\alpha_{12}$ & 0.0001 & 0.0001 & 0.0001 & 0.0001 \\ 
  $\alpha_{13}$ & 0.057 & 0.057 & 0.057 & 0.057 \\ 
  $\alpha_{14}$ & 0.006 & 0.006 & 0.006 & 0.006 \\ 
  $\alpha_{15}$ & -0.007 & -0.007 & -0.007 & -0.007 \\ 
  $\alpha_{16}$ & 0.010 & 0.010 & 0.010 & 0.010 \\ 
   \hline
\end{tabular}
\begin{minipage}{30cm}
\footnotesize
\vspace{0mm}\hspace{54mm}{*Intercepts in all outcome models were modified so that the mean}
\end{minipage}
\begin{minipage}{30cm}
\footnotesize
\vspace{-5mm}\hspace{54mm}{ of Y remained as 0.}
\end{minipage}
\end{table}

\newpage


\begin{table}[ht]
\centering
\begin{footnotesize}
\caption{Parameter values (coefficients) used in the simulation study for the outcome (Complex-1a and Complex-1b scenarios)}
\refstepcounter{SItab}\label{param_vals_8}
\begin{tabular}{crrrrrrrr}
  \hline
Parameters & \multicolumn{4}{c}{\textbf{Complex-1a}} & \multicolumn{4}{c}{\textbf{Complex-1b}} \\
\cmidrule{2-9} & n=200 & n=500 & n=1000 & n=2000 & n=200 & n=500 & n=1000 & n=2000 \\
  \hline
$\alpha_{0}$ & -8.554* & -8.554* & -8.554* & -8.554* & -8.554* & -8.554* & -8.554* & -8.554* \\ 
  $\alpha_{1}$ & -0.900 & -1.250 & -1.500 & -1.650 & -2.000 & -2.750 & -3.125 & -3.425 \\ 
  $\alpha_{2}$ & 1.436 & 1.436 & 1.436 & 1.436 & 1.436 & 1.436 & 1.436 &  1.436\\ 
  $\alpha_{3}$ & 2.596 & 2.596 & 2.596 & 2.596 & 2.596 & 2.596 & 2.596 & 2.596 \\ 
  $\alpha_{4}$ & -0.294 & -0.294 & -0.294 & -0.294 & -0.294 & -0.294  & -0.294  & -0.294 \\ 
  $\alpha_{5}$ & 2.247 & 2.247 & 2.247 & 2.247 & 2.247 & 2.247 & 2.247 & 2.247 \\
\hline
\end{tabular}
\end{footnotesize}
\begin{minipage}{30cm}
\footnotesize
\vspace{0mm}\hspace{29mm}{*Intercepts in all outcome models were modified so that the mean of Y remained as 0.}
\end{minipage}
\end{table}

\newpage

\begin{table}[ht]
\centering
\begin{footnotesize}
\caption{Parameter values (coefficients) used in the simulation study for the outcome (Complex-1a and Complex-1b scenarios) (continued)}
\refstepcounter{SItab}\label{param_vals_8}
\begin{tabular}{crrrrrrrr}
  \hline
Parameters & \multicolumn{4}{c}{\textbf{Complex-1a}} & \multicolumn{4}{c}{\textbf{Complex-1b}} \\
\cmidrule{2-9} & n=200 & n=500 & n=1000 & n=2000 & n=200 & n=500 & n=1000 & n=2000 \\
  \hline
  $\alpha_{51}$ & 0.029 & 0.029 & 0.029 & 0.029 & 0.058 & 0.058 & 0.058 & 0.058 \\ 
  $\alpha_{52}$ & -0.002 & -0.002 & -0.002 & -0.002 & -0.004 & -0.004 & -0.004 & -0.004 \\ 
  $\alpha_{53}$ & 0.149 & 0.149 & 0.149 & 0.149 & 0.297 & 0.297 & 0.297 & 0.297 \\ 
  $\alpha_{54}$ & 0.056 & 0.056 & 0.056 & 0.056 & 0.112 & 0.112 & 0.112 &  0.112\\ 
  $\alpha_{55}$ & -0.344 & -0.344 & -0.344 & -0.344 & -0.688 & -0.688 & -0.688 & -0.688 \\ 
  $\alpha_{56}$ & -1.451 & -1.451 & -1.451 & -1.451 & -2.902 & -2.902 & -2.902 & -2.902 \\
\hline
\end{tabular}
\end{footnotesize}
\end{table}

\newpage


\begin{table}[ht]
\centering
\begin{footnotesize}
\caption{Parameter values (coefficients) used in the simulation study for the outcome (Simple-2 scenarios)}
\refstepcounter{SItab}\label{param_vals_9}
\begin{tabular}[t]{crrrr}
  \hline
Parameters & \multicolumn{4}{c}{\textbf{Simple-2}} \\
\cmidrule{2-5} & n=200 & n=500 & n=1000 & n=2000 \\
  \hline
  $\alpha_{0}$ & -0.585* & -0.585* & -0.585* & -0.585* \\ 
  $\alpha_{1}$ & -0.650 & -0.400 & -0.300 & -0.180 \\ 
  $\alpha_{2}$ & -0.131 & -0.131 & -0.131 & -0.131 \\ 
  $\alpha_{3}$ & -0.089 &  -0.089 &  -0.089 &   -0.089\\ 
  $\alpha_{4}$ & 0.011 & 0.011 & 0.011 & 0.011 \\ 
  $\alpha_{5}$ & 0.161 & 0.161 & 0.161 & 0.161 \\
\hline
        \end{tabular}
  \begin{tabular}[t]{cr}
  \hline
Parameters & \multicolumn{1}{c}{\textbf{Simple-2}} \\
\cmidrule{2-2} & All n \\
\hline
  $\alpha_{45}$ & 0.714 \\ 
 $\alpha_{46}$ & -0.364 \\ 
  $\alpha_{47}$ & 1.952 \\ 
  $\alpha_{48}$ & 1.335 \\ 
  $\alpha_{49}$ & -3.255 \\ 
  $\alpha_{50}$ & -0.130 \\ 
  $\alpha_{51}$ & 2.139 \\ 
  $\alpha_{52}$ & -29.682 \\ 
  $\alpha_{53}$ & 56.226 \\ 
  $\alpha_{54}$ & -56.620 \\ 
  $\alpha_{55}$ & 77.082 \\ 
  $\alpha_{56}$ & 0.020 \\ 
  $\alpha_{57}$ & 1.921 \\ 
  $\alpha_{58}$ & 95.457 \\  
  $\alpha_{59}$ & -104.288 \\ 
  $\alpha_{60}$ & -75.338 \\ 
  $\alpha_{61}$ & 81.208 \\ 
  $\alpha_{62}$ & -54.754 \\ 
  $\alpha_{63}$ & -59.246 \\  
  $\alpha_{64}$ & -1.987 \\ 
  $\alpha_{65}$ & -55.721 \\ 
  $\alpha_{66}$ & 79.113 \\ 
  $\alpha_{67}$ & 40.556 \\ 
  $\alpha_{68}$ & -43.555 \\ 
  $\alpha_{69}$ & 20.110 \\ 
  $\alpha_{70}$ & 18.066 \\ 
  $\alpha_{71}$ & -0.041 \\  
  $\alpha_{72}$ & -0.069 \\ 
  $\alpha_{73}$ & 0.866 \\  
  $\alpha_{74}$ & 0.238 \\ 
  $\alpha_{75}$ & -0.069 \\  
  $\alpha_{76}$ & -0.215 \\ 
  $\alpha_{77}$ & 0.119 \\  
  $\alpha_{78}$ & 0.106 \\ 
  $\alpha_{79}$ & 0.100 \\ 
  $\alpha_{80}$ & -0.294 \\ 
  $\alpha_{81}$ & 1.022 \\ 
  $\alpha_{82}$ & -0.352 \\ 
  $\alpha_{83}$ & -0.019 \\ 
  $\alpha_{84}$ & -0.042 \\ 
  $\alpha_{85}$ & 0.021 \\ 
  $\alpha_{86}$ & -0.065 \\ 
  $\alpha_{87}$ & -0.104 \\ 
  $\alpha_{88}$ & -0.013 \\ 
  $\alpha_{89}$ & -0.050 \\ 
  \hline
\end{tabular}   
\end{footnotesize}
\begin{minipage}{30cm}
\footnotesize
\vspace{0.5mm}\hspace{38.5mm}{*Intercepts in all outcome models were modified so that the mean of Y remained as 0.}
\end{minipage}
\end{table}

\newpage



\begin{table}[ht]
\centering
\begin{footnotesize}
\caption{Parameter values (coefficients) used in the simulation study for the outcome (Complex-2 scenarios)}
\refstepcounter{SItab}\label{param_vals_10}
\begin{tabular}[t]{crrrr}
  \hline
Parameters & \multicolumn{4}{c}{\textbf{Complex-2}} \\
\cmidrule{2-5} & n=200 & n=500 & n=1000 & n=2000 \\
  \hline
  $\alpha_{0}$ & 55.665* & 55.665* & 55.665* & 55.665* \\ 
  $\alpha_{1}$ & 1.700 & 0.200 & -1.400 & -1.600 \\ 
  $\alpha_{2}$ & 9.642 &  9.642 &  9.642 &  9.642 \\ 
  $\alpha_{3}$ & -4.643 & -4.643 & -4.643 &  -4.643\\ 
  $\alpha_{4}$ & 0.764 & 0.764 & 0.764 & 0.764 \\ 
  $\alpha_{5}$ & 2.521 & 2.521 & 2.521 & 2.521 \\
\hline
        \end{tabular}
  \begin{tabular}[t]{cr}
  \hline
Parameters & \multicolumn{1}{c}{\textbf{Complex-2}} \\
\cmidrule{2-2} & All n \\
  \hline
  $\alpha_{43}$ & -8.769  \\ 
  $\alpha_{44}$ & 29.406 \\ 
  $\alpha_{45}$ & 0.908   \\ 
  $\alpha_{46}$ & 0.336  \\ 
  $\alpha_{47}$ & 8.362  \\ 
  $\alpha_{48}$ & 3.739 \\ 
  $\alpha_{49}$ & -1.080 \\ 
  $\alpha_{50}$ & 0.449 \\ 
  $\alpha_{51}$ & -11.492 \\ 
  $\alpha_{52}$ & 77.843  \\ 
  $\alpha_{53}$ & -141.213 \\ 
  $\alpha_{54}$ & 145.671 \\ 
  $\alpha_{55}$ & 9.983 \\ 
  $\alpha_{56}$ & -1.103 \\ 
  $\alpha_{57}$ & 15.394 \\ 
  $\alpha_{58}$ & 561.710 \\ 
  $\alpha_{59}$ & -579.619 \\ 
  $\alpha_{60}$ & -244.248 \\ 
  $\alpha_{61}$ & 134.083 \\ 
  $\alpha_{62}$ & 42.367 \\ 
  $\alpha_{63}$ & -73.747 \\ 
  $\alpha_{64}$ & -13.347 \\ 
  $\alpha_{65}$ & -332.314 \\ 
  $\alpha_{66}$ & 436.301 \\ 
  $\alpha_{67}$ & 136.267 \\ 
  $\alpha_{68}$ & -76.090 \\ 
  $\alpha_{69}$ & -20.368 \\ 
  $\alpha_{70}$ & 21.325 \\ 
  $\alpha_{71}$ & 0.385 \\ 
  $\alpha_{72}$ & 0.870 \\ 
  $\alpha_{73}$ & -6.663 \\ 
  $\alpha_{74}$ & -0.069 \\ 
  $\alpha_{75}$ & -0.435 \\ 
  $\alpha_{76}$ & 0.860 \\ 
  $\alpha_{77}$ & 0.278 \\ 
  $\alpha_{78}$ & -1.344 \\ 
  $\alpha_{79}$ & 0.656 \\ 
  $\alpha_{80}$ & -1.360 \\ 
  $\alpha_{81}$ & 6.218 \\ 
  $\alpha_{82}$ & -2.228 \\ 
  $\alpha_{83}$ & -0.207 \\ 
  $\alpha_{84}$ & -0.179 \\  
  $\alpha_{85}$ & -2.158 \\ 
    \hline
        \end{tabular}
  \begin{tabular}[t]{cr}
  \hline
Parameters & \multicolumn{1}{c}{\textbf{Complex-2}} \\
\cmidrule{2-2} & All n \\
\hline
  $\alpha_{86}$ & -0.270 \\
  $\alpha_{87}$ & 0.646 \\
  $\alpha_{88}$ & -0.534 \\
  $\alpha_{89}$ & -0.122 \\
  $\alpha_{90}$ & 0.0001 \\
  $\alpha_{91}$ & -0.0002 \\ 
  $\alpha_{92}$ & -0.121 \\
  $\alpha_{93}$ & -0.386 \\
  $\alpha_{94}$ & -0.137 \\
  $\alpha_{95}$ & 0.175 \\
  $\alpha_{96}$ & 0.407 \\
  $\alpha_{97}$ & 0.558 \\
  $\alpha_{98}$ & -0.870 \\
  $\alpha_{99}$ & -0.389 \\
  $\alpha_{100}$ & -0.250 \\
  $\alpha_{101}$ & 0.665 \\
  $\alpha_{102}$ & 0.045 \\ 
  $\alpha_{103}$ & 0.674 \\
  $\alpha_{104}$ & -0.450 \\
  $\alpha_{105}$ & 3.542 \\
  $\alpha_{106}$ & -4.306 \\ 
  $\alpha_{107}$ & 1.092 \\
  $\alpha_{108}$ & -0.639 \\
  $\alpha_{109}$ & 0.551 \\
  $\alpha_{110}$ & -0.134 \\
  $\alpha_{111}$ & 4.600 \\
  $\alpha_{112}$ & -4.973 \\
  $\alpha_{113}$ & 1.400 \\
  $\alpha_{114}$ & -1.605 \\
  $\alpha_{115}$ & -18.815 \\
  $\alpha_{116}$ & 6.464 \\
  $\alpha_{117}$ & 0.768 \\
  $\alpha_{118}$ & 17.957 \\
  $\alpha_{119}$ & -4.810 \\
  $\alpha_{120}$ & -0.189 \\
  $\alpha_{121}$ & 0.116 \\
  $\alpha_{122}$ & 0.217 \\
  $\alpha_{123}$ & -4.262 \\
  $\alpha_{124}$ & 4.484 \\
  $\alpha_{125}$ & 0.179 \\
  $\alpha_{126}$ & -0.463 \\
  $\alpha_{127}$ & 0.210 \\
  $\alpha_{128}$ & -0.653 \\
   \hline
\end{tabular}  
\end{footnotesize}
\begin{minipage}{30cm}
\footnotesize
\vspace{0mm}\hspace{19mm}{*Intercepts in all outcome models were modified so that the mean of Y remained as 0.}
\end{minipage}
\end{table}

\newpage

\begin{table}[ht]
\centering
\begin{footnotesize}
\caption{Parameter values (coefficients) used in the simulation study for the outcome (Complex-2 scenarios) (continued)}
\refstepcounter{SItab}\label{param_vals_11}
\begin{tabular}[t]{lr}
  \hline
Parameters & \multicolumn{1}{c}{\textbf{Complex-2}} \\
\cmidrule{2-2} & All n \\
  \hline
  $\alpha_{129}$ & -0.019 \\ 
  $\alpha_{130}$ & 0.273 \\ 
  $\alpha_{131}$ & 0.089 \\ 
  $\alpha_{132}$ & -0.673 \\ 
  $\alpha_{133}$ & 0.330 \\ 
  $\alpha_{134}$ & -0.032 \\ 
  $\alpha_{135}$ & -3.182 \\ 
  $\alpha_{136}$ & -0.037 \\ 
  $\alpha_{137}$ & 0.038 \\ 
  $\alpha_{138}$ & 2.075 \\  
  $\alpha_{139}$ & 0.042 \\ 
  $\alpha_{140}$ & -0.026 \\  
  $\alpha_{141}$ & -0.009 \\ 
  $\alpha_{142}$ & 0.0001 \\ 
  $\alpha_{143}$ & 1.273 \\ 
  $\alpha_{144}$ & 0.093 \\ 
  $\alpha_{145}$ & -0.061 \\ 
  $\alpha_{146}$ & 0.005 \\ 
  $\alpha_{147}$ & -0.025 \\ 
  $\alpha_{148}$ & 0.002 \\ 
  $\alpha_{149}$ & -0.002 \\ 
  $\alpha_{150}$ & 0.001 \\ 
  $\alpha_{151}$ & 0.002 \\ 
  $\alpha_{152}$ & 0.056 \\ 
  $\alpha_{153}$ & -0.040 \\ 
  $\alpha_{154}$ & 0.049 \\ 
  $\alpha_{155}$ & 0.0001 \\ 
  $\alpha_{156}$ & -0.001 \\ 
  $\alpha_{157}$ & 0.049 \\ 
  $\alpha_{158}$ & 0.007 \\ 
  $\alpha_{159}$ & -1.273 \\ 
  $\alpha_{160}$ & 0.0001 \\ 
  $\alpha_{161}$ & -0.006 \\  
  $\alpha_{162}$ & -0.082 \\ 
  $\alpha_{163}$ & -0.034 \\ 
  $\alpha_{164}$ & 0.036  \\ 
  $\alpha_{165}$ & -3.915 \\ 
  $\alpha_{166}$ & 2.134 \\ 
  $\alpha_{167}$ & 1.643 \\ 
  $\alpha_{168}$ & 0.063 \\ 
  $\alpha_{169}$ & -0.082 \\ 
  $\alpha_{170}$ & -0.930 \\ 
  $\alpha_{171}$ & 0.303 \\ 
  $\alpha_{172}$ & -0.002 \\
      \hline
  \end{tabular}
  \begin{tabular}[t]{lc}
  \hline
Parameters & \multicolumn{1}{c}{\textbf{Complex-2}} \\
\cmidrule{2-2} & All n \\
  \hline
  $\alpha_{173}$ & -0.003 \\ 
  $\alpha_{174}$ & -0.466 \\  
  $\alpha_{175}$ & 1.204 \\ 
  $\alpha_{176}$ & 0.077 \\ 
  $\alpha_{177}$ & 0.083 \\ 
  $\alpha_{178}$ & -1.889 \\ 
  $\alpha_{179}$ & -1.461 \\ 
  $\alpha_{180}$ & 1.472 \\ 
  $\alpha_{181}$ & 1.764 \\ 
  $\alpha_{182}$ & -0.044 \\ 
  $\alpha_{183}$ & -0.051 \\ 
  $\alpha_{184}$ & -0.012 \\ 
  $\alpha_{185}$ & 0.048 \\ 
  $\alpha_{186}$ & 0.0001 \\ 
  $\alpha_{187}$ & -0.033 \\ 
  $\alpha_{188}$ & -0.003 \\ 
  $\alpha_{189}$ & -0.040 \\ 
  $\alpha_{190}$ & 0.088 \\ 
  $\alpha_{191}$ & -4.076 \\  
  $\alpha_{192}$ & -0.121 \\ 
  $\alpha_{193}$ & 0.736 \\ 
  $\alpha_{194}$ & -0.005 \\ 
  $\alpha_{195}$ & 1.750 \\ 
  $\alpha_{196}$ & -0.036 \\ 
  $\alpha_{197}$ & -1.473 \\ 
  $\alpha_{198}$ & -1.068 \\ 
  $\alpha_{199}$ & -0.154 \\ 
  $\alpha_{200}$ & -0.989 \\ 
  $\alpha_{201}$ & 0.001 \\ 
  $\alpha_{202}$ & 0.505 \\ 
  $\alpha_{203}$ & 0.025 \\ 
  $\alpha_{204}$ & 2.341 \\ 
  $\alpha_{205}$ & -0.796 \\ 
  $\alpha_{206}$ & 0.180 \\ 
  $\alpha_{207}$ & 0.0001 \\ 
  $\alpha_{208}$ & 0.119 \\ 
  $\alpha_{209}$ & 0.008 \\ 
  $\alpha_{210}$ & 0.119 \\ 
  $\alpha_{211}$ & 0.087 \\ 
  $\alpha_{212}$ & 0.0001 \\ 
  $\alpha_{213}$ & -1.170 \\ 
  $\alpha_{214}$ & -0.032 \\ 
  $\alpha_{215}$ & 2.262 \\ 
  $\alpha_{216}$ & -0.051 \\
    \hline
    \end{tabular}
  \begin{tabular}[t]{lc}
  \hline
Parameters & \multicolumn{1}{c}{\textbf{Complex-2}} \\
\cmidrule{2-2} & All n \\
  \hline
  $\alpha_{217}$ & 0.0001 \\ 
  $\alpha_{218}$ & 0.0001 \\ 
  $\alpha_{219}$ & 0.003 \\ 
  $\alpha_{220}$ & -0.0005 \\ 
  $\alpha_{221}$ & -0.059 \\ 
  $\alpha_{222}$ & 0.777 \\ 
  $\alpha_{223}$ & -0.037 \\ 
  $\alpha_{224}$ & 2.530 \\ 
  $\alpha_{225}$ & 1.118 \\ 
  $\alpha_{226}$ & -1.766 \\  
  $\alpha_{227}$ & 1.109 \\ 
  $\alpha_{228}$ & 0.014 \\ 
  $\alpha_{229}$ & 9.805 \\ 
  $\alpha_{230}$ & 2.544 \\ 
  $\alpha_{231}$ & -2.017 \\ 
  $\alpha_{232}$ & -2.290 \\ 
  $\alpha_{233}$ & -13.214 \\ 
  $\alpha_{234}$ & 15.207 \\ 
  $\alpha_{235}$ & 1.587 \\ 
  $\alpha_{236}$ & -11.014 \\ 
  $\alpha_{237}$ & -0.537 \\ 
  $\alpha_{238}$ & -0.181 \\ 
  $\alpha_{239}$ & -0.979 \\ 
  $\alpha_{240}$ & -1.690 \\ 
  $\alpha_{241}$ & -0.168 \\ 
   \hline
\end{tabular}  
\end{footnotesize}
\end{table}


\clearpage

\section{Super Learner candidate libraries and screening algorithms \label{sec:si5}}

In Super Learner, a (1) \textbf{reduced library} (non-data adaptive and adaptive parametric approaches), and (2) a \textbf{full library} (the reduced library in addition to non-parametric data-adaptive approaches) was considered:

\begin{table}[ht]
\centering
\caption{Overview of learners used within Super Learner}

\begin{footnotesize}
%
} \\

\quad SL.mean (marginal mean) & & \\
\quad SL.glm (GLM) & & \\
\quad SL.glm.interaction (pairwise interactions) & & \\
\quad SL.bayesglm (Bayesian GLM) & & \\
\quad SL.gam (GAM) & & \\

\textbf{b. Parametric, data-adaptive Library:} & & \\
\quad SL.glmnet (Lasso/Elastic net) & & \\

\textbf{c. Non-parametric, data-adaptive (ML) Library:} & & \\
\quad SL.ranger (Random forest) & & \\
\quad SL.rpart (Regression tree) & & \\
\quad SL.nnet (Neural network) & & \\
\quad SL.earth (MARS) & & \\
\quad SL.xgboost (Gradient boosting) & & \\
\quad SL.ipregbragg (Bagging tree) & & \\
\quad SL.svm (Support vector machine) & & \\

\hline

& & \\
\end{tabular}
\end{footnotesize}

\end{table}

\newpage

\section{True values ACE used in the simulation study \label{sec:si6}}

\begin{footnotesize}
\begin{table}[ht]
\centering
\caption{True values of the ACE for each scenario (data-generating mechanism and sample size), that were used when computing the performance measures. A large simulated dataset of size 1000000 was used to obtain the true values.}
\refstepcounter{SItab}\label{SL_ACE}
\begin{tabular}[t]{ccrr}
  \hline
\textbf{Mechanism} & \textbf{Sample size} & \textbf{Main effect of X} & \textbf{true ACE}\\
\hline
 & 200 & 0.45 & 0.45 \\
Simple-1 & 500 & 0.30 & 0.30 \\
 & 1000 & 0.20 & 0.20 \\
 & 2000 & 0.15 & 0.15 \\
\hline
& 200 & -0.90 & 0.95 \\
Complex-1a & 500 & -1.25 & 0.60 \\
 & 1000 & -1.50 & 0.35 \\
 & 2000 & -1.65 & 0.20 \\
\hline
& 200 & -2.00 & 1.70 \\
Complex-1b & 500 & -2.75 & 0.94 \\
 & 1000 & -3.13 & 0.58 \\
 & 2000 & -3.43 & 0.27 \\
\hline
 & 200 & -0.65 & -0.65 \\
Simple-2 & 500 & -0.40 & -0.40 \\
& 1000 & -0.30 & -0.30 \\
& 2000 & -0.18 & -0.18 \\
\hline
 & 200 & 1.70 & 5.08 \\
Complex-2 & 500 & 0.20 & 3.58 \\
 & 1000 & -1.40 & 1.98 \\
& 2000 & -1.60 & 1.78 \\
\hline
\end{tabular}
\end{table}
\end{footnotesize}

\section{Simulation study results\label{sec:si7}} 

AIPW failed to produce sensible results for a few datasets (2-12 datasets across scenarios) when CF was applied. Specifically, the standard error was $> 10$ times the median standard error or the absolute value of the point estimate was $> 5$ times the absolute value of the median point estimate, where the median SE and point estimates were calculated using the results obtained across the $2000$ datasets. These results were excluded from Figures 2-5, and further details regarding the scenarios and number of datasets affected, and the value of performance measures including and excluding results from the affected datasets are presented in Table S26. Full simulation results that include all datasets ($n_{sim} = 2000$) are presented in Tables S27 - S34.


\begin{footnotesize}
\hspace{-6cm}\begin{table}[ht]
\centering
\caption{Scenarios, number of datasets affected, and performance measures including and excluding affected datasets}
\refstepcounter{SItab}\label{Excluded_results}
\begin{tabular}[t]{ccccccccc}
  \hline
\textbf{Mechanism} &  \makecell{\textbf{Sample} \\ \textbf{size}} & \textbf{Library} & \makecell{\textbf{Folds used} \\ \textbf{in CF}} & \makecell{\textbf{No. of} \\ \textbf{datasets\^{}}} & \textbf{Relbias*} & \textbf{EmpSE*} & \textbf{RelModSE*} & \textbf{CP*} \\
\hline
Simple-1 & 1000 & Full & 5 & 2 & \makecell{-1.21 [0.30]} & \makecell{0.12 [0.08]} & \makecell{87.52 [4.53]} & \makecell{95.82 [95.82]} \\
Simple-1 & 2000 & Reduced & 2 & 5 & \makecell{2.30 [1.21]} & \makecell{0.07 [0.06]} & \makecell{59.50 [0.58]} & \makecell{95.09 [95.13]} \\
Simple-2 & 2000 & Full & 10 & 4 & \makecell{1.22 [2.18]} & \makecell{0.14 [0.12]} & \makecell{72.41 [-41.76]} & \makecell{71.83 [71.65]} \\
Complex-1a & 2000 & Full & 2 & 9 & \makecell{-0.67 [-0.32]} & \makecell{0.16 [0.11]} & \makecell{39.50 [-3.55]} & \makecell{93.79 [93.77]} \\
Simple-2 & 2000 & Full & 5 & 12 & \makecell{-10.69 [-6.62]} & \makecell{0.20 [0.12]} & \makecell{15.34 [-40.89]} & \makecell{71.14 [71.17]} \\
Simple-2 & 2000 & Full & 10 & 5 & \makecell{-15.77 [-10.23]} & \makecell{0.31 [0.13]} & \makecell{-2.12 [-37.48]} & \makecell{71.82 [71.64]} \\
\hline
\end{tabular}
\begin{minipage}{30cm}
\footnotesize
\vspace{0mm}\hspace{11mm}{\^{}Number of affected datasets; *Performance measure including results from all datasets [performance measure excluding affected datasets in brackets].}
\end{minipage}
\end{table}
\end{footnotesize}

\begin{footnotesize}
\vspace*{-10mm}\hspace*{2cm}\begin{table*}[ht]
\caption{Simulation study results: Performance measures for the \textbf{Reduced library in SL} and sample size of \textbf{200} ($n_{sim} = 2000$)}
\refstepcounter{SItab}\label{PM_Red_200}
\fontsize{9pt}{9pt}\selectfont
\begingroup
\hspace{-0.5cm}
\begin{tabular}{ccccccccccccccc}
  \hline
Method & CF & Mechanism & Bias & \makecell{Bias \\ (MCSE)} & RelBias & EmpSE & \makecell{EmpSE \\ (MCSE)} & ModSE & \makecell{ModSE \\ (MCSE)} & RelModSE & \makecell{RelModSE \\ (MCSE)} & CP & \makecell{CP \\ (MCSE)} \\ 
  \hline
AIPW & No CF & {} & 0.0116 & 0.009 & 2.573 & 0.4070 & 0.006 & 0.42 & 0.00010 & 2.17 & 1.62 & 91.07 & 0.0064 \\
  TMLE & No CF & {}  & -0.0009 & 0.004 & -0.194 & 0.1760 & 0.003 & 0.15 & 0.00010 & -14.00 & 1.36 & 90.85 & 0.0064 \\ 
  AIPW & 2-folds & {}  & 0.0039 & 0.023 & 0.875 & 0.9950 & 0.016 & 1.17 & 0.00020 & 17.21 & 1.88 & 94.64 & 0.0051 \\ 
  TMLE & 2-folds & Simple-1 & 0.0038 & 0.005 & 0.847 & 0.2260 & 0.004 & 0.22 & 0.00010 & -2.28 & 1.54 & 95.05 & 0.0048 \\ 
  AIPW & 5-folds & {} & 0.0033 & 0.017 & 0.734 & 0.7690 & 0.012 & 0.81 & 0.00020 & 5.17 & 1.69 & 95.66 & 0.0046 \\ 
  TMLE & 5-folds & {} & 0.0014 & 0.004 & 0.309 & 0.1880 & 0.003 & 0.19 & \makecell{\textless 0.0001} & 2.33 & 1.62 & 95.45 & 0.0047 \\ 
  AIPW & 10-folds & {} & 0.0121 & 0.014 & 2.685 & 0.6360 & 0.010 & 0.73 & 0.00010 & 14.33 & 1.84 & 95.67 & 0.0046 \\
\hline
\end{tabular}
\endgroup
\end{table*}
\end{footnotesize}

\newpage

\hspace*{-2.8cm}\begin{table}[h!]
\centering
\vspace{2cm}\caption{Simulation study results: Performance measures for the \textbf{Full library in SL} and sample size of \textbf{200} ($n_{sim} = 2000$)}
\refstepcounter{SItab}\label{PM_Full_200}
\fontsize{9pt}{9pt}\selectfont
\begingroup
\hspace*{-0.5cm}\begin{tabular}{ccccccccccccccc}
  \hline
Method & CF & Mechanism & Bias & \makecell{Bias \\ (MCSE)} & RelBias & EmpSE & \makecell{EmpSE \\ (MCSE)} & ModSE & \makecell{ModSE \\ (MCSE)} & RelModSE & \makecell{RelModSE \\ (MCSE)} & CP & \makecell{CP \\ (MCSE)} \\ 
  \hline
  AIPW & No CF & {} & 0.0160 & 0.007 & 3.548 & 0.3210 & 0.005 & 0.25 & 0.00010 & -23.49 & 1.21 & 83.41 & 0.0083 \\ 
  TMLE & No CF & {} & 0.0016 & 0.004 & 0.346 & 0.1760 & 0.003 & 0.13 & 0.00010 & -25.61 & 1.18 & 83.35 & 0.0083 \\ 
  AIPW & 2-folds & {} & -0.0080 & 0.017 & -1.784 & 0.7460 & 0.012 & 0.84 & 0.00020 & 12.22 & 1.79 & 95.05 & 0.0049 \\ 
  TMLE & 2-folds & Simple-1 & -0.0009 & 0.005 & -0.195 & 0.2250 & 0.004 & 0.24 & 0.00010 & 8.78 & 1.72 & 96.30 & 0.0042 \\ 
  AIPW & 5-folds & {} & -0.0211 & 0.015 & -4.694 & 0.6710 & 0.011 & 0.67 & 0.00020 & -0.05 & 1.60 & 95.96 & 0.0045 \\ 
  TMLE & 5-folds & {} & 0.0008 & 0.004 & 0.176 & 0.1920 & 0.003 & 0.21 & \makecell{\textless 0.0001}& 9.74 & 1.74 & 96.45 & 0.0041 \\ 
  AIPW & 10-folds & {} & 0.0143 & 0.013 & 3.187 & 0.5750 & 0.009 & 0.65 & 0.00010 & 12.80 & 1.80 & 96.73 & 0.0040 \\
\hline
\end{tabular}
\endgroup
\end{table}

\newpage

\hspace*{-2.8cm}\begin{table}[h!]
\centering
\vspace{2cm}\caption{Simulation study results: Performance measures for the \textbf{Reduced library in SL} and sample size of \textbf{500} ($n_{sim} = 2000$)}
\refstepcounter{SItab}\label{PM_Red_500}
\fontsize{9pt}{9pt}\selectfont
\begingroup
\hspace*{-0.5cm}\begin{tabular}{ccccccccccccccc}
  \hline
Method & CF & Mechanism & Bias & \makecell{Bias \\ (MCSE)} & RelBias & EmpSE & \makecell{EmpSE \\ (MCSE)} & ModSE & \makecell{ModSE \\ (MCSE)} & RelModSE & \makecell{RelModSE \\ (MCSE)} & CP & \makecell{CP \\ (MCSE)} \\ 
  \hline
AIPW & No CF & {} & 0.0022 & 0.004 & 0.749 & 0.1580 & 0.002 & 0.19 & \makecell{\textless 0.0001}& 20.72 & 1.91 & 92.43 & 0.0059 \\ 
  TMLE &  No CF & {} & -0.0005 & 0.002 & -0.167 & 0.1100 & 0.002 & 0.10 & \makecell{\textless 0.0001}& -9.63 & 1.43 & 92.40 & 0.0059 \\ 
  AIPW & 2-folds & {} & 0.0159 & 0.013 & 5.310 & 0.5920 & 0.009 & 0.81 & 0.00010 & 36.19 & 2.18 & 94.83 & 0.0050 \\ 
  TMLE & 2-folds & Simple-1 & 0.0025 & 0.003 & 0.831 & 0.1240 & 0.002 & 0.12 & \makecell{\textless 0.0001}& -0.54 & 1.57 & 95.10 & 0.0048 \\ 
  AIPW & 5-folds & {} & 0.0060 & 0.010 & 2.001 & 0.4250 & 0.007 & 0.44 & 0.00010 & 3.47 & 1.64 & 95.81 & 0.0045 \\ 
  TMLE & 5-folds & {} & 0.0001 & 0.003 & 0.023 & 0.1130 & 0.002 & 0.11 & \makecell{\textless 0.0001}& -1.37 & 1.56 & 94.95 & 0.0049 \\ 
  AIPW & 10-folds & {} & 0.0012 & 0.007 & 0.412 & 0.3140 & 0.005 & 0.35 & 0.00010 & 10.05 & 1.76 & 95.25 & 0.0048 \\
\hline
\end{tabular}
\endgroup
\end{table}

\newpage

\hspace*{-2.8cm}\begin{table}[h!]
\centering
\vspace{2cm}\caption{Simulation study results: Performance measures for the \textbf{Full library in SL} and sample size of \textbf{500} ($n_{sim} = 2000$)}
\refstepcounter{SItab}\label{PM_Full_500}
\fontsize{9pt}{9pt}\selectfont
\begingroup
\hspace*{-0.5cm}\begin{tabular}{ccccccccccccccc}
  \hline
Method & CF & Mechanism & Bias & \makecell{Bias \\ (MCSE)} & RelBias & EmpSE & \makecell{EmpSE \\ (MCSE)} & ModSE & \makecell{ModSE \\ (MCSE)} & RelModSE & \makecell{RelModSE \\ (MCSE)} & CP & \makecell{CP \\ (MCSE)} \\ 
  \hline
  AIPW & No CF & {} & -0.0005 & 0.003 & -0.180 & 0.1190 & 0.002 & 0.12 & \makecell{\textless 0.0001}& -1.77 & 1.56 & 86.62 & 0.0076 \\ 
  TMLE & No CF & {} & 0.0006 & 0.002 & 0.182 & 0.1100 & 0.002 & 0.09 & \makecell{\textless 0.0001}& -19.80 & 1.27 & 86.85 & 0.0076 \\ 
  AIPW & 2-folds & {} & 0.0066 & 0.008 & 2.186 & 0.3620 & 0.006 & 0.46 & 0.00010 & 27.03 & 2.02 & 94.73 & 0.0050 \\ 
  TMLE & 2-folds & Simple-1 & -0.0009 & 0.003 & -0.310 & 0.1270 & 0.002 & 0.13 & \makecell{\textless 0.0001}& 0.31 & 1.59 & 95.15 & 0.0048 \\ 
  AIPW & 5-folds & {} & 0.0045 & 0.005 & 1.505 & 0.2330 & 0.004 & 0.30 & \makecell{\textless 0.0001}& 30.86 & 2.08 & 95.15 & 0.0048 \\ 
  TMLE & 5-folds & {} & 0.0022 & 0.003 & 0.745 & 0.1140 & 0.002 & 0.12 & \makecell{\textless 0.0001}& 3.17 & 1.63 & 95.30 & 0.0047 \\ 
  AIPW & 10-folds & {} & -0.0000 & 0.004 & -0.017 & 0.1960 & 0.003 & 0.21 & \makecell{\textless 0.0001}& 4.85 & 1.67 & 95.59 & 0.0046 \\
\hline
\end{tabular}
\endgroup
\end{table}

\hspace*{-2.8cm}\begin{table}[h!]
\centering
\vspace{2cm}\caption{Simulation study results: Performance measures for the \textbf{Reduced library in SL} and sample size of \textbf{1000} ($n_{sim} = 2000$)}
\refstepcounter{SItab}\label{PM_Red_1000}
\fontsize{9pt}{9pt}\selectfont
\begingroup
\hspace*{-0.5cm}\begin{tabular}{ccccccccccccccc}
  \hline
Method & CF & Mechanism & Bias & \makecell{Bias \\ (MCSE)} & RelBias & EmpSE & \makecell{EmpSE \\ (MCSE)} & ModSE & \makecell{ModSE \\ (MCSE)} & RelModSE & \makecell{RelModSE \\ (MCSE)} & CP & \makecell{CP \\ (MCSE)} \\ 
  \hline
AIPW &  No CF & {} & 0.0009 & 0.002 & 0.469 & 0.0910 & 0.001 & 0.07 & \makecell{\textless 0.0001}& -19.12 & 1.28 & 94.24 & 0.0052 \\ 
  TMLE &  No CF & {} & -0.0002 & 0.002 & -0.095 & 0.0740 & 0.001 & 0.07 & \makecell{\textless 0.0001}& -2.99 & 1.53 & 94.25 & 0.0052 \\ 
  AIPW &  2-folds & {} & 0.0026 & 0.005 & 1.290 & 0.2030 & 0.003 & 0.27 & \makecell{\textless 0.0001}& 32.67 & 2.11 & 94.91 & 0.0049 \\ 
  TMLE &  2-folds & Simple-1 & -0.0013 & 0.002 & -0.629 & 0.0790 & 0.001 & 0.08 & \makecell{\textless 0.0001}& 1.94 & 1.61 & 95.25 & 0.0048 \\ 
  AIPW &  5-folds & {} & 0.0002 & 0.003 & 0.099 & 0.1420 & 0.002 & 0.18 & \makecell{\textless 0.0001}& 26.96 & 2.01 & 95.78 & 0.0045 \\ 
  TMLE &  5-folds & {} & -0.0001 & 0.002 & -0.068 & 0.0750 & 0.001 & 0.08 & \makecell{\textless 0.0001}& 2.23 & 1.62 & 95.55 & 0.0046 \\ 
  AIPW &  10-folds & {} & -0.0052 & 0.005 & -2.596 & 0.2340 & 0.004 & 0.22 & 0.00010 & -4.14 & 1.52 & 95.87 & 0.0045 \\
\hline
\end{tabular}
\endgroup
\end{table}

\hspace*{-2.8cm}\begin{table}[h!]
\centering
\vspace{2cm}\caption{Simulation study results: Performance measures for the \textbf{Full library in SL} and sample size of \textbf{1000} ($n_{sim} = 2000$)}
\refstepcounter{SItab}\label{PM_Full_1000}
\fontsize{9pt}{9pt}\selectfont
\begingroup
\hspace*{-0.5cm}\begin{tabular}{ccccccccccccccc}
  \hline
Method & CF & Mechanism & Bias & \makecell{Bias \\ (MCSE)} & RelBias & EmpSE & \makecell{EmpSE \\ (MCSE)} & ModSE & \makecell{ModSE \\ (MCSE)} & RelModSE & \makecell{RelModSE \\ (MCSE)} & CP & \makecell{CP \\ (MCSE)} \\ 
  \hline  
  AIPW &  No CF & {} & -0.0004 & 0.002 & -0.203 & 0.0920 & 0.002 & 0.07 & \makecell{\textless 0.0001}& -27.05 & 1.16 & 90.69 & 0.0065 \\ 
  TMLE &  No CF & {} & 0.0005 & 0.002 & 0.230 & 0.0740 & 0.001 & 0.06 & \makecell{\textless 0.0001}& -12.31 & 1.39 & 90.50 & 0.0066 \\ 
  AIPW &  2-folds & {} & 0.0010 & 0.003 & 0.509 & 0.1120 & 0.002 & 0.14 & \makecell{\textless 0.0001}& 27.02 & 2.02 & 95.52 & 0.0046 \\ 
  TMLE &  2-folds & Simple-1 & 0.0005 & 0.002 & 0.258 & 0.0790 & 0.001 & 0.08 & \makecell{\textless 0.0001}& 4.42 & 1.65 & 95.85 & 0.0045 \\ 
  AIPW &  5-folds & {} & -0.0022 & 0.003 & -1.122 & 0.1200 & 0.002 & 0.23 & \makecell{\textless 0.0001}& 87.52 & 2.98 & 95.82 & 0.0045 \\ 
  TMLE &  5-folds & {} & 0.0005 & 0.002 & 0.257 & 0.0760 & 0.001 & 0.08 & \makecell{\textless 0.0001}& 4.58 & 1.65 & 96.00 & 0.0044 \\ 
  AIPW &  10-folds & {} & 0.0009 & 0.002 & 0.460 & 0.0810 & 0.001 & 0.08 & \makecell{\textless 0.0001}& 2.13 & 1.62 & 96.32 & 0.0042 \\
\hline
\end{tabular}
\endgroup
\end{table}

\hspace*{-2.8cm}\begin{table}[h!]
\centering
\vspace{2cm}\caption{Simulation study results: Performance measures for the \textbf{Reduced library in SL} and sample size of \textbf{2000} ($n_{sim} = 2000$)}
\refstepcounter{SItab}\label{PM_Red_1000}
\fontsize{9pt}{9pt}\selectfont
\begingroup
\hspace*{-0.5cm}\begin{tabular}{ccccccccccccccc}
  \hline
Method & CF & Mechanism & Bias & \makecell{Bias \\ (MCSE)} & RelBias & EmpSE & \makecell{EmpSE \\ (MCSE)} & ModSE & \makecell{ModSE \\ (MCSE)} & RelModSE & \makecell{RelModSE \\ (MCSE)} & CP & \makecell{CP \\ (MCSE)} \\ 
  \hline  
AIPW &  No CF & {} & 0.0013 & 0.001 & 0.850 & 0.0540 & 0.001 & 0.05 & \makecell{\textless 0.0001}& -4.31 & 1.51 & 93.60 & 0.0055 \\ 
  TMLE &  No CF & {}  & 0.0012 & 0.001 & 0.793 & 0.0540 & 0.001 & 0.05 & \makecell{\textless 0.0001}& -4.49 & 1.51 & 93.50 & 0.0055 \\ 
  AIPW &  2-folds & {}  & 0.0014 & 0.001 & 0.964 & 0.0630 & 0.001 & 0.08 & \makecell{\textless 0.0001}& 19.78 & 1.90 & 95.29 & 0.0047 \\ 
  TMLE &  2-folds & Simple-1 & 0.0007 & 0.001 & 0.465 & 0.0560 & 0.001 & 0.06 & \makecell{\textless 0.0001}& -1.50 & 1.56 & 95.10 & 0.0048 \\ 
  AIPW &  5-folds & {}  & 0.0013 & 0.002 & 0.880 & 0.0680 & 0.001 & 0.07 & \makecell{\textless 0.0001}& 6.70 & 1.69 & 94.39 & 0.0051 \\ 
  TMLE &  5-folds & {}  & 0.0013 & 0.001 & 0.867 & 0.0550 & 0.001 & 0.05 & \makecell{\textless 0.0001}& -1.34 & 1.56 & 94.35 & 0.0052 \\ 
  AIPW &  10-folds & {}  & 0.0024 & 0.002 & 1.629 & 0.0830 & 0.001 & 0.07 & \makecell{\textless 0.0001}& -15.00 & 1.34 & 95.05 & 0.0049 \\
\hline
\end{tabular}
\endgroup
\end{table}

\hspace*{-2.8cm}\begin{table}[h!]
\centering
\vspace{2cm}\caption{Simulation study results: Performance measures for the \textbf{Full library in SL} and sample size of \textbf{2000} ($n_{sim} = 2000$)}
\refstepcounter{SItab}\label{PM_Full_2000}
\fontsize{9pt}{9pt}\selectfont
\begingroup
\hspace*{-0.5cm}\begin{tabular}{ccccccccccccccc}
  \hline
Method & CF & Mechanism & Bias & \makecell{Bias \\ (MCSE)} & RelBias & EmpSE & \makecell{EmpSE \\ (MCSE)} & ModSE & \makecell{ModSE \\ (MCSE)} & RelModSE & \makecell{RelModSE \\ (MCSE)} & CP & \makecell{CP \\ (MCSE)} \\ 
  \hline  
  AIPW &  No CF & {}  & 0.0014 & 0.001 & 0.922 & 0.0550 & 0.001 & 0.05 & \makecell{\textless 0.0001}& -13.50 & 1.37 & 91.00 & 0.0064 \\ 
  TMLE &  No CF & {}  & 0.0015 & 0.001 & 1.015 & 0.0540 & 0.001 & 0.05 & \makecell{\textless 0.0001}& -12.13 & 1.39 & 90.95 & 0.0064 \\ 
  AIPW &  2-folds & {}  & 0.0034 & 0.002 & 2.297 & 0.0690 & 0.001 & 0.11 & \makecell{\textless 0.0001}& 59.50 & 2.52 & 95.09 & 0.0048 \\ 
  TMLE &  2-folds & Simple-1 & 0.0017 & 0.001 & 1.132 & 0.0560 & 0.001 & 0.06 & \makecell{\textless 0.0001}& 0.88 & 1.60 & 95.10 & 0.0048 \\ 
  AIPW &  5-folds & {}  & 0.0015 & 0.001 & 1.003 & 0.0550 & 0.001 & 0.06 & \makecell{\textless 0.0001}& -0.21 & 1.58 & 95.04 & 0.0049 \\ 
  TMLE &  5-folds & {}  & 0.0018 & 0.001 & 1.165 & 0.0550 & 0.001 & 0.06 & \makecell{\textless 0.0001}& 0.20 & 1.58 & 95.15 & 0.0048 \\ 
  AIPW &  10-folds & {}  & 0.0032 & 0.002 & 2.167 & 0.0710 & 0.001 & 0.06 & \makecell{\textless 0.0001}& -8.82 & 1.44 & 95.54 & 0.0046 \\
\hline
\end{tabular}
\endgroup
\end{table}

\clearpage

\section{Further results: application to the BIS case study \label{sec:si8}}

\begin{spacing}{0.5}
\vspace{10mm}
\begin{table}[h]
\centering 
\caption{BIS case study results} \label{table:BIS_case_study_complete} 
\begingroup\normalsize
\begin{tabular}{cccrr}
  \hline\\
Method & Library & {CF used} & \makecell{Partially adjusted$^*$ \\ (small confounder set)} & \makecell{Fully adjusted$^{\#}$\\ (large confounder set)} \\\\
  \hline\\\\
   AIPW & Reduced & No CF &  1.050 (-0.063,2.163) &  0.401 (-0.044,0.846) \\ 
  TMLE &  &  &  0.029 (-0.273,0.331) &  0.554 ( 0.179,0.928) \\\\
  \hline\\\\
  AIPW & Full & No CF & -0.007 (-0.260,0.245) &  0.011 (-0.388,0.409) \\ 
  TMLE &  &  &  0.026 (-0.217,0.270) &  0.170 (-0.169,0.509) \\\\ 
  \hline\\\\
  AIPW & Reduced & 2-folds & -0.037 (-0.570,0.496) &  0.408 (-0.039,0.854) \\ 
  TMLE &  &  & -0.004 (-0.532,0.524) &  0.560 ( 0.184,0.936) \\\\ 
  \hline\\\\
  AIPW & Full & 2-folds & -0.002 (-0.534,0.531) &  0.063 (-0.413,0.538) \\ 
  TMLE &  &  &  0.030 (-0.499,0.558) &  0.061 (-0.412,0.533) \\\\ 
  \hline\\\\
  AIPW & Reduced & 5-folds & -0.072 (-0.597,0.452) &  0.394 (-0.051,0.838) \\ 
  TMLE &  &  & -0.037 (-0.556,0.481) &  0.549 ( 0.174,0.924) \\\\ 
  \hline\\\\
  AIPW & Full & 5-folds &  0.095 (-0.777,0.967) &  0.136 (-0.944,1.217) \\ 
  TMLE &  &  &  0.113 (-0.753,0.978) &  0.102 (-0.577,0.780) \\\\ 
  \hline\\\\
  AIPW & Reduced & 10-folds &  0.044 (-0.477,0.565) &  0.412 (-0.035,0.858) \\ 
  TMLE &  &  &  0.072 (-0.445,0.589) &  0.564 (0.188,0.940) \\\\
  \hline\\\\
  AIPW & Full & 10-folds &  0.282 (-0.483,1.047) & -3.702 (-17.178,9.774) \\ 
  TMLE &  &  &  0.268 (-0.476,1.012) &  0.079 (-0.487,0.645) \\\\ 
   \hline

\end{tabular}

\begin{minipage}{\linewidth}
\vspace{0.5mm}\hspace{40mm}\footnotesize Results displayed are the estimated average causal effect (ACE) with accompanying 95\% CI,
\end{minipage}
\begin{minipage}{\linewidth}
\vspace{-3mm}\hspace{40mm}\footnotesize of inflammation (GlycA) in 1-year old infants on pulse wave velocity (PWV) at 4 years of age 
\end{minipage}
\begin{minipage}{\linewidth}
\vspace{-6mm}\hspace{40mm}\footnotesize (standardised), obtained by applying the methods to the BIS motivating example.
\end{minipage}
\begin{minipage}{\linewidth}
\vspace{-9mm}\hspace{40mm}\footnotesize\textsuperscript{*}Adjusted for demographic and background factors (i.e. excluding metabolomic measures).
\end{minipage}
\begin{minipage}{\linewidth}
\vspace{-12mm}\hspace{40mm}\footnotesize\textsuperscript{\#}Adjusted for demographic and metabolites.
\end{minipage}

\endgroup
\end{table}
\end{spacing}

\newpage

\begin{figure}[h]
\centering\includegraphics[scale=0.5]{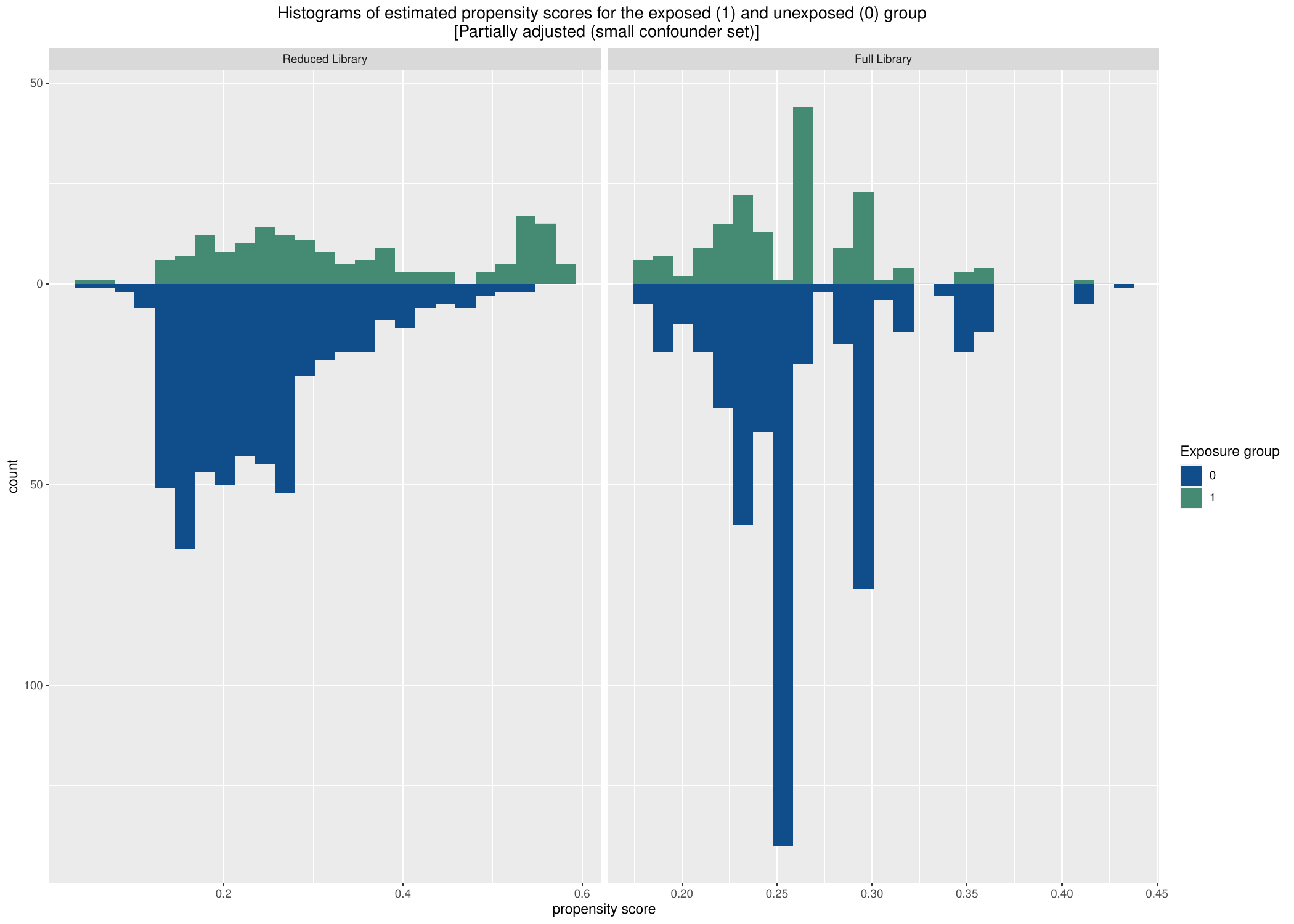}
\caption{Distribution of estimated propensity scores for the BIS case study by exposure group (exposed and unexposed), for the partially adjusted (small confounder set).}
\refstepcounter{SIfig}\label{fig:Fig_prop_small}
\end{figure}  

\begin{figure}[h]
\centering\includegraphics[scale=0.5]{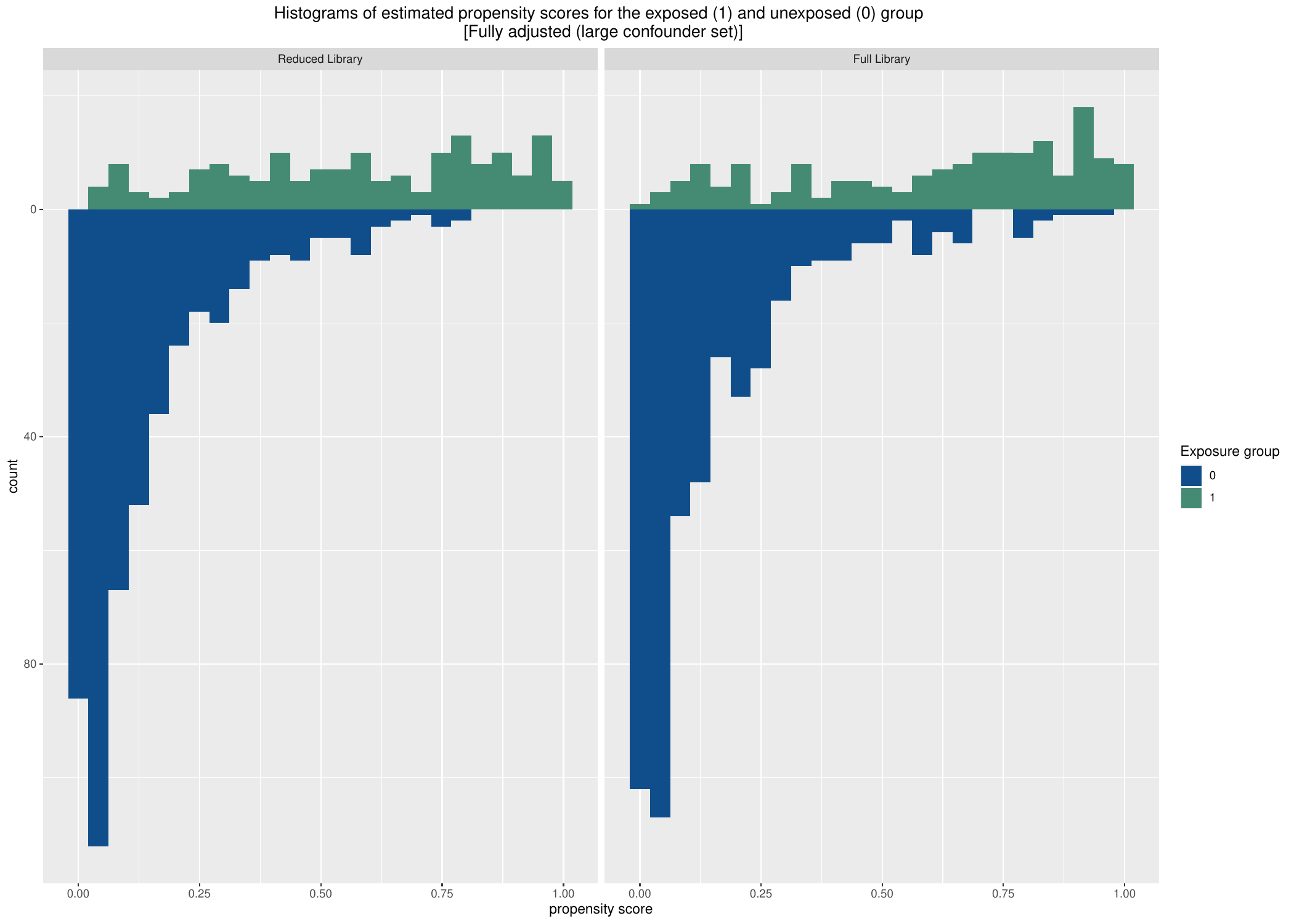}
\caption{Distribution of estimated propensity scores for the BIS case study by exposure group (exposed and unexposed), for the fully adjusted (large confounder set).}
\refstepcounter{SIfig}\label{fig:Fig_prop_large}
\end{figure}  

\clearpage

\section{Further details: SL coefficients and propensity scores \label{sec:si9}}

\begin{figure}[h]
\centering\includegraphics[scale=0.5]{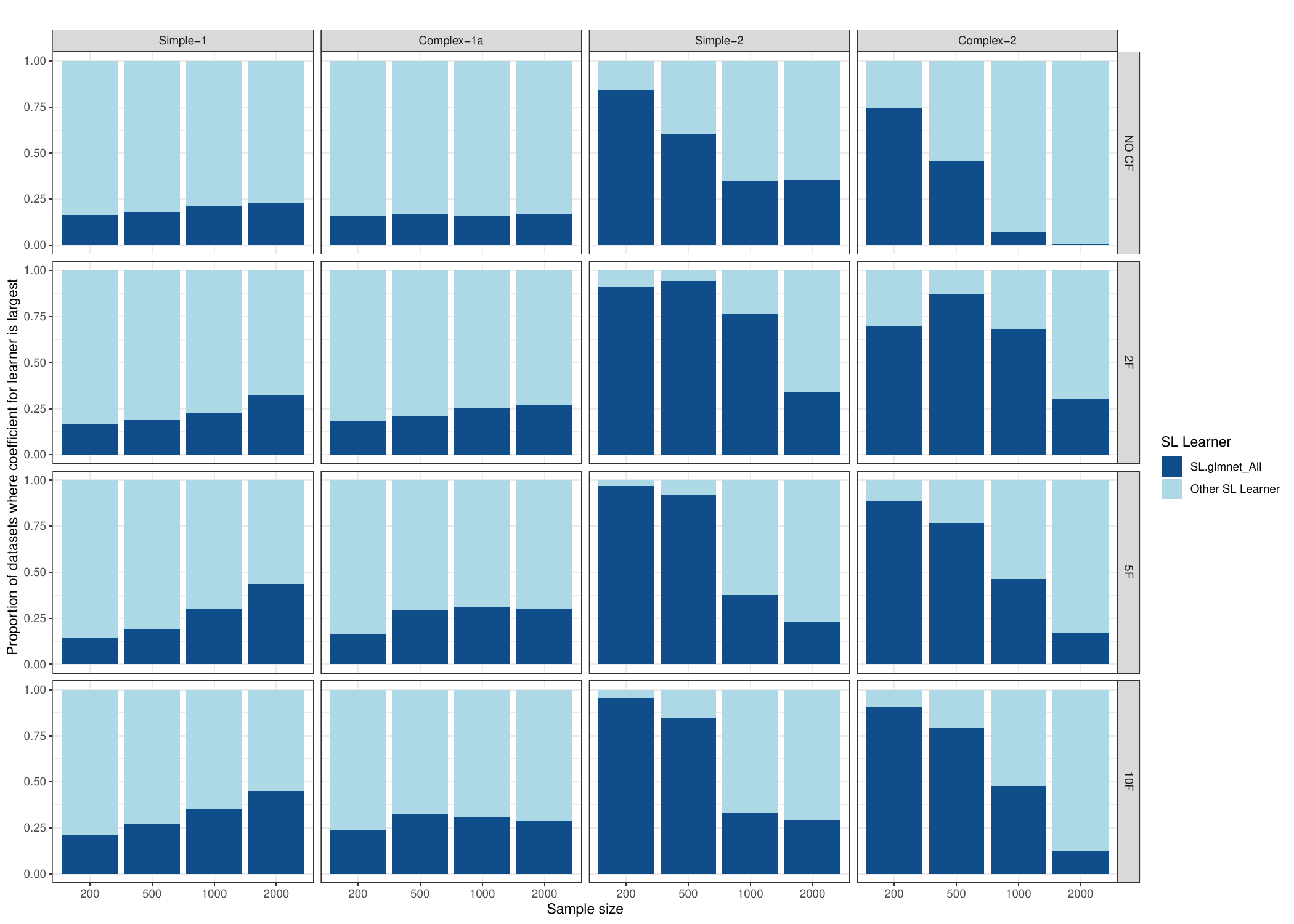}
\caption{Summary of Super Learner (SL) coefficients for the Reduced Library when using SL to obtain estimated propensity scores, across scenarios (sample size and data-generating mechanism), without and with cross-fitting (2, 5, 10 folds). Dark blue shading refers the the proportion of datasets where the glmnet Learner had the largest coefficient. Lighter blue shading indicates the proportion of datasets where a learner other than glmnet had the largest coefficient.}
\refstepcounter{SIfig}\label{fig:Fig_prop_SL_red}
\end{figure}  

\begin{figure}[h]
\centering\includegraphics[scale=0.5]{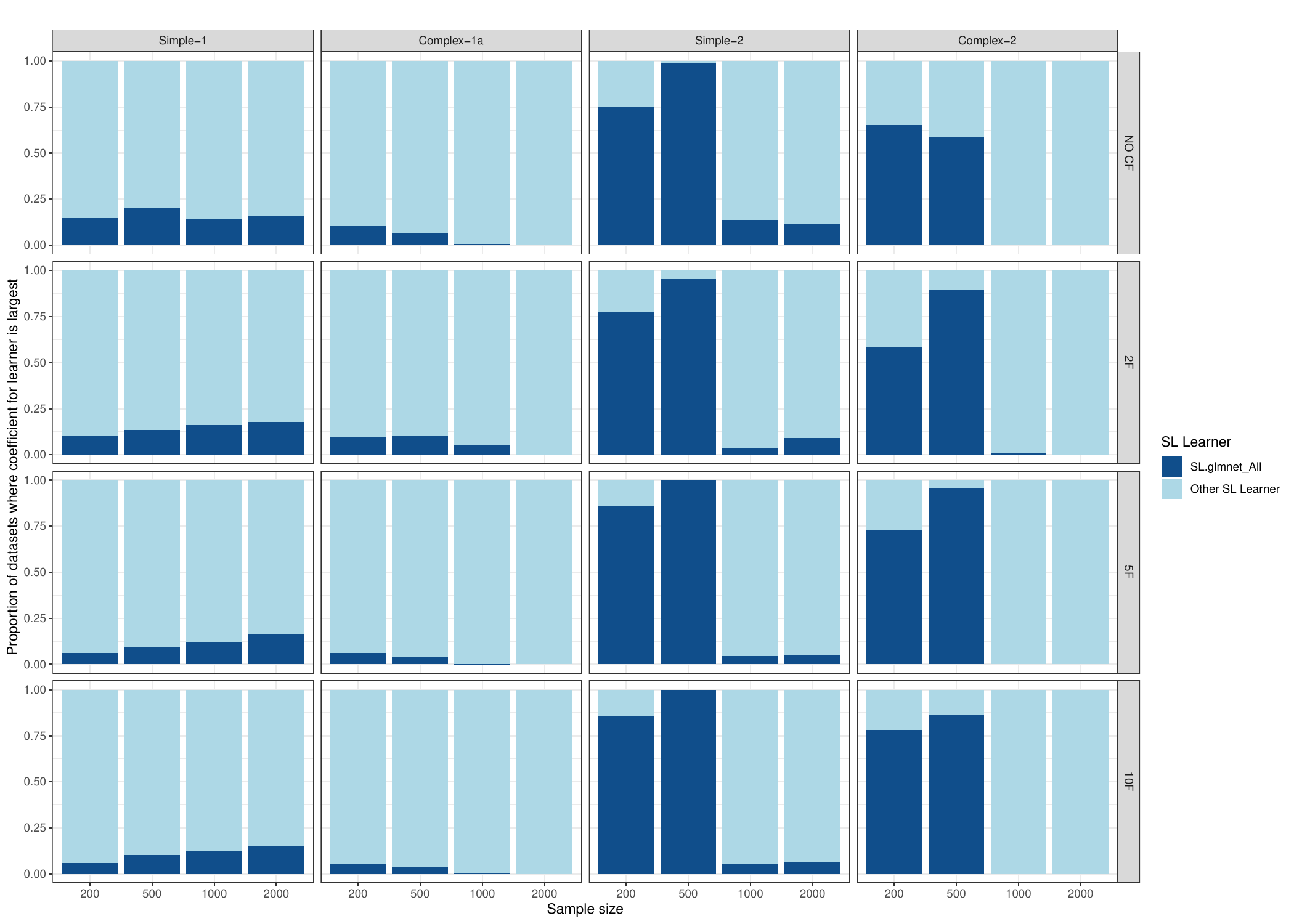}
\caption{Summary of Super Learner (SL) coefficients for the Full Library when using SL to obtain estimated propensity scores, across scenarios (sample size and data-generating mechanism), without and with cross-fitting (2, 5, 10 folds). Dark blue shading refers the the proportion of datasets where the glmnet Learner had the largest coefficient. Lighter blue shading indicates the proportion of datasets where a learner other than glmnet had the largest coefficient.}
\refstepcounter{SIfig}\label{fig:Fig_prop_SL_full}
\end{figure}  

\begin{figure}[h]
\centering\includegraphics[scale=0.5]{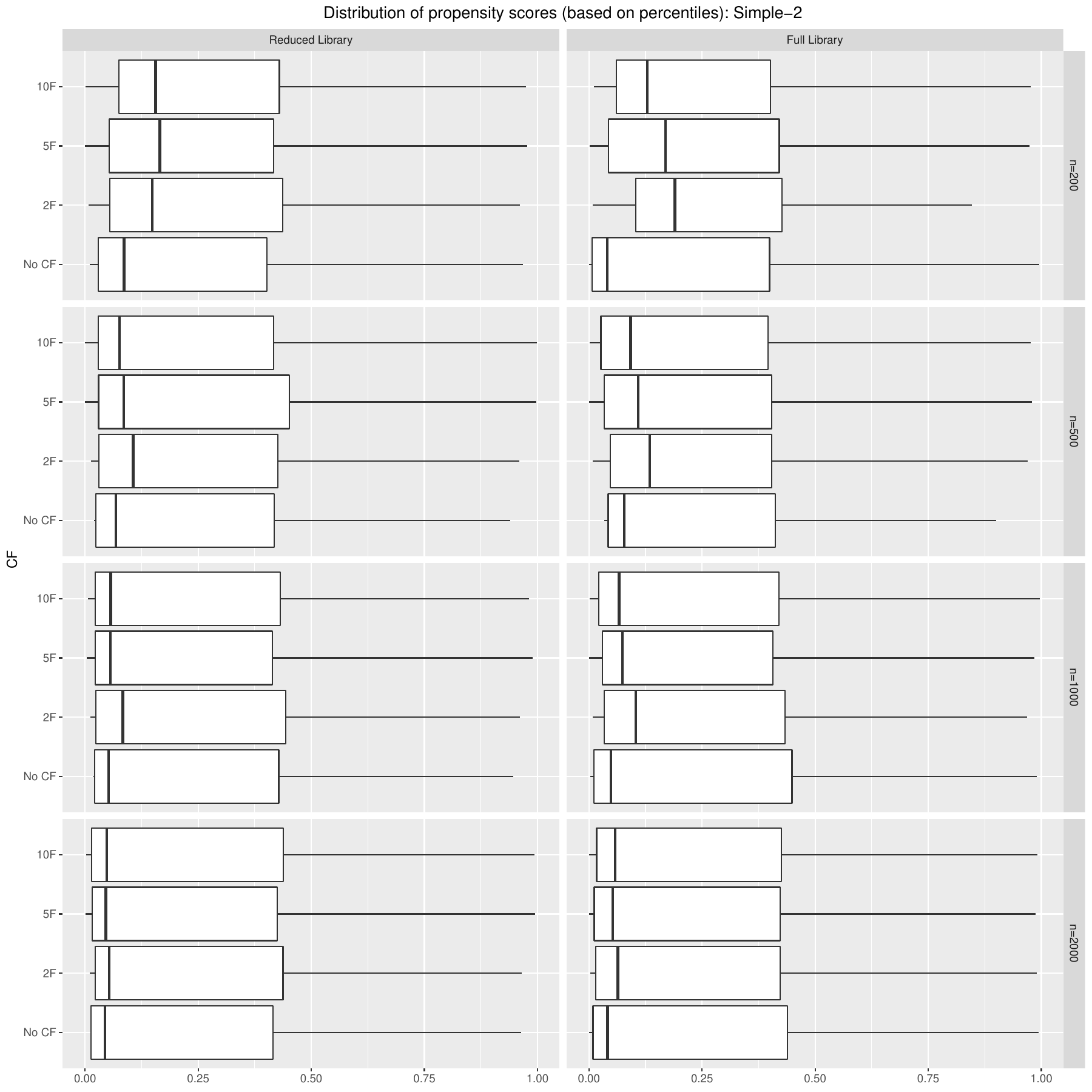}
\caption{Example of the propensity score distribution for Simple-2 by library and sample size, without and with cross-fitting (2,5,10 folds). Box plots in the figure were plotted based on given percentiles of estimated propensity scores.}
\refstepcounter{SIfig}\label{fig:Fig_prop_simple2}
\end{figure}  

\begin{figure}[h]
\centering\includegraphics[scale=0.5]{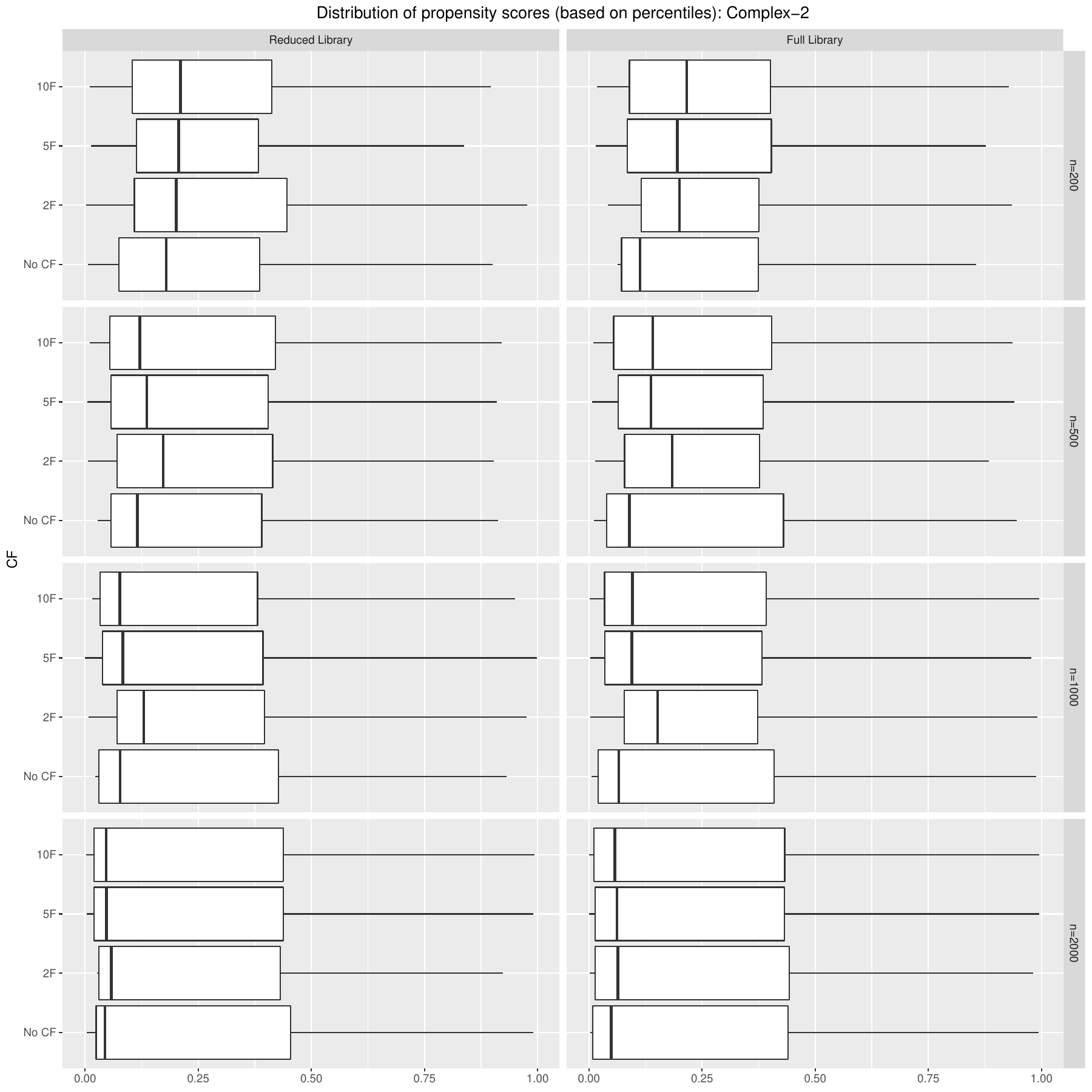}
\caption{Example of the propensity score distribution for Complex-2 by library and sample size, without and with cross-fitting (2,5,10 folds). Box plots in the figure were plotted based on given percentiles of estimated propensity scores.}
\refstepcounter{SIfig}\label{fig:Fig_prop_Complex2}
\end{figure}  

\clearpage

\section{Further exploration: sensitivity and dependence on seeds in cross-fitting \label{sec:si10}}

\begin{figure}[h]
\centering\includegraphics[scale=0.5]{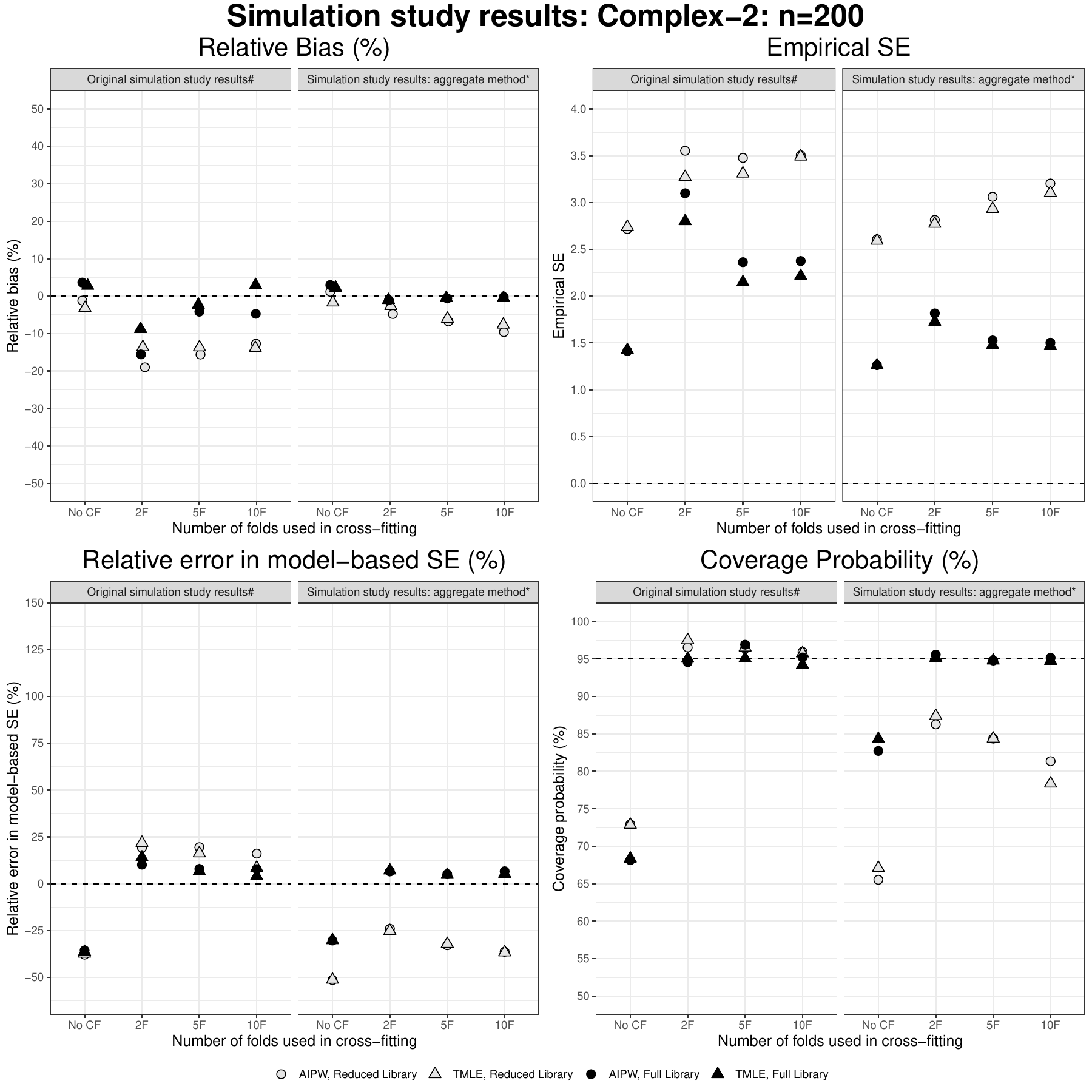}
\caption{Simulation study results for the relative bias of point estimates (\%), Empirical SE, bias in model-based SE (\%), and coverage probability (\%) for AIPW and TMLE with varying use of cross-fitting for complex-2 with sample size of n=200 (and using 2000 simulated datasets). \#Original simulation study results reported in the main paper. *Results provided are based on 50 replications (different sample splits), where estimates for the methods were obtained across the replications using the median method (Chernozhukov et al.$^{23}$). I.e. point estimates were obtained using $\tilde{\theta}_0^{median} = \text{median}\{\tilde{\theta}_0^s\}_{s=1}^{50}$ and variance obtained using $\hat{\sigma}^{2,median} = \text{median}\{\hat{\sigma}_s^2 + (( \hat{\sigma}_s - \tilde{\theta}_{median}) ( \hat{\sigma}_s - \tilde{\theta}_{median} )' ) \}_{s=1}^{50}$, where $\tilde{\theta}_0^s \text{ and } \hat{\sigma}_s^2 \text{ represent the point and variance estimates obtained for sample split } s \, (s=1,....,50) \text{ respectively}$. The performance measures were then calculated as usual.}
\refstepcounter{SIfig}\label{fig:Fig_200_complex_2_sim_seed_comp}
\end{figure}  

\begin{figure}[h]
\centering\includegraphics[scale=0.5]{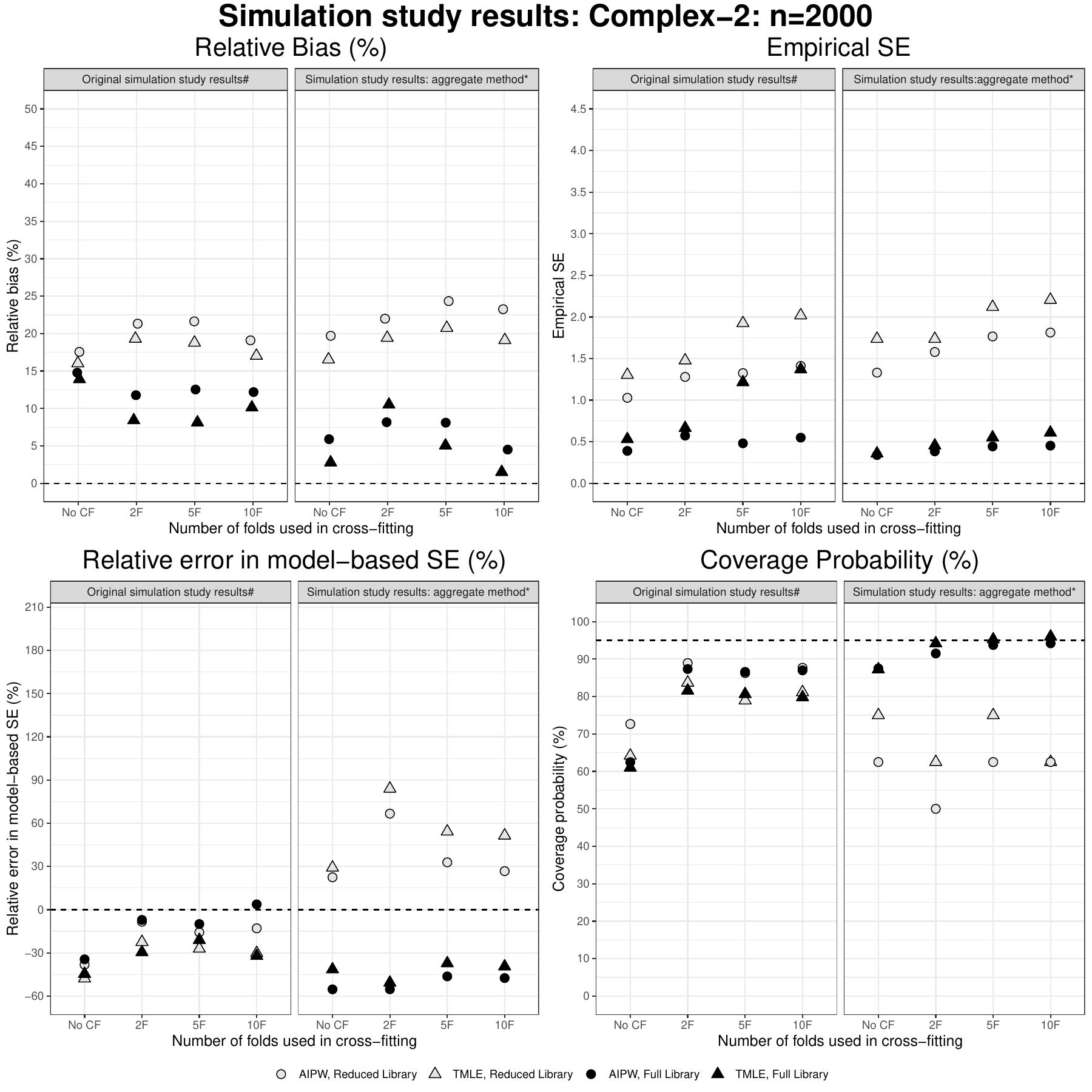}
\caption{Simulation study results for the relative bias of point estimates (\%), Empirical SE, bias in model-based SE (\%), and coverage probability (\%) for AIPW and TMLE with varying use of cross-fitting for complex-2 with sample size of n=2000 (and using 2000 simulated datasets). \#Original simulation study results reported in the main paper. *Results provided are based on 50 replications (different sample splits), where estimates for the methods were obtained across the replications using the median method (Chernozhukov et al.$^{23}$). I.e. point estimates were obtained using $\tilde{\theta}_0^{median} = \text{median}\{\tilde{\theta}_0^s\}_{s=1}^{50}$ and variance obtained using $\hat{\sigma}^{2,median} = \text{median}\{\hat{\sigma}_s^2 + (( \hat{\sigma}_s - \tilde{\theta}_{median}) ( \hat{\sigma}_s - \tilde{\theta}_{median} )' ) \}_{s=1}^{50}$, where $\tilde{\theta}_0^s \text{ and } \hat{\sigma}_s^2 \text{ represent the point and variance estimates obtained for sample split } s \, (s=1,....,50) \text{ respectively}$. The performance measures were then calculated as usual.}
\refstepcounter{SIfig}\label{fig:Fig_2000_complex_2_sim_seed_comp}
\end{figure}  

\begin{figure}[h]
\centering\includegraphics[scale=0.5]{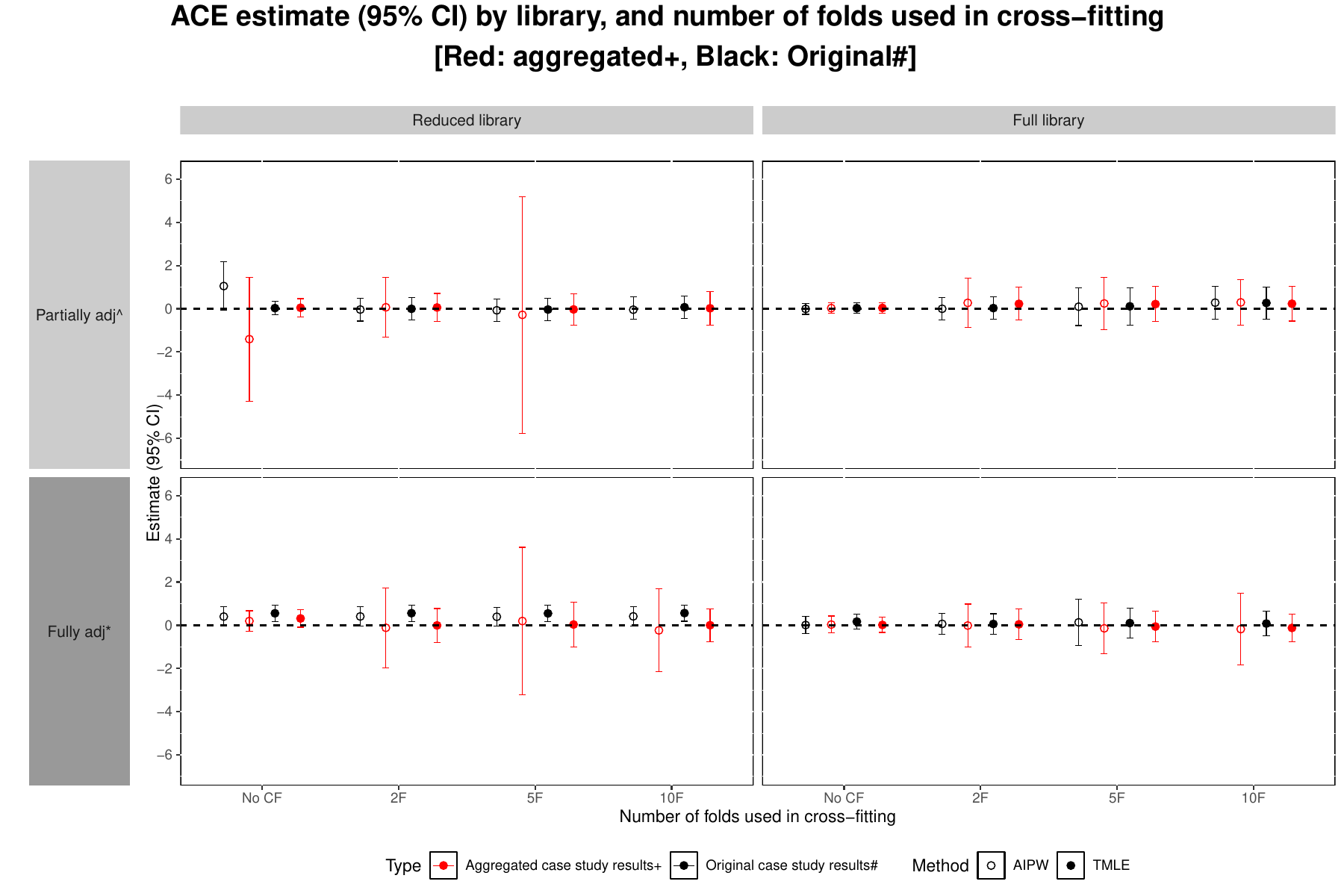}
\caption{Estimated average causal effect (ACE) with accompanying $95\%$ CI, of inflammation (GlycA) in 1-year old infants on Pulse Wave velocity (PWV) at 4 years of age (standardised), obtained by applying the methods to the BIS motivating example. $\hat{}$ demographic and background confounders only; $^*$demographic, background, and metabolomic confounders. \\ Note: Original estimate for AIPW using fully adjusted confounder set and full library in SL with 10 fold cross-fitting not shown in figure as too large to be sensible (refer to Table 35 in Section 8 of the Supporting Information). \#Original results are presented in black. $^+$Results presented in red are based on 50 replications (different sample splits), where estimates for the methods were obtained across the replications using the median method (Chernozhukov et al.$^{23}$). I.e. point estimates were obtained using $\tilde{\theta}_0^{median} = \text{median}\{\tilde{\theta}_0^s\}_{s=1}^{50}$ and variance obtained using $\hat{\sigma}^{2,median} = \text{median}\{\hat{\sigma}_s^2 + (( \hat{\sigma}_s - \tilde{\theta}_{median}) ( \hat{\sigma}_s - \tilde{\theta}_{median} )' ) \}_{s=1}^{50}$, where $\tilde{\theta}_0^s \text{ and } \hat{\sigma}_s^2 \text{ represent the point and variance estimates obtained for sample split } s \, (s=1,....,50) \text{ respectively}$.}
\refstepcounter{SIfig}\label{fig:Fig_CS_seed_depend}
\end{figure}

\end{document}